\documentclass[11pt,a4paper,twoside]{article}
\usepackage{graphicx} % Required for inserting images

\usepackage[headheight=25mm,top=30mm,inner=30mm,outer=30mm,marginparwidth=50pt,centering]{geometry}

\usepackage{amsmath,amsfonts,mathrsfs,bbm}
\usepackage{amssymb}
\usepackage{cancel}
\usepackage{slashed}
\usepackage{braket}
\usepackage[normalem]{ulem}

\usepackage{tabularray}
\usepackage{makecell}

\usepackage{comment}
\usepackage{color}
\usepackage[colorlinks=true]{hyperref}
\hypersetup{
    breaklinks=true,
    colorlinks,
    linkcolor={black},
    citecolor={black},
    urlcolor={black}
}

\numberwithin{equation}{section} 

\usepackage{tikz} 
\usetikzlibrary{calc}
\usetikzlibrary{arrows}
\usetikzlibrary{trees}
\usetikzlibrary{decorations.pathmorphing}
\usetikzlibrary{decorations.markings}

\newcommand{\Ad}{\operatorname{Ad}}

\begin{document}

\vspace{36pt}

\begin{center}
{\huge{\bf Jordanian deformation of the non-compact and $\mathfrak{sl}_2 $-invariant 
\\
\vspace{8pt}
 $XXX_{-1/2}$ spin-chain}} 

\vspace{36pt}

Riccardo Borsato and Miguel Garc\'ia Fern\'andez

\vspace{24pt}

{
\small {\it 
Instituto Galego de F\'isica de Altas Enerx\'ias (IGFAE),\\[2pt]
and Departamento de F\'\i sica de Part\'\i culas,\\[2pt]
Universidade de  Santiago de Compostela,\\[2pt]
15705 Santiago de Compostela,  Spain\\[4pt]}
\vspace{12pt}
\texttt{riccardo.borsato@usc.es}, \qquad \texttt{miguelg.fernandez@usc.es}}\\

\vspace{36pt}

{\bf Abstract}
\end{center}
\noindent
Using a Drinfeld twist of Jordanian type, we construct a deformation of the non-compact and $\mathfrak{sl}_2$-invariant $XXX_{-1/2}$ spin-chain. Before the deformation, the seed model can be understood as a sector of the $\mathfrak{psu}(2,2|4)$-invariant spin-chain encoding the spectral problem of $\mathcal{N}=4$ super Yang-Mills at one loop in the planar limit. The deformation gives rise to interesting features because, while being integrable, the Hamiltonian is non-hermitian and non-diagonalisable, so that it only admits a Jordan decomposition. Moreover, the eigenvalues of the deformed Hamiltonian coincide with those of the original undeformed spin-chain. We use explicit examples as well as the techniques of the coordinate and of the algebraic Bethe ansatz to discuss the construction of the (generalised) eigenvectors of the deformed model. We also show that the deformed spin-chain is equivalent to an undeformed one with twisted boundary conditions, and that it may be derived from a scaling limit of the non-compact $U_q(\mathfrak{sl}_2)$-invariant $XXZ_{-1/2} $ spin-chain.

\newpage 

%\baselineskip=24pt
%\pagebreak 

\tableofcontents

%%%%%%%%%%%%%%%%%%%%%%%%%%%%%%%%%%%%%%%%%%%%%%%%%%%%%%%%%%%%%%%%%%%%%%%%%%%%%%%%
%%%%%%%%%%%%%%%%%%%%%%%%%%%%%%%%%%%%%%%%%%%%%%%%%%%%%%%%%%%%%%%%%%%%%%%%%%%%%%%%
\section{Introduction}

Integrable spin-chains were initially studied to understand ferromagnetic materials~\cite{Bethe:1931hc}, but in more recent years they proved to have important applications also to describe the planar spectrum of gauge theories appearing in the AdS/CFT correspondence (see~\cite{Beisert:2010jr} for a review), and  even for applications in quantum computing (see for example~\cite{Dyke:2021vkq,VanDyke:2021nuz,Li:2022czv,Sopena:2022ntq,Ruiz:2023rew,Ruiz:2024lsl,Ruiz:2024mei}).

In this paper, we study an integrable spin-chain that is a deformation of a nearest-neighbour $XXX$ Hamiltonian, which is invariant under the $\mathfrak{sl}_2$ symmetry algebra. Although the original paper of Bethe~\cite{Bethe:1931hc} deals with spins transforming in the $1/2$ representation of $\mathfrak{sl}_2$ (i.e.~at each site the states can  only be ``spin-up'' or ``spin-down''), one can generalise the construction to higher-spin representations~\cite{Faddeev:1994nk}. The starting point of our construction is the \emph{non-compact} $\mathfrak{sl}_2$ spin-chain, where each site is a state in the infinite dimensional $j=-1/2$ representation (see e.g.~\cite{Faddeev:1994zg} and~\cite{Hao:2019cfu}). The reason to consider this case is two-fold. On the one hand, this non-compact spin-chain appears as a sector of the $\mathfrak{psu}(2,2|4)$-invariant spin-chain that encodes the spectral problem of $\mathcal N=4$ super Yang-Mills at one loop in the planar limit~\cite{Beisert:2003jj,Beisert:2003yb}. On the other hand,  on this spin-chain we can introduce a deformation parameter via a Drinfeld twist of ``Jordanian'' type.

Interestingly, Drinfeld twists can be used to introduce deformation parameters in quasi-triangular Hopf algebras~\cite{drinfeld1983constant} (see also~\cite{Reshetikhin:1990ep} and~\cite{Giaquinto:1994jx}). In particular, given a quasi-triangular Hopf algebra where $R$ satisfies the quantum Yang-Baxter equation, the application of a Drinfeld twist allows one to generate another quasi-triangular Hopf algebra with a new $\tilde R$ also satisfying the quantum Yang-Baxter equation. Drinfeld twists can be applied also at the level of the Hamiltonian $H$ of the spin-chain, to give rise to a new Hamiltonian $\tilde H$ that is integrable by construction. We are interested in twists that are continuously connected to the identity via a deformation parameter, so that $\tilde H$ can be interpreted as a deformation of $H$. If  we call $\mathfrak{g}$ the Lie algebra of the Hopf algebra, Drinfeld proved that (equivalence classes of) twists are in one-to-one correspondence with antisymmetric solutions of the classical Yang-Baxter equation on $\mathfrak{g}$~\cite{drinfeld1983constant} (see also~\cite{Giaquinto:1994jx}). 

When the symmetry algebra $\mathfrak{g}$ under which $H$ is invariant is compact, the only available twists are of ``abelian'' type, and they are often called ``Drinfeld-Reshetikhin'' twists~\cite{Reshetikhin:1990ep}. In these cases, the twist is constructed from generators that close into an abelian subalgebra of $\mathfrak{g}$. More options are available for non-compact Lie algebras, and Jordanian twists~\cite{10.1007/BFb0101176,Ogievetsky:1992ph,Kulish_1997} are perhaps the simplest and the most studied class of ``non-abelian'' twists. They can be applied every time $\mathfrak{g}$ has a subalgebra spanned by generators $h,e$ satisfying $[h,e]=e$, but ``extended Jordanian'' twists exist for higher-rank Lie algebras~\cite{Kulish:1998be,Kulish:1999ua}, and even twists involving generators of odd grading in superalgebras~\cite{tolstoy2004chainsextendedjordaniantwists}.

One of our main motivations is the construction of integrable deformations in the context of the AdS/CFT correspondence. In recent years, a family of ``homogeneous Yang-Baxter'' deformations of sigma-models was constructed~\cite{Kawaguchi:2014qwa,vanTongeren:2015soa} (see also~\cite{Klimcik:2002zj,Klimcik:2008eq,Delduc:2013qra}). They can be used to deform, for example, the $AdS_5\times S^5$ superstring background that is holographically dual to $\mathcal N=4$ super Yang-Mills. The deformed sigma-models in this family are classified by antisymmetric solutions of the classical Yang-Baxter equation, where for example $\mathfrak{g}=\mathfrak{psu}(2,2|4)$ when deforming $AdS_5\times S^5$. It is then natural to expect that their manifestation on the other side of the holographic duality should be implemented via Drinfeld twists~\cite{vanTongeren:2015uha}, see also~\cite{vanTongeren:2016eeb,Araujo:2017jkb,Araujo:2017jap,Meier:2023kzt,Meier:2023lku}. Some of the deformations in this family have a known holographic interpretation that confirms this expectation, most notably the Lunin-Maldacena deformation that predates the more general construction of Yang-Baxter deformations, and that is dual to the $\beta$-deformation of the gauge theory~\cite{Lunin:2005jy}. But many options are still mysterious from the point of view of AdS/CFT, and Jordanian deformations are among them. Our strategy to shed some light on this is to construct a Jordanian deformation of the non-compact $\mathfrak{sl}_2$ spin-chain, with the aim of getting insights on possible Jordanian deformations of the gauge theory. The work in this paper is a first step in that direction, because the construction performed here should correspond to the deformation of a sector of the full spin-chain, which we hope to address in a future work.

Jordanian and other kinds of Yang-Baxter deformations have attracted a lot of attention because of the possibility to deform AdS/CFT while preserving integrability, opening then the door to integrable deformations of $\mathcal N=4$ super Yang-Mills in the planar limit while breaking conformal invariance and supersymmetry. Some attempts to understand Jordanian deformations in the context of the AdS/CFT correspondence may be found in~\cite{vanTongeren:2015uha}. Jordanian deformations of $AdS_5\times S^5$ were classified in~\cite{Borsato:2022ubq}, where an explicit list of options can be found. Thanks to semiclassical integrability techniques and to the reformulation of the deformed sigma models as twisted models~\cite{Borsato:2021fuy}, a Jordanian deformation of $AdS_5\times S^5$ was studied in~\cite{Borsato:2022drc}. Jordanian deformations were also used to regularise the spectral curves of other kinds of deformations~\cite{Driezen:2024mcn}. 

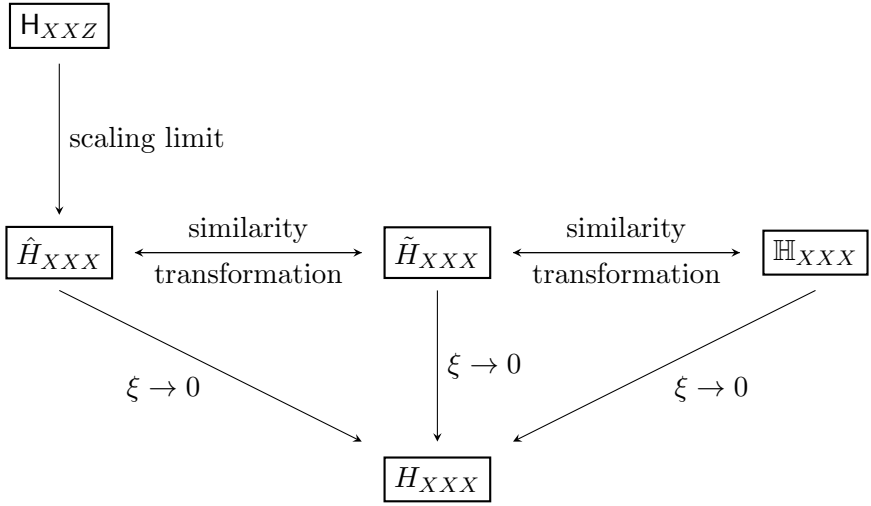
\begin{figure}
    \centering
  \begin{tikzpicture}
%    \fill[gray!20] (0,0) -- (2,2) -- (3,0) -- (2,-2) -- cycle;
%    \fill[gray!20] (0,0) -- (-2,2) -- (-3,0) -- (-2,-2) -- cycle;
 %   \draw [-stealth](-3,0) -- (3,0) node[below right]{$x$};
  %  \draw [-stealth](0,-2) -- (0,2) node [right]{$t$};
   % \draw[thick] (-2,-2) -- (2,2);
    %\draw[thick] (-2,2) -- (2,-2);
    \draw (-5,3) node[shape=rectangle,draw,thick] {$\mathsf H_{XXZ}$};
    \draw (0,0) node[shape=rectangle,draw,thick] {$\tilde H_{XXX}$};
    \draw (5,0) node[shape=rectangle,draw,thick] {$\mathbb H_{XXX}$};
    \draw (-5,0) node[shape=rectangle,draw,thick] {$\hat H_{XXX}$};
    \draw (0,-3) node[shape=rectangle,draw,thick] {$H_{XXX}$};
    \draw [-stealth](-5,2.5) --node[right]{scaling limit} (-5,0.5) ;
    \draw [-stealth](-4,0) -- node[above]{similarity} (-1,0) ;
    \draw [-stealth](-1,0) -- node[below]{transformation} (-4,0) ;
    \draw [-stealth](1,0) -- node[above]{similarity} (4,0) ;
    \draw [-stealth](4,0) -- node[below]{transformation} (1,0) ;
    \draw [-stealth](0,-0.5) --node[right]{$\xi\to 0$} (0,-2.5) ;
    \draw [-stealth](5,-0.5) --node[below right]{$\xi\to 0$} (1,-2.5) ;
    \draw [-stealth](-5,-0.5) --node[below left]{$\xi\to 0$} (-1,-2.5) ;
\end{tikzpicture}
    \caption{Diagram summarising the relations between the models considered in this paper. $\mathsf H_{XXZ}$ is the Hamiltonian of the $XXZ$ spin-chain which, after a scaling limit, gives rise to the deformed Hamiltonian $\hat H_{XXX}$. This is equivalent through a similarity transformation to another deformed Hamiltonian $\tilde H_{XXX}$ and to an undeformed yet twisted Hamiltonian $\mathbb H_{XXX}$. All these reduce to the undeformed and periodic Hamiltonian $H_{XXX}$ when sending the parameter $\xi\to 0$.}
    \label{fig:diagram}
\end{figure}

In this paper we use a Jordanian twist to construct the Hamiltonian $\tilde H_{XXX}$ (the middle node in the diagram of Figure~\ref{fig:diagram}) for the periodic spin-chain that reduces to the undeformed and periodic Hamiltonian $H_{XXX}$ (the lowest node in the diagram) when the deformation parameter $\xi$ is sent to zero. We will also show that $\tilde H_{XXX}$ is equivalent to the Hamiltonian $\mathbb H_{XXX}$ (right node) of an \emph{undeformed} spin-chain  with \emph{twisted} boundary conditions. The equivalence is shown by relating the two Hamiltonians by an explicit similarity transformation. At the same time, we will show that there is a similarity transformation also with the Hamiltonian $\hat H_{XXX}$ (left node). This can be obtained from a scaling limit of the Hamiltonian $\mathsf H_{XXZ}$ (highest node) of the $XXZ$ spin-chain that is invariant under the $U_q(\mathfrak{sl}_2)$ symmetry.

To avoid confusion, in the rest of the paper we will always refer to $\tilde H_{XXX}$, $\hat H_{XXX}$ and the corresponding models as being ``deformed''. The reason is that their Hamiltonian densities have an explicit dependence on the deformation parameter, while the boundary conditions on the chain are taken to be periodic. The Hamiltonian $\mathbb H_{XXX}$ and the corresponding model, instead, will be referred to as ``twisted''. In that case the Hamiltonian density is the \emph{undeformed} one, and the dependence on the deformation parameter only enters through the \emph{twisted} boundary conditions.

Jordanian twists for the $\mathfrak{sl}_2$-invariant spin-chain in the $j=+1/2$ \emph{finite} dimensional representation were already considered, see in particular in~\cite{Kulish_1997,Kulish2009}. We do not  work with complex Lie algebras, and we must pay attention to the real form of $\mathfrak{g}$ to ensure the unitarity of the deformed string sigma models and gauge theories corresponding to the deformation of the spin-chain. Nevertheless, our results qualitatively mirror the outcomes of the study for the case of the $j=1/2$ representation, see in particular~\cite{Kulish2009}. On top of the relation between the deformed, twisted and $q$-deformed model, we also find that because of the Jordanian deformation  the Hamiltonian $\tilde H_{XXX}$ ceases to be diagonalisable. It admits, instead, a Jordan block decomposition. Moreover, the eigenvalues remain those of the undeformed Hamiltonian $H_{XXX}$, and the only effect of the deformation is in the expressions for the (generalised) eigenvectors. Working out explicitly these results in the non-compact case is technically more difficult compared to the compact case. In fact, calling $J^3,J^\pm$ the $\mathfrak{sl}_2$ generators, in the case of $j=+1/2$ one can work with a $2\times 2 $ matrix realisation and drastically simplify formulas using the fact that for $J^-=\sigma^-$ one has $(\sigma^-)^2=0$. Similarly, for higher-spin representations of dimension $n$ one may still use  that $(J^-)^n=0$ on the representation. For the non-compact case, instead, one does not have such simplifications.

The paper is organised as follows. In section~\ref{sec:undef-sp-ch} we review some material on the non-compact $\mathfrak{sl}_2 $ $XXX_{-1/2}$ spin-chain, to introduce our notation and fix our conventions. In section~\ref{sec:Jord-sp-ch}, after writing the Jordanian twist in our conventions, we construct the deformed spin-chain and we show that it is equivalent to an undeformed one with twisted boundary conditions. In section~\ref{sec:examples} we present some explicit examples of spin-chains of length $L=2$ with up to $N=2,3$ excitations, working out the generalised eigenvectors and  eigenvalues. In section~\ref{section:CBA} we apply the method of the coordinate Bethe ansatz to generalise the results observed on the explicit examples, and to discuss  the Jordan-block decomposition of the Hamiltonian. In section~\ref{sec:XXZ} we obtain a Jordanian deformation from a scaling limit of the $XXZ$ spin-chain, and we show that the scaling limit is also able to reproduce the desired (generalised) eigenvectors. In section~\ref{sec:ABA} we apply the method of the algebraic Bethe ansatz to the Jordanian spin-chain. In section~\ref{sec:gen-arbitrary-spin} we explain how to generalise the deformed spin-chain to arbitrary negative values of the spin. Finally, in section~\ref{sec:conclusions} we end with some conclusions.

\section{The non-compact $\mathfrak{sl}_2 $ $XXX_{-1/2}$ spin-chain}\label{sec:undef-sp-ch}
The $XXX_{-1/2}$ Heisenberg spin-chain is a quantum mechanical model of interacting spins located at fixed sites along a 1-dimensional chain of length $L$, and with an $\mathfrak{sl}_2$ symmetry algebra in the infinite dimensional $j=-1/2$ representation. Before introducing the Hamiltonian of the model, let us review some basic concepts about $\mathfrak{sl}_2$ and its representations. We will adopt the conventions of~\cite{Beisert:2003jj}.

Consider the $\mathfrak{sl}_2$ Lie algebra spanned by the basis generators $\{J^3,J^\pm\}$, subject to the commutation relations 
\begin{align}
    [J^+,J^-]=-2J^3, \quad [J^3,J^\pm]=\pm J^\pm. \label{eq:sl2}
\end{align}
The quadratic Casimir element is given by
\begin{align}
    C = \left(J^3\right)^2-\frac{1}{2}\{J^+,J^-\}. \label{eq:sl2-casimir}
\end{align}
The representations of $\mathfrak{sl}_2$ are labelled by the number $j$,  implicitly defined by the Dynkin label $[2j]$ of the representation. We consider the  infinite dimensional representation $j=-1/2$, which can be realised by introducing a set of bosonic oscillators $a, a^\dagger$ with the commutation relation $[a,a^\dagger]=1$, such that
\begin{align}
    J^3 = \frac{1}{2}+a^\dagger a, \quad J^+ = a^\dagger + a^\dagger a^\dagger a, \quad J^- = a.
\end{align}
The module $V_F$, on which the representation acts, is spanned by all states constructed by applying an arbitrary number of times the raising generator $J^+$ to the lowest-weight state $\ket{0}$,
\begin{align}
    V_F = \text{span}\{\left(a^+\right)^n \ket{0},\ n=0,1,\ldots \}, \quad \text{where} \quad a \ket{0} = 0. \label{eq:fundamental-module}
\end{align}
We will be interested in the tensor product of two modules $V_F$, on which the algebra acts via the coproduct  that we take to be the trivial one
\begin{align}
    \Delta(J) = J \otimes \mathbb{I} + \mathbb{I} \otimes J, \quad J \in \{J^3,J^\pm\}. \label{eq:sl2-coproduct}
\end{align}
Under this convention, the irreducible decomposition of two modules $V_F $ is given by
\begin{align}
    V_F \otimes V_F = \bigoplus_{j=0}^\infty V_j, \label{eq:sl2-irr-decom}
\end{align}
where $V_j$ is a module with lowest-weight state  $\ket{\Phi_j}$, such that
\begin{align}
    \Delta(J^-) \ket{\Phi_j} &= 0, \\
    \Delta(C) \ket{\Phi_j} &= j (j+1) \ket{\Phi_j}.
\end{align}
An explicit calculation shows that 
\begin{align}
    \ket{\Phi_j} = (a_1^\dagger-a_2^\dagger)^j \ket{00} ,
\end{align}
where the  sub-indices denote on which site of the tensor product the operators act.

In the $XXX_{-1/2}$ chain, each site is taken to be the module $V_F$, which implies that the Hilbert space of the model is $\bigotimes_{m=1}^L \left(V_F\right)_m$. Let us now specify the dynamics. The interactions between the spins are of nearest-neighbour type, so that the Hamiltonian is constructed out of operators that have support only on consecutive sites
\begin{align}
    H_{XXX} = \sum_{m=1}^L h_{m,m+1} \label{eq: xxx-h12},
\end{align}
where the subindices indicate the sites on which $h_{m,m+1}$ is acting.
We will be interested in imposing periodic boundary conditions, that are implemented by the identification $h_{L,L+1}=h_{L1}$. The Hamiltonian density $h_{m,m+1}$ is diagonal on each module $V_j$ 
\begin{align}
    h_{12}=\sum_{j=0}^\infty 2h(j) P_{12,j}, \label{eq:xxx-ham-density}
\end{align}
where the eigenvalues are given by the harmonic numbers $h(j)=\sum_{k=1}^j \frac{1}{k}$ and $P_{12,j}$ denotes the projector onto $V_j$ acting on the sites $1$ and $2$.\footnote{We specify $m=1$ and write formulas for $h_{12}$ rather than $h_{m,m+1}$ just to simplify the notation. The generalisation is obvious.} These projectors can be represented as\footnote{It is straightforward to see that $P_{12,j}V_l=0$ if $j\neq k$ because there is a zero in the product coming from $(\Delta(J^2)-l(l+1))V_l=0$. When acting on $V_j$, instead, $P_{12,j}$ clearly acts as the identity operator.}
\begin{align}
    P_{12,j}=\underset{k\neq j}{\prod_{k=0}^\infty} \frac{\Delta(C)-k(k+1)}{j(j+1)-k(k+1)}.
\end{align}
In terms of the bosonic oscillators defined above, the Casimir acting on two sites can be represented by
\begin{align}
    \Delta(C) = -\left(a_1^\dagger - a_2^\dagger\right)^2 a_1 a_2 + \left(a_1^\dagger - a_2^\dagger\right)\left(a_1 - a_2\right). \label{eq:sl2-Corpoduct-Casimir}
\end{align}
In~\cite{Beisert:2003jj} it was found that the action of the Hamiltonian density onto the tensor product of two modules $V_F$ takes the form
\begin{align}
    h_{12} \ket{n_1,n_2} = \sum_{k=0}^{n_1+n_2}\left(\delta_{kn_1}\left( h(n_1) + h(n_2) \right) - \frac{1-\delta_{kn_1}}{|n_1 - k|}\right) \ket{k,n_1+n_2-k},
\end{align}
where $\ket{n,m}=(a_1^\dagger)^{n} (a_2^\dagger)^{m}\ket{0,0}$.
Notice that the action of the Hamiltonian density maintains the total number of excitations $n_1+n_2$ fixed---in other words, it commutes with the number operator $N_{12}=a_1^\dagger a_1+a_2^\dagger a_2$---and it distributes them in all possible combinations. 

The $XXX_{-1/2}$ Hamiltonian turns out to be integrable, and to prove it one may construct an explicit $R$-matrix. Let  the $\mathfrak{sl}_2$ $R$-matrix in the spin $j=-1/2$  representation be given by~\cite{Beisert:2003yb}
\begin{align}
     R_{12}(u) = \sum_{j=0}^\infty R_j(u) P_{12,j}, \quad \text{with} \quad R_j(u) = (-1)^{j+1} \frac{\Gamma(-j+u)}{\Gamma(-j-u)} \frac{\Gamma(1-u)}{\Gamma(1+u)},
\end{align}
and consider the following operator called the monodromy matrix
\begin{align}
     T_a = \prod_{m=1}^L R_{am}. \label{eq:xxx-trasnfer-matrix}
\end{align}
This operator acts on $\left(V_F\right)_a \otimes \left(V_F\right)_L \otimes \cdot \cdot \cdot \otimes \left(V_F\right)_1$, where $\left(V_F\right)_a$ denotes an auxiliary space chosen to be  another copy of $V_F$.  A set of mutually commuting conserved quantities $I_n$ can be obtained through the Taylor expansion around the point $u=0$ of the logarithm of the trace of the monodromy matrix over the auxiliary space,
\begin{align}
    I_n = \left(\frac{d^n}{du^n} \log tr_a T_a \right)_{u=0}. 
\end{align}
The evaluation of the $R$-matrix at the point $u=0$ yields the permutation operator $P$. Therefore, the shift operator $U=P_{12} \cdots P_{L-1,L}$ is a symmetry of the model. The higher conserved quantities $I_n$, for $n > 0$, are local operators of range $n+1$. It is easy to show that the Hamiltonian $H_{XXX}$ corresponds precisely to the conserved quantity $I_1$,
\begin{align}
    I_1 = \sum_{m=1}^{L-1} P_{m,m+1}\left(\frac{\mathrm{d}}{\mathrm{d}u} R_{m,m+1}(u)\right)_{u=0} + P_{L,1}\left(\frac{\mathrm{d}}{\mathrm{d}u} R_{L,1}(u)\right)_{u=0}.
\end{align} 
Integrability allows one to completely solve the spectral problem of the $XXX_{-1/2}$ spin-chain. The eigenstates are labelled by the total number of excitations  $N$ along the spin-chain, which is the eigenvalue of the number operator 
\begin{align}
    N= \sum_{m=1}^L a^\dagger_m a_m. \label{eq:number-oper} 
\end{align}
This quantity is conserved due to the $\mathfrak{sl}_2$ symmetry of the model. For a given value of $N$, the eigenvectors take the form~\cite{Staudacher_2005}
\begin{align}
    \ket{\Psi_{xxx}}_{(N)} &= \sum_{1\leq x_1\leq ...\leq x_N \leq L} \left(\sum_{\pi \in S_N} A_{\pi} \exp \left(i \sum_{j=1}^N p_{\pi(j)} x_j \right)\right) \left|x_1,\cdot \cdot \cdot, x_N\right) , \nonumber\\
    A_{id} &= 1, \nonumber\\
    A_{\pi} &= \prod_{\substack{1\leq j < k \leq N \\ \pi(j)>\pi(k)}} S_{xxx}\left(p_k,p_j\right), \quad S_{xxx}(p,k) = - \frac{e^{i(p+k)}-2e^{ik}+1}{e^{i(p+k)}-2e^{ip}+1}, \label{eq:undef-solution}
\end{align}
 where $\pi$ is an element of the symmetric group of degree $N$ and $x_i$ labels the \emph{position} of the excitation\footnote{We warn the reader about a possible confusion with the notation. Previously we used $\ket{n_1n_2}$ to denote a state with $n_1$ excitations in the first site and $n_2$ excitations in the second site. Now we use a ``rounded ket'' and write $\left|x_1,\cdot \cdot \cdot, x_N\right)$ to indicate the position $x_m$ where just one excitation is created. One can easily go from one notation to the other, for example $\ket{21}=\left|1,1,2\right)$.} 
\begin{align}
    \left|x_1,\cdot \cdot \cdot, x_N\right) = a_{x_1}^\dagger \cdot \cdot \cdot a_{x_N}^\dagger \ket{0}^{\otimes L}.
\end{align}
Additionally, the periodic boundary conditions imply that the set of pseudo-momenta $\{p_j\}$ must satisfy a quantisation condition, known as the Bethe equations
\begin{align}
    e^{ip_k L} = \prod_{\substack{j=1\\j \neq k}}^N S_{xxx}(p_k,p_j), \quad k=1, \cdot \cdot \cdot ,N.
\end{align}
The energy of the eigenstates~\eqref{eq:undef-solution} is given by the sum of the energies of the individual excitations
\begin{align}
    E_{(N)}=\sum_{j=1}^N\epsilon\left(p_j\right), \quad \text{where} \quad \epsilon\left(p_j\right) = 4 \sin^2\left(\frac{p_j}{2}\right). \label{eq:undef-energy}
\end{align}
Observe that the solution of the eigenvalue problem is invariant under the transformation\footnote{We will adopt the convention that the momenta take values in the range $[0,2\pi)$.} $p_j \rightarrow p_j + 2 \pi \mathbb{Z}$. To eliminate this redundancy, the following parameterisation of the momenta may be introduced,
\begin{align}
    \lambda_j = \frac{1}{2} \cot{\frac{p_j}{2}}.
\end{align}
In terms of the pseudo-rapidity $\lambda$, the energy and the Bethe equations take the form
\begin{align}
    \epsilon\left(p_j\left(\lambda_j\right)\right) = \frac{1}{\lambda_j^2+\frac{1}{4}}, \qquad\quad
    \left(\frac{\lambda_k+\frac{i}{2}}{\lambda_k-\frac{i}{2}}\right)^L = \prod_{\substack{j=1\\j \neq k}}^N \frac{\lambda_k-\lambda_j-i}{\lambda_k-\lambda_j+i}.\label{eq:xxx-Bethe-eqs}
\end{align}

\section{The Jordanian $XXX_{-1/2}$ spin-chain}\label{sec:Jord-sp-ch}
Drinfeld twist deformations of quasi-triangular Hopf algebras represent a large class of transformations used to generate new solutions of the quantum Yang-Baxter equation. Therefore, they provide a natural framework to derive new integrable systems as deformations of  existing models. In this section, we construct the Jordanian twist deformation of the $XXX_{-1/2}$ spin-chain.
\subsection{The Jordanian twist}
The starting point in the construction of the model is the Jordanian twist operator, defined as~\cite{10.1007/BFb0101176,Ogievetsky:1992ph}
\begin{align}
    F_{12} = e^{-J^3 \otimes \log\left(1+2\xi J^- \right)} \label{eq:twist},
\end{align}
where $J^3$ and $J^-$ are $\mathfrak{sl}_2$ generators. The parameter $\xi$ is a continuos variable that controls the  deformation, so that when $\xi \to 0$ the twist becomes the identity operator. 

Notice that in the $\mathfrak{sl}_2$ algebra there exists an outer automorphism given by
\begin{align}
    J^3 \to -J^3, \quad J^\pm \to J^\mp.
\end{align}
Therefore, it is also possible to consider the following  Jordanian twist constructed with $J^+$,
\begin{align}
    F'_{12} = e^{J^3 \otimes \log\left(1+2\xi J^+ \right)}. \label{eq:twist-raising}
\end{align}
Because of the presence of $J^+$, the action of the twist $F'_{12}$ on a vector of $V_F \otimes V_F$ results in an  infinite sum of states. Therefore, the two twists $F_{12}$ and $F'_{12}$ are physically inequivalent. The vacuum state of the Hamiltonian that we will deform with $F_{12}$ (i.e. the one constructed with $J^-$) coincides with the vacuum of the undeformed Hamiltonian, namely the state $\ket{0}^{\otimes L}$. Moreover, we will see that the rest of its eigenvectors can be understood as superpositions of vectors with a number of excitations that is always bounded by a maximum value. This value depends on the eigenvector under consideration. By contrast, the vacuum state of the Hamiltonian deformed with $F'_{12}$ (i.e. the twist constructed with $J^+$) is an infinite linear combination of vectors that can have an arbitrary numbers of excitations. Moreover, the rest of its eigenvectors will be labelled by indicating the \emph{minimum} number of excitations that appear in the superposition of the vectors. In what follows and for purely computational simplicity, we will mainly consider the twist defined in~\eqref{eq:twist}, but we will comment also on the similar results for~\eqref{eq:twist-raising}.

The Jordanian twist~\eqref{eq:twist} admits the following power series expansion in the parameter $\xi$~\cite{Giaquinto:1994jx} 
\begin{equation}
    F_{12} = \sum_{k=0}^\infty  \frac{ (-2\xi)^k}{k!}  \left(J^3 \right)^{\langle k \rangle} \otimes \left(J^-\right)^k, \label{eq:twist-expa}
\end{equation}
where $\left(J^3\right)^{\langle k \rangle}$ is the raising factorial defined as
\begin{align}
    \left(J^3\right)^{\langle k \rangle} = \prod_{n=0}^{k-1} \left(J^3+n\right) = \sum_{n=0}^k \left[\genfrac{}{}{0pt}{}{k}{n}\right] \left(J^3 \right)^n,
\end{align}
with $\left[\genfrac{}{}{0pt}{}{k}{n}\right]$  the absolute values of the Stirling numbers of the first kind. The antisymmetrisation of the first order contribution, namely 
\begin{equation}
    r=2 \xi \left(J^- \wedge J^3\right),    
\end{equation}
is a classical $r$-matrix of Jordanian type solving the classical Yang-Baxter equation. 

\subsection{The Jordanian deformed $XXX_{-1/2}$ Hamiltonian }
According to the standard construction of Drinfeld twists (see~\cite{drinfeld1983constant}), the transformation rule for the $R$-matrix is given by
\begin{align}
    R_{12} \to \Tilde{R}_{12} = F_{21} R_{12} F_{12}^{-1},
\end{align}
where $F_{21} = P_{12} F_{12} P_{12}$ is the permuted twist operator. This deformed $R$-matrix can be used to construct the deformed Hamiltonian. First, notice that the twist preserves the regularity property of the $R$-matrix, $\Tilde{R}_{12}(0) = R_{12}(0) = P_{12}$. As a result, the shift operator $U$ belongs to the set of mutually commuting conserved quantities generated from the twisted monodromy matrix. This makes it possible to write a local, shift-invariant (i.e. boundary periodic) deformed Hamiltonian by identifying it with the first-order derivative of the logarithm of the trace of the twisted monodromy matrix evaluated at the value of the spectral parameter $u=0$
\begin{align}
    \Tilde{H}_{XXX} = \sum_{m=1}^L \Tilde{h}_{m,m+1}, \quad \text{with} \quad \Tilde{h}_{12} =P_{12}\left(\frac{\mathrm{d}}{\mathrm{d}u} \Tilde{R}_{12}(u)\right)_{u=0} \quad \text{and} \quad \Tilde{h}_{L,L+1} = \Tilde{h}_{L,1}. \label{eq:deformed-ham}
\end{align}
From this definition, it immediately follows that, at the level of the Hamiltonian density, the twist acts as a similarity  transformation, $h_{12} \to \Tilde{h}_{12} = F_{12} h_{12} F_{12}^{-1}$. Therefore, using~\eqref{eq:xxx-ham-density}, the deformed $XXX_{-1/2}$ Hamiltonian density takes the form
\begin{align}
    \Tilde{h}_{12} = \sum_{j=0}^\infty 2h(j) \Tilde{P}_{12,j},  
\end{align}
where $\Tilde{P}_{12,j}$ are the twisted projectors given by
\begin{align}
    \Tilde{P}_{12,j} = F_{12} P_{12,j} F_{12}^{-1} = \underset{k\neq j}{\prod_{k=0}^\infty} \frac{\Tilde{\Delta}(C)-k(k+1)}{j(j+1)-k(k+1)}, \quad \text{with} \quad \Tilde{\Delta}(C) = F_{12} \Delta(C) F_{12}^{-1}.
\end{align}
The deformed Hamiltonian \emph{density} is invariant under the twisted $\mathfrak{sl}_\xi(2)$ algebra,  which is obtained by deforming the trivial coproduct~\eqref{eq:sl2-coproduct} by the adjoint action of the twist~\eqref{eq:twist} (see~\cite{Kulish2009}),
\begin{align}
  \Tilde{\Delta}(J^3) &= J^3 \otimes e^{-w} + \mathbb{I} \otimes J^3, \nonumber\\
  \Tilde{\Delta}(J^-) &= J^- \otimes e^w + \mathbb{I} \otimes J^-, \nonumber\\
  \Tilde{\Delta}(J^+) &= J^+ \otimes e^{-w} + \mathbb{I} \otimes J^+ -4\xi J^3 \otimes J^3 e^{-w} + \label{eq:twisted-coproduct} \\
  &+2 \xi\left(J^3 + \left(J^3\right)^2\right) \otimes \left(e^{-w} - e^{-2w}\right), \nonumber\\
  \Tilde{\Delta}(w) &= w \otimes \mathbb{I} + \mathbb{I} \otimes w, \nonumber
\end{align}
where $w = \log\left(1 + 2\xi J^-\right)$. The eigenstates of the Hamiltonian density are arranged in the irreducible decomposition of the tensor product of two fundamental modules~\eqref{eq:fundamental-module}
\begin{align}
    V_F \otimes V_F = \bigoplus_{j=0}^\infty \Tilde{V}_j,
\end{align}
where the modules $\Tilde{V}_j$ are related to the $\mathfrak{sl}_2$ modules $V_j$~\eqref{eq:sl2-irr-decom} by the twist action, $\Tilde{V}_j = F_{12} V_j$.

It is important to note that the deformed $XXX_{-1/2}$ Hamiltonian~\eqref{eq:deformed-ham} is non-hermitian due to the triangular nature of the twist~\eqref{eq:twist}. This is a direct consequence of the fact that  $F_{12}$ is constructed with the lowering operator $J^-$ without pairing it with the raising operator $J^+$. Moreover, we will argue in the following sections that the Hamiltonian is also not diagonalisable, but instead admits a Jordan decomposition. Despite these two peculiar properties, this Hamiltonian describes an integrable model.

\subsection{The Jordanian model as an undeformed chain   with twisted-boundary conditions}
\label{subsec:equivalence-with-boundary-twisted}
In this subsection we show that the Jordanian-deformed $XXX_{-1/2}$ Hamiltonian~\eqref{eq:deformed-ham} is equivalent to an undeformed Hamiltonian with twisted boundary conditions. We  will denote by $\mathbb{H}_{XXX}$ the Hamiltonian of the undeformed yet twisted model, that we can then write as
\begin{align}
   \mathbb{H}_{XXX} = \sum_{m=1}^{L-1} h_{m,m+1} + S^{-1} h_{L,1} S, \label{eq:ham-non-periodic}
\end{align}
where $h_{12}$ is the Hamiltonian density of the \emph{undeformed} $XXX_{-1/2}$ model~\eqref{eq:xxx-ham-density}. The operator $S$ implements the boundary conditions by imposing that the Hamiltonian $\mathbb{H}_{XXX}$ acting on  sites $L,L+1$ of the chain is related to the action on $L,1$ as $S^{-1} h_{L,1} S$.   

In the following, we will always be careful referring to the Hamiltonian $\tilde H_{XXX}$ in~\eqref{eq:deformed-ham} and the corresponding model as ``deformed'', and to the Hamiltonian $\mathbb{H}_{XXX}$ in~\eqref{eq:ham-non-periodic} and the corresponding model as ``twisted''. The reason is that in the former case the Hamiltonian density has an explicit dependence on a deformation parameter, while in the latter case the Hamiltonian density is undeformed and the parameter enters only in the boundary conditions. The two are a priory different modifications of the original $\mathfrak{sl}_2$ spin-chain, but they can be shown to be equivalent thanks to the fact that the modifications are generated by a Drinfeld twist. In particular, the equivalence is proved by showing that the two Hamiltonians are related by a similarity transformation \begin{equation}\label{eq:equivalence-def-twist}
    \Tilde{H}_{XXX} = \Omega \mathbb{H}_{XXX} \Omega^{-1}, 
\end{equation}
where the operator  $\Omega$ can be constructed explicitly.
Similar constructions have been developed for  spin-chains with abelian twists,\footnote{See~\cite{Guica:2017mtd} for applications in the AdS/CFT correspondence and for some useful formulas.} but those formulas are not directly applicable to non-abelian twists such as the Jordanian one. We refer to the appendix~\ref{ap:boundary-twist-ham} for the details on the demonstration of the equivalence between both models, while here we only collect the main points. For the Jordanian twisted model~\eqref{eq:deformed-ham}, we find that $\Omega$ is given by the global intertwiner~\cite{kulish1998twistingsolutionsyangbaxterequation}, 
\begin{align}
    \Omega = \left(F_{12} ... F_{1L} \right)\left(F_{23} ... F_{2L} \right)...F_{L-1,L}, \label{eq:global-intertwiner}
\end{align}
while $S$ takes the form 
\begin{align}
    S=F_{L1}^{-1} \Omega. \label{eq:twist-boundary}
\end{align}
The operator $\Omega$ is non-local, in the sense that it acts on all sites of the spin-chain. Therefore, the boundary conditions of the model~\eqref{eq:ham-non-periodic} depend on the state configurations of all spin sites.
 In fact, as done in appendix~\ref{ap:boundary-twist-ham}, one can prove that also at the level of the Hamiltonian densities we have
\begin{equation}
    \Tilde{h}_{m,m+1} = \Omega h_{m,m+1} \Omega^{-1}, \quad m = 1,\ldots,L-1. \label{eq:equivalence-bulk}
\end{equation}

Let us now consider the bulk terms of the deformed Hamiltonian~\eqref{eq:deformed-ham} and of the undeformed untwisted one~\eqref{eq: xxx-h12}. Here by ``bulk'' we mean that we restrict the sum in the expression of the Hamiltonian up to the site $L-1$, so that we do not include the contribution of the ``boundary term'' with $m=L$,
\begin{align}
    H_{XXX}^{\text{bulk}} &= \sum_{m=1}^{L-1} h_{m,m+1}, \label{eq:bulk-ham} \\
    \tilde{H}_{XXX}^{\text{bulk}} &= \sum_{m=1}^{L-1} \tilde{h}_{m,m+1}. \label{eq:bulk-ham-twisted}
\end{align}
The relation~\eqref{eq:equivalence-bulk} implies that the bulk Hamiltonian~\eqref{eq:bulk-ham} is  equivalent to~\eqref{eq:bulk-ham-twisted} by a similarity transformation given again by $\Omega$~\cite{Kulish2009} 
\begin{align}
     \tilde{H}_{XXX}^{\text{bulk}} = \Omega  H_{XXX}^{\text{bulk}} \Omega^{-1} .
\end{align}
Therefore,~\eqref{eq:bulk-ham-twisted} is diagonalisable with a spectrum that is equal to that of the open underformed  spin-chain~\eqref{eq:bulk-ham}.

Note that from the definition of the Hamiltonian density~\eqref{eq:xxx-ham-density}, it is immediate to verify that the Hamiltonian~\eqref{eq:bulk-ham} is invariant under a global $\mathfrak{sl}_2$ algebra
\begin{align}
    [\Delta^{(L-1)}(x), H_{XXX}^{\text{bulk}}] = 0, \quad x \in \mathfrak{sl}_2, \label{eq:sym_bulk-ham}
\end{align}
with $\Delta^{(L-1)}$ the $L-1$ composition of the coproduct~\eqref{eq:sl2-coproduct}, where the composition of the coproduct $\Delta^{(n)}:V_F \to V_F^{\otimes n+1}$ is defined as 
\begin{align}
    \Delta^{(n)} = \left(\Delta \otimes \mathbb{I}\right) \Delta^{(n-1)}, \quad \text{with} \quad \Delta^{(1)}=\Delta.
\end{align}
Similarly, one can show that the Hamiltonian~\eqref{eq:bulk-ham-twisted} possesses a global twisted $\mathfrak{sl}_\xi(2)$ symmetry. To prove this statement, first note that the composition of the coproduct transforms under the adjoint action of a Drinfeld twist as follows (see e.g.~\cite{Maillet:1996yy,Kulish2009})
\begin{align}
    \tilde{\Delta}^{(L-1)} = F_{12\ldots L}  \Delta^{(L-1)} F_{12\ldots L}^{-1}, \label{eq:composition-twisted-coproduct}
\end{align}
where 
\begin{align}
    F_{12\ldots L} = F_{12} F_{(12),3}\cdots F_{(12\ldots L-1),L}, \label{eq:global-twist}
\end{align}
with $F_{(12\ldots n),n+1}= \left(\Delta^{(n-1)} \otimes \mathbb{I}\right)\left(F\right)$.  
Moreover, let us note that, using $\Delta(J^3)=J^3\otimes\mathbb{I}+\mathbb{I}\otimes J^3$, one can immediately verify that the Jordanian twist~\eqref{eq:twist} satisfies the following factorisation property
\begin{align}
    \left(\Delta \otimes \mathbb{I}\right)\left(F\right) = F_{13} F_{23}.
\end{align}
This allows one to rewrite~\eqref{eq:global-twist} as follows
\begin{align}
    F_{12\ldots L} = F_{12} \left(F_{13} F_{23}\right)\cdots \left(F_{1L} \cdots F_{L-1,L}\right).
\end{align}
Therefore, the operator $F_{12\ldots L}$ coincides with $\Omega$ in~\eqref{eq:global-intertwiner}. Now, using equation~\eqref{eq:composition-twisted-coproduct} together with~\eqref{eq:sym_bulk-ham}, it follows that
\begin{align}
    [\tilde{\Delta}^{(L-1)}(x), \tilde{H}_{XXX}^{\text{bulk}}] = \Omega \; [\Delta^{(L-1)}(x), H_{XXX}^{\text{bulk}}] \; \Omega^{-1} = 0, \quad x \in \mathfrak{sl}_2,
\end{align}
which proves the invariance of the bulk Hamiltonian~\eqref{eq:bulk-ham-twisted} under the global twisted $\mathfrak{sl}_{\xi}(2)$ algebra.

Importantly, the inclusion of the boundary term $\tilde{h}_{L,L+1}=\tilde{h}_{L1}$ breaks the global $\mathfrak{sl}_\xi(2)$ symmetry. In fact, thanks to the above equation, we obviously have $[\tilde{\Delta}^{(L-1)}(x), \tilde{H}_{XXX}]=[\tilde{\Delta}^{(L-1)}(x), \tilde{H}_{XXX}^{\text{bulk}}]+[\tilde{\Delta}^{(L-1)}(x), \tilde h_{L,L+1}]=[\tilde{\Delta}^{(L-1)}(x), \tilde h_{L,L+1}]$.
This last commutator is in general non-zero, since    the $(L-1)$-composition $\tilde{\Delta}^{(L-1)}$ of the twisted coproduct~\eqref{eq:twisted-coproduct}  is in general  not invariant under the shift operator $U=P_{12}\cdots P_{L-1,L}$. In particular, one may write
\begin{align}
    [\tilde{\Delta}^{(L-1)}(x), \tilde{h}_{L,L+1}] = U[ U^{-1}\tilde{\Delta}^{(L-1)}(x) U, \tilde{h}_{L-1,L}] U^{-1}, \quad x \in \mathfrak{sl}_2, \label{eq:commutator-boundary}
\end{align}
and, as mentioned previously, $\tilde{h}_{L-1,L}$ is invariant under the global $\mathfrak{sl}(2)_\xi$ symmetry as
\begin{align}
    [\tilde{\Delta}^{(L-1)}(x),\tilde{h}_{L-1,L}] = 0 \quad x \in \mathfrak{sl}_2.
\end{align}
Therefore, the commutator~\eqref{eq:commutator-boundary} is in general non-zero, except for those elements $x$ of $\mathfrak{sl}_2$ for which $U^{-1}\tilde{\Delta}^{(L-1)}(x) U$ is also of the form of  $\tilde{\Delta}^{(L-1)}(x')$ for some $x'\in\mathfrak{sl}_2$. Interestingly, the twisted coproduct of the operator $w$ is primitive, hence $U^{-1}\tilde{\Delta}^{(L-1)} (w) U = \tilde{\Delta}^{(L-1)} (w)$  and it commutes with  $\tilde{H}_{XXX}$.

Given the equivalence relation between the deformed Hamiltonian~\eqref{eq:deformed-ham} and the undeformed one with twisted boundary conditions~\eqref{eq:ham-non-periodic}, there must exist a monodromy matrix $\mathbb{T}_a$ associated to the $\mathbb{H}_{XXX}$ Hamiltonian~\eqref{eq:ham-non-periodic}, such that
\begin{align}
    \Tilde{T}_a = \Omega \mathbb{T}_a \Omega^{-1},
\end{align}
where $\Tilde{T}_a$ is the transfer matrix of the deformed spin-chain~\eqref{eq:deformed-ham}. The model with twisted boundary conditions is not translationally invariant. This is a consequence of the fact that the shift operator $U$ does not belong to the conserved quantities generated by $\mathbb{T}_a$. Instead, the conserved quantity obtained from the zero-order coefficient of the Taylor expansion of the trace of the transfer matrix $\mathbb{T}_a$ is a twisted version of the shift operator, namely $\Omega^{-1} U \Omega$.

Since the bulk terms of the $\mathbb{H}_{XXX}$ Hamiltonian are undeformed, it must be possible to write the monodromy matrix $\mathbb{T}_a$, or equivalently $\Tilde{T}_a$, as some function of the untwisted $XXX_{-1/2}$ monodromy matrix~\eqref{eq:xxx-trasnfer-matrix}. We find (see the appendix~\ref{ap:twisted-transfer-matrix} for the derivation) that 
\begin{align}
    \Tilde{T}_a =\Omega F_{a} T_{a} \Omega^{-1}\left(F_{a}\right)_{op}^{-1}, \label{eq:trasnfer-matrix}
\end{align}
where we have defined
\begin{align}
    F_{a} = F_{1a}\cdots F_{La}, \\
    \left(F_{a}\right)_{op}^{-1} = F_{aL}^{-1}\cdots F_{a1}^{-1}. 
\end{align}
Therefore, we can write $\mathbb{T}_a$ as follows
\begin{equation}
    \mathbb{T}_a =  F_a T_a \Omega^{-1} \left(F_{a}\right)_{op}^{-1} \Omega. \label{eq:non-periodic-trasnfer-matrix}
\end{equation}
Requiring additional properties for the Drinfeld twist, the relation between the twisted and the undeformed untwisted monodromy matrices may be further simplified. For example, if the twist commutes for every index position
\begin{align}
    [F_{ab},F_{cd}]=0 \quad \forall a,b,c,d=1,\ldots,L,
\end{align}
and satisfies $F_{ba} = F_{ab}^{-1}$, as it happens for example for abelian twists, then~\eqref{eq:non-periodic-trasnfer-matrix} can be simplified to $\mathbb{T}_a =  F_a T_a F_a$, and we recover the result given in~\cite{Guica:2017mtd}.

\section{Some useful explicit examples}\label{sec:examples}
In this section, we discuss the form of the eigenvalues and (generalised) eigenvectors of~\eqref{eq:deformed-ham} by computing in some concrete examples the matrix form of the Hamiltonian~\eqref{eq:deformed-ham}. As already anticipated in the previous sections, the (generalised) eigenstates of~\eqref{eq:deformed-ham} will be labelled by an integer number $N$ accounting for the maximum number of excitations that appear in the superposition of states. This implies that the Hamiltonian~\eqref{eq:deformed-ham} can be diagonalised within each finite dimensional subspace identified by fixing $N$.  When writing down the examples, we will restrict to the subspaces of the Hamiltonian with $N=2,3$. For simplicity reasons and to keep the explicit matrices of the examples of reasonable size, we focus on a spin-chain of length two.

\subsection{Example I: $N=2$}
Let us consider the following basis of states
\begin{align}
    \{\ket{00},\ket{10},\ket{01},\ket{20},\ket{02},\ket{11}\}. \label{eq:basis-2ex}
\end{align}
In this basis, the Hamiltonian~\eqref{eq:deformed-ham} takes the form
\begin{equation}
\left(
\begin{array}{cccccc}
 0 & 0 & 0 & \xi ^2 & \xi ^2 & 0 \\
 0 & 2 & -2 & 2 \xi  & 2 \xi  & -4 \xi  \\
 0 & -2 & 2 & 2 \xi  & 2 \xi  & -4 \xi  \\
 0 & 0 & 0 & 3 & -1 & -2 \\
 0 & 0 & 0 & -1 & 3 & -2 \\
 0 & 0 & 0 & -2 & -2 & 4 \\
\end{array}
\right). \label{eq:ham-length2-2ex}    
\end{equation}
If we set the parameter deformation $\xi$ to zero, we recover the $XXX_{-1/2}$ spin-chain~\eqref{eq: xxx-h12}. In that limit, the matrix~\eqref{eq:ham-length2-2ex} becomes block diagonal where each block is the restriction of the Hamiltonian~\eqref{eq: xxx-h12} to the subspaces  with zero, one and two excitations respectively. That is to say, thanks to the $J^3$ symmetry of the $XXX_{-1/2}$, the Hamiltonian can be diagonalised within each subspace with fixed number of excitations. In Table~\ref{tab:eigen-ham-length2-2ex} we list the spectrum and a possible set of eigenvectors of the matrix~\eqref{eq:ham-length2-2ex} in the $\xi \to 0$ limit, together with the momenta of the Bethe wavefunctions associated to each eigenstate. The eigenvectors with only non-zero momenta are the highest-weight states of the $\mathfrak{sl}_2$ multiplets of the irreducible decomposition of $V_F \otimes V_F$, while eigenstates with at least one vanishing momentum are descendants.

\begin{table}[h!]
    \centering
    \begin{tabular}{c|c|c}
 Eigenvalue & Momenta & Eigenvector \\
 \hline
 $0$ & $-$ & $\ket{00}$ \\
 \hline
 $0$&$(0)$& $\ket{10} + \ket{01}$ \\
 \hline
 $0$&$(0,0)$& $\ket{20} + \ket{02}+ \ket{11}$ \\
 \hline
 $4$ & $(\pi)$ & $\ket{10}-\ket{01}$ \\
 \hline
 $4$ & $(\pi,0)$ & $\ket{20}-\ket{02}$ \\
 \hline
 $6$ & $(2\pi/3,4\pi/3)$ & $3 \left(\ket{20} + \ket{02}\right) - 6 \ket{11}$ \\  
    \end{tabular}
\caption{Eigenvalues and eigenvectors of the matrix~\eqref{eq:ham-length2-2ex} in the $\xi \to 0$ limit. Notice that the eigenvectors coincide with the Bethe states~\eqref{eq:undef-solution} provided we choose the appropriate normalisation factor. We also list the momenta of the Bethe wavefunctions associated to each eigenstate.  }
\label{tab:eigen-ham-length2-2ex}
\end{table}
    
Suppose now that the parameter deformation $\xi$ is non-zero. Then the matrix~\eqref{eq:ham-length2-2ex} is not diagonalisable. What one can do is to find a basis where~\eqref{eq:ham-length2-2ex} is in a Jordan normal form. We find the following Jordan decomposition
\begin{equation}
    \left(
\begin{array}{cccccc}
 0 & 1 & 0 & 0 & 0 & 0 \\
 0 & 0 & 0 & 0 & 0 & 0 \\
 0 & 0 & 0 & 0 & 0 & 0 \\
 0 & 0 & 0 & 4 & 0 & 0 \\
 0 & 0 & 0 & 0 & 4 & 0 \\
 0 & 0 & 0 & 0 & 0 & 6 \\
\end{array}
\right). \label{eq:Jordan-form-ham-length2-2ex}
\end{equation}
Notice that the Jordan form of the Hamiltonian is independent of the deformation parameter $\xi$. In other words, the eigenvalues of the matrix~\eqref{eq:ham-length2-2ex} for $\xi\neq 0$ coincide with the eigenvalues of~\eqref{eq:ham-length2-2ex} in the $\xi \to 0$ limit. Moreover, the matrix~\eqref{eq:Jordan-form-ham-length2-2ex} consists of four Jordan chains of length one  and one chain of length two. Therefore,~\eqref{eq:ham-length2-2ex} possesses only one generalised eigenvector, which is in the Jordan chain of the vacuum. The (generalised) eigenvectors of~\eqref{eq:ham-length2-2ex} are listed in Table~\ref{tab:gen-eigen-ham-length2-2ex}.
\begin{table}[h!]
    \centering
    \begin{tabular}{c|c}
 Eigenvalue & (Generalised) eigenvector \\
 \hline
 $0$ & $\ket{00}$ \\
 \hline
 $0$& $\frac{1}{2\xi^2}\left(\ket{20} + \ket{02}+ \ket{11}\right)$ \\
 \hline
 $0$&$\ket{10} + \ket{01}$ \\
 \hline
 $4$ &  $\ket{10}-\ket{01}$ \\
 \hline
 $4$ &  $\ket{20}-\ket{02}$ \\
 \hline
 $6$ &  $3 \left(\ket{20} + \ket{02}\right) - 6 \ket{11}+6 \xi \left(\ket{10}+\ket{01}\right) + \xi^2 \ket{00}$ \\  
    \end{tabular}
\caption{Eigenvalues and (generalised) eigenvectors of the matrix~\eqref{eq:ham-length2-2ex} for non zero $\xi$ . The generalised eigenvector is the state $\frac{1}{2\xi^2}\left(\ket{20} + \ket{02}+ \ket{11}\right)$. The normalisation factor is chosen so that the action of~\eqref{eq:ham-length2-2ex} on it gives the vacuum $\ket{00}$.   }
\label{tab:gen-eigen-ham-length2-2ex}
\end{table}
Let us now discuss the changes induced by the twist on the $XXX_{-1/2}$ model, or in other words the difference between the matrix~\eqref{eq:ham-length2-2ex} with zero and non-zero $\xi$. First, we remark that the spectrum is not modified, in the sense that the eigenvalues of~\eqref{eq:ham-length2-2ex} for non zero $\xi$ coincides with the eigenvalues of~\eqref{eq:ham-length2-2ex} in the $\xi \to 0$ limit. In addition, the eigenstate corresponding to the descendant solution with two vanishing momenta, i.e the state $\ket{20}+\ket{02}+\ket{11}$, turns into a generalised eigenvector (up to some normalisation factor). Moreover, some of the eigenstates receive corrections in $\xi$ given by states with fewer and fewer excitations down to the vacuum. Interestingly, the eigenstates constructed with states of one excitation do not receive any correction. This implies that the restriction of~\eqref{eq:ham-length2-2ex} to the subspace of states with one excitation coincides with the $XXX_{-1/2}$ model. This can already be seen in the matrix~\eqref{eq:ham-length2-2ex} by noting that the block corresponding to the restriction to the subspace spanned by $\{\ket{00},\ket{10},\ket{01}\}$ does not receive corrections.

\subsection{Example II: $N=3$}
Now we repeat the same calculation in the subspace of states with at most three excitations. Let us consider the following basis
\begin{align}
    \{\ket{00},\ket{10},\ket{01},\ket{20},\ket{02},\ket{11},\ket{30},\ket{03},\ket{21},\ket{12}\}.
\end{align}
The restriction of the Hamiltonian~\eqref{eq:deformed-ham} to the subspace spanned by the previous basis is given by
\begin{equation}
    \left(
\begin{array}{cccccccccc}
 0 & 0 & 0 & \xi ^2 & \xi ^2 & 0 & -6 \xi ^3 & -6 \xi ^3 & 0 & 0 \\
 0 & 2 & -2 & 2 \xi  & 2 \xi  & -4 \xi  & -3 \xi ^2 & -6 \xi ^2 & 0 & 5 \xi ^2 \\
 0 & -2 & 2 & 2 \xi  & 2 \xi  & -4 \xi  & -6 \xi ^2 & -3 \xi ^2 & 5 \xi ^2 & 0 \\
 0 & 0 & 0 & 3 & -1 & -2 & 4 \xi  & 2 \xi  & -8 \xi  & -2 \xi  \\
 0 & 0 & 0 & -1 & 3 & -2 & 2 \xi  & 4 \xi  & -2 \xi  & -8 \xi  \\
 0 & 0 & 0 & -2 & -2 & 4 & 6 \xi  & 6 \xi  & -2 \xi  & -2 \xi  \\
 0 & 0 & 0 & 0 & 0 & 0 & \frac{11}{3} & -\frac{2}{3} & -2 & -1 \\
 0 & 0 & 0 & 0 & 0 & 0 & -\frac{2}{3} & \frac{11}{3} & -1 & -2 \\
 0 & 0 & 0 & 0 & 0 & 0 & -2 & -1 & 5 & -2 \\
 0 & 0 & 0 & 0 & 0 & 0 & -1 & -2 & -2 & 5 \\
\end{array}
\right). \label{eq:ham-length2-3ex}
\end{equation}
If we take the limit $\xi \to 0$, the matrix~\eqref{eq:ham-length2-3ex} becomes block diagonal, where each block is the restriction of the $XXX_{-1/2}$ Hamiltonian~\eqref{eq: xxx-h12} to the subspaces  with zero, one, two and three excitations respectively. The spectrum and eigenvectors of the $XXX_{-1/2}$ Hamiltonian in the subspace of states with zero, one and two excitations were already discussed in the previous subsection, and they are listed in Table~\ref{tab:eigen-ham-length2-2ex}. In Table~\ref{tab:eigen-ham-length2-3ex} we list the eigenvalues and eigenvectors of the $XXX_{-1/2}$ Hamiltonian in the subspace  with three excitations, which corresponds to the lower-right diagonal block of~\eqref{eq:ham-length2-3ex}.  
\begin{table}[h!]
    \centering
    \begin{tabular}{c|c|c}
 Eigenvalue & Momenta & Eigenvector \\
 \hline
 $0$ & $(0,0,0)$ & $\ket{30}+\ket{03}+\ket{21}+\ket{12}$ \\
 \hline
 $4$&$(\pi,0,0)$& $3 \left(\ket{30} - \ket{03}\right) + \ket{21}-\ket{12}$ \\
 \hline
 $6$&$(2\pi/3,4\pi/3,0)$& $3 \left(\ket{30} + \ket{03}\right) -3\left( \ket{21}+\ket{12}\right)$ \\
 \hline
 $22/3$ & $(\pi,\arctan{\sqrt{35}},2\pi-\arctan{\sqrt{35}})$ & $5 \left(\ket{30} - \ket{03}\right) -15\left( \ket{21}-\ket{12}\right)$ \\
   \end{tabular}
\caption{Eigenvalues, momenta and eigenvectors of the lower-right diagonal block of the matrix~\eqref{eq:ham-length2-3ex} in the $\xi \to 0$ limit, which corresponds to the $XXX_{-1/2}$ Hamiltonian~\eqref{eq: xxx-h12} in the subspace of states with three excitations. The eigenvectors coincide with the Bethe states~\eqref{eq:undef-solution} provided we choose the appropriate normalisation factor.}
\label{tab:eigen-ham-length2-3ex}
\end{table}

Let us now consider the case when $\xi$ is non zero. Then the matrix~\eqref{eq:ham-length2-3ex} is not diagonalisable. Its Jordan normal form is given by
\begin{equation}
    \left(
\begin{array}{cccccccccc}
 0 & 1 & 0 & 0 & 0 & 0 & 0 & 0 & 0 & 0 \\
 0 & 0 & 0 & 0 & 0 & 0 & 0 & 0 & 0 & 0 \\
 0 & 0 & 0 & 1 & 0 & 0 & 0 & 0 & 0 & 0 \\
 0 & 0 & 0 & 0 & 0 & 0 & 0 & 0 & 0 & 0 \\
 0 & 0 & 0 & 0 & 4 & 1 & 0 & 0 & 0 & 0 \\
 0 & 0 & 0 & 0 & 0 & 4 & 0 & 0 & 0 & 0 \\
 0 & 0 & 0 & 0 & 0 & 0 & 4 & 0 & 0 & 0 \\
 0 & 0 & 0 & 0 & 0 & 0 & 0 & 6 & 0 & 0 \\
 0 & 0 & 0 & 0 & 0 & 0 & 0 & 0 & 6 & 0 \\
 0 & 0 & 0 & 0 & 0 & 0 & 0 & 0 & 0 & \frac{22}{3} \\
\end{array}
\right). \label{eq:Jordan-form-ham-length2-3ex}
\end{equation}
Again, we see no $\xi$-dependence, so that the eigenvalues do not receive twist corrections, and they coincide with the eigenvalues of the matrix~\eqref{eq:ham-length2-3ex} in the $\xi \to 0$ limit. In addition, the matrix~\eqref{eq:Jordan-form-ham-length2-3ex} possesses four Jordan chains of length one, and three chains of length two, i.e., there is a total of three generalised eigenvectors.

In Table~\ref{tab:gen-eigen-ham-length2-3ex}, we list the (generalised) eigenvectors of the matrix~\eqref{eq:ham-length2-3ex} for a non-zero parameter deformation $\xi$. Once again, we see that there is a one-to-one correspondence between the (generalised) eigenvectors of ~\eqref{eq:ham-length2-3ex} and the eigenstates of the $XXX_{-1/2}$. In fact, the state with maximum number of excitations in the superposition coincides with the eigenstate of the $\xi \to 0$ limit of~\eqref{eq:ham-length2-3ex} (up to some  normalisation factor that is singular in $\xi\to 0$ for the generalised eigenvectors), see Table~\ref{tab:eigen-ham-length2-2ex} and~\ref{tab:eigen-ham-length2-3ex} .

\begin{table}[h!]
    \centering
    \begin{tabular}{c|c}
 Eigenvalue  & (Generalised)  \\
   & eigenvector\\
 \hline
 $0$ & $\ket{00}$ \\
 \hline
 $0$& $\frac{1}{2\xi^2}\left(\ket{20} + \ket{02}+ \ket{11}\right)$ \\
 \hline
 $0$&$\ket{10} + \ket{01}$ \\
 \hline
 $0$&$\frac{1}{4 \xi^2}\left(\ket{30}+\ket{03}+\ket{21}+\ket{12}\right) + \frac{1}{\xi}\left(\frac{3}{2}\ket{20}+\frac{3}{2}\ket{02}+\ket{11}\right)$ \\
 \hline
 $4$ &  $\ket{10}-\ket{01}$ \\
 \hline
 $4$ &  $\frac{1}{4 \xi^2} \left(3 \ket{30} - 3\ket{03} + \ket{21}-\ket{12}\right)$ \\
 \hline
 $4$ &  $\ket{20}-\ket{02}$ \\
 \hline
 $6$ &  $3 \left(\ket{20} + \ket{02}\right) - 6 \ket{11}+6 \xi \left(\ket{10}+\ket{01}\right) + \xi^2 \ket{00}$ \\
 \hline
 $6$ &  \makecell{$ 3 \left(\ket{30} + \ket{03}\right) -3\left( \ket{21}+\ket{12}\right) + 12 \xi \left(\ket{20}+\ket{02}\right) + $\\ $+\xi^2 \left(\ket{10}+\ket{01}\right) - 2 \xi^3 \ket{00}$} \\
 \hline
 $22/3$ &  \makecell{$5 \left(\ket{30} - \ket{03}\right) -15\left( \ket{21}-\ket{12}\right) + 30 \xi \left(\ket{20} - \ket{02}\right) + $ \\ $ + 27 \xi^2 \left(\ket{10}-\ket{01}\right)$ }  \\
    \end{tabular}
\caption{Eigenvalues and (generalised) eigenvectors of the matrix~\eqref{eq:ham-length2-3ex} for non zero value of the parameter deformation $\xi$. The generalised eigenvectors are the states $\frac{1}{2\xi^2}\left(\ket{20} + \ket{02}+ \ket{11}\right)$, $\frac{1}{4 \xi^2}\left(\ket{30}+\ket{03}+\ket{21}+\ket{12}\right) + \frac{1}{\xi}\left(\frac{3}{2}\ket{20}+\frac{3}{2}\ket{02}+\ket{11}\right)$ and $\frac{1}{4 \xi^2} \left(3 \ket{30} - 3\ket{03} + \ket{21}-\ket{12}\right)$.    }
\label{tab:gen-eigen-ham-length2-3ex}
\end{table}

The generalised eigenvectors of~\eqref{eq:ham-length2-3ex} are the state $\frac{1}{2\xi^2}\left(\ket{20} + \ket{02}+ \ket{11}\right)$ which is in the Jordan chain of the vacuum, the state \begin{equation}
    \frac{1}{4 \xi^2}\left(\ket{30}+\ket{03}+\ket{21}+\ket{12}\right) + \frac{1}{\xi}\left(\frac{3}{2}\ket{20}+\frac{3}{2}\ket{02}+\ket{11}\right),
\end{equation}
in the chain of the first descendant of the vacuum, $\ket{10} + \ket{01}$; and the state 
\begin{equation}
    \frac{1}{4 \xi^2} \left(3 \ket{30} - 3\ket{03} + \ket{21}-\ket{12}\right),
\end{equation}
in the Jordan chain of $\ket{10} - \ket{01}$. Interestingly, using the one-to-one correspondence observed above between the (generalised) eigenvectors of the deformed model and the eigenvectors of the undeformed one, all the generalised eigenvectors of the system are related to the descendant solutions of the $XXX_{-1/2}$ with two or more vanishing momenta. Notice, however, that not al descendants of the undeformed model will correspond to generalised eigenvectors after the deformation. We suspect that this distribution of generalised eigenvectors is a general property of the deformed $XXX_{-1/2}$ given in~\eqref{eq:deformed-ham}, however, we were not able to prove it.

\section{Coordinate Bethe Ansatz for the deformed $XXX_{-1/2}$ spin-chain}
\label{section:CBA}
In this section we discuss the construction of the coordinate Bethe Ansatz from the point of view of the deformed model. Having proved the equivalence to the twisted model, this discussion automatically provides the spectrum also in that picture.

In the original $XXX_{-1/2}$ model, the coordinate Bethe Ansatz method is based on the existence of a $\mathfrak{u}_1$ symmetry generated by $J^3$. This symmetry enables to construct the eigenstates as  linear combinations of states with a fixed number of excitations, where this number is the eigenvalue of the operator defined in~\eqref{eq:number-oper}.

In the Jordanian  $XXX_{-1/2}$ spin-chain with $F_{12}$ given by~\eqref{eq:twist}, the state with no excitations $\ket{0}^{\otimes L}$ is still the ground state of the deformed Hamiltonian~\eqref{eq:deformed-ham}. However, the number of excitations is not preserved by the action of~\eqref{eq:deformed-ham}, as a consequence of the fact that the operator $J^-$ in the twist~\eqref{eq:twist} annihilates the excitations, 
\begin{align}
    F_{12} \ket{n_1,n_2} = \sum_{k=0}^{n_2} \left(-2\xi \right)^k \binom{n_2}{k} \left(\frac{1}{2}+n_1\right)^{\langle k \rangle} \ket{n_1,n_2-k}. 
\end{align} 
Therefore, the spectrum of the twisted theory does not decompose into subspaces of a fixed number of excitations. Nevertheless, we can still label the eigenstates by eigenvalues of the number operator $N$, if we now just focus on the  maximum number of excitations that are used in the linear combination to construct the eigenstates. With abuse of notation, we will call $N$ also the maximum number of excitations. To summarise, the eigenstates are given by a superposition of states ranging from the vacuum up to a state with $N$ excitations,\footnote{Interestingly, a similar type of ansatz was employed also for an open spin-chain with boundary conditions implemented via triangular matrices~\cite{Crampe:2011fm}.}
\begin{equation}
    \ket{\Psi}_{(N)} = \sum_{r=0}^N \sum_{\Vec{x}_{(r)}} \psi_{(r)} (\Vec{x}_{(r)})|\Vec{x}_{(r)}), \label{eq:twisted-ansatz}
\end{equation}
where we have used the following notation
\begin{align}
    |\Vec{x}_{(0)}) = \ket{0}^{\otimes L}, \quad \psi_{(0)} (\Vec{x}_{(0)}) = \psi_{(0)} = \text{constant}, \\
    \sum_{\Vec{x}_{(r)}} = \sum_{1\leq x_1\leq \cdots\leq x_r \leq L}, \qquad 
    |\Vec{x}_{(r)}) = |x_1,x_2,\ldots,x_r), \quad r>0.
\end{align}
The set of wavefunctions $\{\psi_{(r)}(\Vec{x}_{(r)})\}$ is determined by imposing that the state~\eqref{eq:twisted-ansatz} is an eigenstate  of the twisted Hamiltonian~\eqref{eq:deformed-ham} with some eigenvalue $E_{(N)}$\footnote{We denote the energy of the deformed eigenstate~\eqref{eq:twisted-ansatz} the same as the energy of the undeformed and untwisted model~\eqref{eq:undef-energy}, because as we will see they coincide.}. Since all states $|\Vec{x}_{(r)})$ are linearly independent for each label $r$, we obtain the following system of coupled equations
\begin{align}
    E_{(N)} \sum_{\Vec{x}_{(r)}} \psi_{(r)} (\Vec{x}_{(r)})|\Vec{x}_{(r)}) &= \sum_{p=r}^N \sum_{\Vec{x}_{(p)}} \psi_{(p)} (\Vec{x}_{(p)})\Tilde{H}_{XXX}^{(p-r)}|\Vec{x}_{(p)}), \qquad r=0, \ldots,N, \label{eq:sist-wfs}
\end{align}
where $\Tilde{H}_{XXX}^{(q)}$ are the terms of the expansion of~\eqref{eq:deformed-ham} that exclusively contain a number of $q$ powers of $J^-$. From the expansion of the Jordanian twist~\eqref{eq:twist-expa}, it follows
\begin{align}
    \Tilde{H}_{XXX}^{(q)} = \sum_{m=1}^L \Tilde{h}_{m,m+1}^{(q)},
\end{align}
with
\begin{align}
    \Tilde{h}_{12}^{(q)} = \sum_{k=0}^q \frac{(-2\xi)^q}{k!(q-k)!}\left(\left(J^3\right)^{\langle q-k \rangle}\otimes\left(J^-\right)^{q-k}\right) h_{12} \left(\left(-J^3\right)^{\langle k \rangle}\otimes\left(J^-\right)^k\right), \label{eq:expansion-ham-density}
\end{align}
where $h_{12}$ is the untwisted $XXX_{-1/2}$ Hamiltonian density~\eqref{eq:xxx-ham-density}.

In addition to satisfying the above system of equations, the solution must be compatible with the periodic boundary conditions, which requires the wavefunctions to satisfy the following identity,
\begin{align}
    \psi_{(r)}(x_1,\ldots,x_r) = \psi_{(r)}(x_2,\ldots,x_r,x_1+L), \qquad r=1,\ldots ,N.
\end{align}

For a given $r$, the right-hand side of the equations~\eqref{eq:sist-wfs}, contains only the sum of wavefunctions ranging from $r$ to $N$. Therefore, the system of equations can be solved recursively. In other words, from the equation with $r=N$ one can solve for the wavefunction $ \psi_{(N)} (\Vec{x}_{(N)})$. Once this is known, the wavefunction $ \psi_{(N-1)} (\Vec{x}_{(N-1)})$ is obtained by solving the equation with $r=N-1$, and so on. 

Notice, that the zeroth-order Hamiltonian $\Tilde{H}_{XXX}^{(0)}$ is equal to the undeformed and untwisted  Hamiltonian~\eqref{eq: xxx-h12}. Therefore, the equation with $r=N$ is just the eigenvalue equation for a state with $N$ excitations in the original $XXX_{-1/2}$ model. This implies that the wavefunction $\psi_{(N)}(\Vec{x}_{(N)})$ is given by the untwisted solution~\eqref{eq:undef-solution} and, most importantly, the energy $E_{(N)}$ of the eigenstate $\ket{\Psi}_{(N)}$ of the deformed Hamiltonian coincides with the energy of an eigenstate of the undeformed and untwisted Hamiltonian with $N$ excitations, see the equation~\eqref{eq:undef-energy}. All these considerations are valid provided that the equations can be recursively solved also for $r<N$.

Importantly, for some eigenstates of the undeformed spin-chain the equations with $r<N$ in the system~\eqref{eq:sist-wfs} do not have  a solution.  In order to prove this claim, suppose for a moment that the system of equations~\eqref{eq:sist-wfs} has a solution for every value of the allowed momenta. This would imply that the Hamiltonian~\eqref{eq:deformed-ham} is diagonalisable with the same eigenvalues and multiplicities as the undeformed model with periodic boundary conditions~\eqref{eq: xxx-h12}. Consequently, the deformed Hamiltonian~\eqref{eq:deformed-ham} would be related to the $XXX_{-1/2}$ model~\eqref{eq: xxx-h12} by a similarity transformation. However, from the discussion in  subsection~\ref{subsec:equivalence-with-boundary-twisted}, we know that the deformed Hamiltonian is similar to an undeformed spin-chain  with \emph{twisted} boundary conditions~\eqref{eq:ham-non-periodic}. Since the operator~\eqref{eq:twist-boundary} which implements the boundary conditions is non-trivial, it follows that for some values of the momenta the system~\eqref{eq:sist-wfs} does not have a solution.
Since  eigenvectors of the deformed Hamiltonian~\eqref{eq:deformed-ham} must be a solutions of~\eqref{eq:sist-wfs}, we conclude that the only possibility is that the deformed Hamiltonian~\eqref{eq:deformed-ham} is not diagonalisable, and that generalised eigenvectors appear whenever there is a wavefunction $\psi_{(N)}(\Vec{x}_{(N)})$ for which the equations with $r<N$ in~\eqref{eq:sist-wfs} have no solution. Obviously, this claim is also supported by the findings of the explicit examples worked out in the previous section.

In summary, the twist preserves the eigenvalues of the untwisted spin-chain, however it makes the Hamiltonian non-diagonalisable. Importantly, the eigenvalues of the Hamiltonian are still determined by the original undeformed Bethe equations. Some of the eigenvectors of the $XXX_{-1/2}$ model become generalised eigenvectors and may receive $\xi$-dependent corrections from states with fewer excitations. Conversely, other eigenstates of the undeformed and untwisted chain remain eigenvectors of the deformed model and may receive corrections from states with fewer and fewer excitations down to the vacuum. The corrections to the deformed eigenstates are controlled by the  deformation parameter $\xi$, so that
\begin{equation}
    \psi_{r} \left(\Vec{x}_{(r)}\right) \propto \xi^{N-r}, \qquad r = 0,\ldots,N. 
\end{equation}

It is instructive to discuss the structure of the eigenvalues and (generalised) eigenvectors of the Hamiltonian deformed by the twist $F'_{12}$, which is constructed with the raising generator $J^+$, as defined in~\eqref{eq:twist-raising}. The situation is in a sense the inverse of the one described above. Now, the operator $J^+$ in the twist~\eqref{eq:twist-raising} creates excitations. Therefore, we can label the (generalised) eigenstates by an integer which accounts for the minimum number of excitations in the superposition of the states. Moreover, projecting the eigenvalue equation to the states with the minimum number of excitations yields the undeformed eigenvalue equation, which allows us to conclude that also for the Hamiltonian deformed with $F'_{12}$ in~\eqref{eq:twist-raising} the eigenvalues and Bethe equations do not receive twist corrections. Moreover, with a similar reasoning as above, we can conclude that also in this case the Hamiltonian is non-diagonalisable and that generalised eigenvectors appear.

In what follows, we explicitly solve~\eqref{eq:sist-wfs} for a generic length $L$ of the chain and  for the lowest values of the maximum number of excitations. Together with the explicit examples already worked out in the previous section, this will confirm the above expectation for the eigenvalues and (generalised) eigenvectors) of the Hamiltonian~\eqref{eq:deformed-ham} deformed by the twist $F_{12}~\eqref{eq:twist}$. 

\vspace{12pt}

\textbf{N=1 eigenstates}. \, In this case, the ansatz~\eqref{eq:twisted-ansatz} reduces to
\begin{equation}
    \ket{\Psi}_{(1)} = \psi_{(0)} \ket{0}^{\otimes L} + \sum_{1 \leq x \leq L} \psi_{(1)} (x)|x).
\end{equation}
In order to obtain the constant $\psi_{(0)}$, we have to solve the following equation obtained from~\eqref{eq:sist-wfs}
\begin{align}
    E_{(1)} \psi_{(0)} \ket{0}^{\otimes L} = \Tilde{H}_{XXX}^{(1)} \sum_{1 \leq x \leq L} \psi_{(1)} (x)|x).
\end{align}
From the equation~\eqref{eq:expansion-ham-density}, we can compute the first order twisted Hamiltonian density,
\begin{align}
    \Tilde{h}_{12}^{(1)} = 2 \xi [h_{12},J^3 \otimes J^-]. \label{eq:first-ord-ham}
\end{align}
Its action on all the possible two-site states with exactly one excitation is given by
\begin{align}
     \Tilde{h}_{12}^{(1)} \ket{01} &= -\xi \ket{00}, \\
      \Tilde{h}_{12}^{(1)} \ket{10} &= \xi \ket{00}.
\end{align}
Therefore, the action of the first order twisted Hamiltonian on the one-excitation state is identically zero,
\begin{align}
    \Tilde{H}_{XXX}^{(1)} \sum_{1 \leq x \leq L} \psi_{(1)} (x)|x) = \sum_{1 \leq x \leq L} \psi_{(1)}(x) (\Tilde{h}_{x,x-1}^{(1)}+\Tilde{h}_{x,x+1}^{(1)}) |x) = 0,
\end{align}
which implies that the zero-order constant $\psi_{(0)}$ must vanish. Then, in the sector with at most one excitation, the energy spectrum and the eigenstates  of the deformed  Hamiltonian  coincide with those of the undeformed model, namely
\begin{align}
     \ket{\Psi}_{(1)} = \sum_{1\leq x \leq L} e^{i p x} |x),\quad 
     E_{(1)} = 4 \sin^2\left(\frac{p}{2}\right),\quad 
     p = \frac{2\pi n}{L}, \, n = 0,\ldots,L-1. \label{eq:psi1-def-xxx}
\end{align}
In the sector with maximum one excitation, the deformed $XXX_{-1/2}$ Hamiltonian~\eqref{eq:deformed-ham} reduces to the undeformed $XXX_{-1/2}$ Hamiltonian~\eqref{eq: xxx-h12}.

\vspace{12pt}

\textbf{N=2 eigenstates}. \, In this case, the initial ansatz is
\begin{equation}
    \ket{\Psi}_{(2)} = \psi_{(0)} \ket{0}^{\otimes L} + \sum_{1 \leq x \leq L} \psi_{(1)} (x)|x) +\sum_{1 \leq x_1 \leq x_2 \leq L} \psi_{(2)} (x_1,x_2)|x_1,x_2) . \label{eq:twsited-eigenstate-2ex}
\end{equation}
The wavefunction $\psi_{(2)} (x_1,x_2)$ for the state with two excitations and the energy $E_{(2)}$ are fixed by the undeformed spectral problem~\eqref{eq:undef-solution}
\begin{align}
    \psi_{(2)} (x_1,x_2) &= e^{ip_1x_1+ip_2x_2} + S_{xxx}(p_2,p_1) e^{ip_2x_1+ip_1x_2}, \label{eq:undeformed-wf2}\\
    E_{(2)} &= \epsilon\left(p_1\right)+\epsilon\left(p_2\right), \; \epsilon\left(p_j\right)= 4 \sin^2{\frac{p_j}{2}}, \\
    e^{i p_1 L} &= S_{xxx}(p_1,p_2), \quad
    e^{i p_2 L} = S_{xxx}(p_2,p_1).
\end{align}
To obtain the constant $\psi_{(0)}$ and the wavefunction $\psi_{(1)}(x)$, we must solve the following system of equations derived from~\eqref{eq:sist-wfs}
\begin{align}
    \left(E_{(2)}-H_{XXX}\right)\sum_{1 \leq x \leq L} \psi_{(1)} (x)|x) = \Tilde{H}_{XXX}^{(1)}\sum_{1 \leq x_1 \leq x_2 \leq L} \psi_{(2)} (x_1,x_2)|x_1,x_2), \label{eq:2wf1}\\
    E_{(2)} \psi_{(0)} \ket{0}^{\otimes L} =  \Tilde{H}_{XXX}^{(1)}\sum_{1 \leq x \leq L} \psi_{(1)} (x)|x) +\Tilde{H}_{XXX}^{(2)} \sum_{1 \leq x_1 \leq x_2 \leq L} \psi_{(2)} (x_1,x_2)|x_1,x_2).\label{eq:2wf0}
\end{align}
In order to solve the above equations, we need the action of the Hamiltonian density (at first and second-order in the expansion of the deformation parameter) on all possible two-site states containing up to two excitations. From the equation~\eqref{eq:first-ord-ham}, we can compute the action of the deformed Hamiltonian density at order one
\begin{align}
    \Tilde{h}_{12}^{(1)} \ket{10} &= -\Tilde{h}_{12}^{(1)} \ket{01} =\xi \ket{00} , &
    \Tilde{h}_{12}^{(1)} \ket{11} &= - 3 \xi \ket{10} - \xi \ket {01}, \\
    \Tilde{h}_{12}^{(1)} \ket{20} &= 3 \xi \ket{10} + \xi \ket {01}, &
    \Tilde{h}_{12}^{(1)} \ket{02} &=  \xi \ket{10} - \xi \ket {01},
\end{align}
On the other hand, the deformed Hamiltonian density at order two takes the form
\begin{align}
    \Tilde{h}_{12}^{(2)} &= 2 \xi^2 \left(J^3 \otimes \left(J^-\right)^2  + \left(J^3\right)^2 \otimes \left(J^-\right)^2  \right) h_{12} \nonumber -4 \xi^2 \left(J^3 \otimes J^-\right) h_{12} \left(J^3 \otimes J^-\right) \nonumber \\
    &+2 \xi ^2 h_{12} \left(\left(J^3\right)^2 \otimes \left(J^-\right)^2 - J^3 \otimes \left(J^-\right)^2  \right).
\end{align}
Its action on  all possible two-site states with exactly two excitations is given by
\begin{align}
    \Tilde{h}_{12}^{(2)} \ket{20} = -\frac{3}{2} \xi^2 \ket{00}, \quad
    \Tilde{h}_{12}^{(2)} \ket{02} = \frac{5}{2} \xi^2 \ket{00}, \quad 
    \Tilde{h}_{12}^{(2)} \ket{11} = 0.
\end{align}
With this and with $\Tilde{h}_{12}^{(2)} \ket{10}=\Tilde{h}_{12}^{(2)} \ket{01}=0$, equations~\eqref{eq:2wf1} and~\eqref{eq:2wf0} can be rewritten as the following equations for the wavefunction $\psi_{(1)}(x)$ and the constant $\psi_{(0)}$, 
\begin{align}
    \left(E_{(2)} - 2\right) \psi_{(1)}(x) &+ \psi_{(1)}(x+1) + \psi_{(1)}(x-1)= \nonumber \\
    &= \xi \left[ 2 \psi_{(2)}(x,x) + \psi_{(2)}(x+1,x+1) \right. \nonumber \\
    &\quad + \psi_{(2)}(x-1,x-1) - 2 \psi_{(2)}(x,x+1) \nonumber \\
    &\quad \left. - 2 \psi_{(2)}(x-1,x) \right], \\
    E_{(2)} \psi_{(0)} &= \xi^2 \sum_{1 \leq x \leq L} \psi_{(2)}(x,x).
\end{align}
It is natural to try to solve the first of the above equations with an ansatz of the form $\psi_{(1)}(x) = A_{(1)} e^{i k^{(1)} x}$. Substituting back this ansatz and the solution~\eqref{eq:undeformed-wf2} for the wavefunction $\psi_{(2)}(x_1,x_2)$, we obtain 
\begin{align}
    \left( E_{(2)} - \epsilon\left(k^{(1)}\right) \right) A_{(1)} e^{ik^{(1)}x} 
    &= \xi e^{i(p_1 + p_2)x} \Big[B(p_1,p_2) + S_{xxx}(p_2, p_1) B(p_2,p_1) \Big], \label{eq:for-A1}\\
    E_{(2)}\psi_{(0)} 
    &= \xi^2 \left( 1 + S_{xxx}(p_2, p_1) \right) \left(\sum_{1 \leq x \leq L} e^{i(p_1 + p_2)x}\right)\label{eq:for-psi0},
\end{align}
where
\begin{align}
    B(p_1,p_2) = 2 - 2 e^{-ip_1} - 2 e^{ip_2} + e^{i(p_1 + p_2)} + e^{-i(p_1 + p_2)}.
\end{align}
Since the equation~\eqref{eq:for-A1} should be valid for every $x$, it must be satisfied that $ k^{(1)} = p_1 + p_2 \pmod{2\pi}$. At the same time, the periodic boundary conditions on $\psi_{(1)}(x)$ imply the following condition on the momentum $k^{(1)}$,
\begin{align}
    k^{(1)} = p_1 + p_2 \pmod{2\pi} = \frac{2\pi n}{L}, \qquad n=0,\ldots,L-1
\end{align}
The above identity is automatically satisfied  because we obviously want the wavefunction $\psi_{(2)}(x_1,x_2)$ to be compatible with the periodic boundary conditions. In fact, the Bethe equations on the momenta $p_1,p_2$ imply
\begin{align}
    e^{i(p_1+p_2)L} = S_{xxx}(p_1,p_2)S_{xxx}(p_2,p_1) = 1 \Rightarrow p_1 + p_2 \pmod{2\pi} = \frac{2\pi n}{L}, \; n=0,\ldots,L-1.
\end{align}
 Let us now distinguish between two scenarios. First, consider the case when $\epsilon\left(p_1\right)+\epsilon\left(p_2\right) \neq \epsilon\left(p_1+p_2\right)$. Then, the left-hand side of the equation~\eqref{eq:for-A1} never vanishes. Moreover, the former condition implies that the total energy $\epsilon\left(p_1\right)+\epsilon\left(p_2\right)$ is non-zero, and the left hand side of the equation~\eqref{eq:for-psi0} does not cancel either. Therefore, after some simplifications we arrive at the following solution
\begin{align}
    A_{(1)} &= 4 \xi \left(\frac{e^{i p_1}-e^{ip_2}}{1-2e^{ip_2}+e^{i(p_1+p_2)}}\right), \\
    \psi_{(0)} &= \delta_{0,k^{(1)}}\frac{2 \xi^2}{\epsilon\left(p_1\right)+\epsilon\left(p_2\right)} \left(\frac{e^{i p_1}-e^{ip_2}}{1-2e^{ip_2}+e^{i(p_1+p_2)}}\right) L .  
\end{align}
Interestingly, $\psi_{(1)}(x)$ is proportional to the $XXX_{-1/2}$ wavefunction of two excitations evaluated at the same position 
\begin{align}
    \sum_{1 \leq x \leq L} A_{(1)} e^{i(k_1+k_2)x} |x) = 2\xi \sum_{1 \leq x \leq L} \Psi_{(2)}(x,x) |x). \label{eq:first-order-twisted-eigenstate-2ex}
\end{align}
Suppose now that $\epsilon\left(p_1\right)+\epsilon\left(p_2\right) = \epsilon\left(p_1+p_2\right)$. This condition forces the pseudomomenta to satisfy $p_1+p_2=\pi$, or that one of the two momenta---and only one of them---is zero. The option $p_1+p_2=\pi$ is only possible if the length of the spin-chain is an even number. In this case, the constant $\psi_{(0)}$ equals zero. Moreover, both sides of~\eqref{eq:for-A1} vanish, which implies that $A_{(1)}$ is an arbitrary constant. This can be understood by  noting that for this value of the total momentum, both the state $\sum_{1 \leq x \leq L} e^{i(p_1+p_2)}|x)$  and $ \sum_{1 \leq x_1 \leq x_2 \leq L} \left(e^{ip_1x_1+ip_2x_2} + S_{xxx}(p_2,p_1) e^{ip_2x_1+ip_1x_2}\right) |x_1,x_2)$  are eigenstates of the deformed Hamiltonian with the same eigenvalue. Therefore, any linear combination of both states remains an eigenstate of the deformed Hamiltonian. Without loss of generality, we can set the constant $A_{(1)}$ to zero, so that the eigenstate $\ket{\Psi}_{(2)}$ coincides with the undeformed eigenstate for two excitations. The same is true when only one of the two momenta is zero. When both momenta are zero,~\eqref{eq:for-psi0} becomes
\begin{align}
    0 \cdot \psi_{(0)} = 2 \xi^2 L \neq 0.
\end{align}
Therefore, there is no solution to~\eqref{eq:for-psi0} when the momenta are $p_1=p_2=0$. Observe that the fact that there is no solution to~\eqref{eq:for-psi0} for two vanishing momenta has nothing to do with the particular ansatz chosen for $\psi_{(1)}(x)$, since the one-excitation component of~\eqref{eq:twsited-eigenstate-2ex} does not enter equation~\eqref{eq:for-psi0}. Importantly, one can verify that the undeformed state for this value of the momenta is a generalised eigenstate of rank two in the chain of the zero eigenvalue,
\begin{align}
    \Tilde{H}_{XXX} \sum_{1\leq x_1 \leq x_2 \leq L} |x_1,x_2) & = \xi^2 L   \ket{0}^{\otimes L},\\
    \Tilde{H}_{XXX}^2 \sum_{1\leq x_1 \leq x_2 \leq L} |x_1,x_2) & =  0.
\end{align}
In Table~\ref{tab:gen-eigen-ham-2ex} we summarise the set of (generalised) eigenvectors of the deformed Hamiltonian~\eqref{eq:deformed-ham} in the subspace with at most two excitations, for arbitrary length $L$.  When the spin-chain length is set to $L=2$, we recover the (generalised) eigenvectors listed in Table~\ref{tab:gen-eigen-ham-length2-2ex}, up to a normalisation factor.

\begin{table}[h!]
    \centering
    \begin{tabular}{c|c}
 Momenta  & (Generalised)  \\
   & eigenvector\\
 \hline
 $-$ & $\ket{0}^{\otimes L}$ \\
 \hline
 $p=\frac{2\pi n}{L}, \, n=0,\ldots, L-1$& $\ket{\Psi_{xxx}}_{(1)}$ \\
 \hline
 \makecell{$p_1+p_2 \pmod{2\pi} = \pi$ \\ \text{or}\\
 $p_1=0, p_2 \neq 0$ } & $\ket{\Psi_{xxx}}_{(2)}$ \\
 \hline
 $p_1 = p_2 = 0$ & $\frac{1}{\xi^2 
 L}\ket{\Psi_{xxx}}_{(2)}$\\
 \hline
 \makecell{$p_1+p_2 \pmod{2\pi} = 0$ \\ \text{with} $p_1 \neq 0, p_2 \neq 0$} & \makecell{$\ket{\Psi_{xxx}}_{(2)} + 2\xi \sum_{1 \leq x \leq L} \psi_{(2)}(x,x) |x)$\\ $+ \frac{\xi^2 L}{4\left(1-\cos p_1\right)} \left(1+e^{i p_1}\right) \ket{0}^{\otimes L}$} \\
 \hline
\text{rest of} $p_1, p_2$  &  $\ket{\Psi_{xxx}}_{(2)} + 2\xi \sum_{1 \leq x \leq L} \psi_{(2)}(x,x) |x)$ \\
 
    \end{tabular}
\caption{(Generalised) eigenvectors and momenta of the deformed Hamiltonian~\eqref{eq:deformed-ham} in the subspace with at most two excitations for arbitrary length $L$. The states $\ket{\Psi_{xxx}}_{N}$ are the eigenstates of the $XXX_{-1/2}$ with $N$ excitations~\eqref{eq:undef-solution}, while $\psi_{(N)}(x_1,...x_N)$ denotes its associated wavefunction. The generalised eigenvector is the state $\frac{1}{\xi^2 
 L}\ket{\Psi_{XXX}}_{(2)}$ with $p_1 = p_2 =0$. The normalisation condition is chosen so that the action of~\eqref{eq:deformed-ham} on it gives the vacuum $\ket{0}^{\otimes L}$. The momenta $p_1,p_2$ are always subject to the undeformed Bethe equations. }
\label{tab:gen-eigen-ham-2ex}
\end{table}

\section{The Jordanian $XXX$ spin-chain from a scaling limit of the $XXZ$ spin-chain}\label{sec:XXZ}
It was proven in~\cite{Kulish2009} that the Jordanian twist of the $XXX_{1/2}$ spin-chain can be obtained by applying a similarity transformation to the $XXZ_{1/2}$ model followed by a singular scaling limit. In this section, we provide  evidences that support this statement also in the $s=-1/2$ representation.  
\subsection{The $XXZ_{-1/2}$ Hamiltonian}
Let us briefly review the construction of the $XXZ_{-1/2}$ Hamiltonian. We adopt the conventions of~\cite{Frassek:2019isa}.
The $XXZ_{-1/2}$ Hamiltonian density is an  operator invariant under the quantum algebra $U_{q}(\mathfrak{sl}_2)$. This algebra is spanned by the generators $\{S^\pm,S^3\}$ with  commutation relations 
\begin{align}
    [S^+,S^-] = - [2S^3]_q, \qquad [S^3,S^\pm] = \pm S^\pm,
\end{align}
where the $q$-number takes the form, $\left[n\right]_q =\frac{q^n-q^{-n}}{q-q^{-1}}$. The quadratic Casimir is defined to be
\begin{equation}
     \mathsf{C} = \left[S^3\right]_q \left[S^3-1\right]_q-S^+S^-, 
\end{equation}
while the coproduct is taken to be
\begin{align}
    \Delta_{q}(S^3) &= S^3 \otimes \mathbb{I} + \mathbb{I} \otimes S^3, \\
    \Delta_{q}(S^\pm) &= S^\pm \otimes q^{-S^3} + q^{S^3} \otimes S^\pm .\label{eq:xxz_cop_s+}
\end{align}
In the $s=-1/2$ representation, the generators read
\begin{align}
    S^3 &= \sum _{m=0}^\infty \left(\frac{1}{2}+m\right) \ket{m} \bra{m}, \\
    S^+ &= \sum_{m=0}^\infty \left[1+m\right]_q \ket{m+1} \bra{m}, \\
    S^- &= \sum_{m=0}^\infty \left[1+m\right]_q \ket{m} \bra{m+1}, 
\end{align}
The Hamiltonian density acts on the tensor product of two fundamental $s=-1/2$ modules, $D_F \otimes D_F$. This product decomposes in an infinite sum of irreducible modules,
\begin{align}
    D_{F} \otimes D_{F} = \bigoplus_{j=0}^\infty D_{j}. \label{eq:irreducible-decomposition-xxz}
\end{align}
Each module $D_{j}$ is an eigenspace of the coproduct of the quadratic Casimir, with eigenvalue $\left[j+1\right]_q \left[j\right]_q$, and possesses a lowest-weight state $\ket{\chi_{j}}$ that satisfies
\begin{align}
    \Delta_{q}(S^-) \ket{\chi_{j}} = 0, \quad
    \Delta_{q}(S^3)\ket{\chi_{j}} = (j+1) \ket{\chi_{j}}.
\end{align}
The $XXZ_{-1/2}$ Hamiltonian density is diagonal on the modules $D_{j}$~\cite{Bytsko:2001uh},
\begin{align}
    \mathsf{h}_{12} = 2 \sinh{\eta} \sum_{j=0}^\infty \left(\prod_{k=1}^j  \frac{\cosh{k \eta}}{\sinh{k \eta }}\right) \mathsf{P}_{12,j}, \label{eq:xxz-ham-density}
\end{align}
where $\eta$ is the anisotropic parameter defined by $q=e^{ \eta}$ and $\mathsf{P}_{12,j}$ is the projector to the module $D_{j}$,
\begin{align}
    \mathsf{P}_{12,j} = \prod_{\substack{r=0 \\r \neq j}}^\infty \frac{\Delta_q(\mathcal{C})-\left[r+1\right]_q\left[r\right]_q}{\left[j+1\right]_q\left[j\right]_q-\left[r+1\right]_q\left[r\right]_q}.
\end{align}
A direct computation shows that the coproduct of the quadratic Casimir takes the form \footnote{In our conventions,~\eqref{eq:q-coproduct-quadratic-casimir} coincides with the result given in~\cite{Bytsko:2001uh} provided we make the substitution $S^+ \to -S^+$ (or $S^- \to -S^-$) and extend the spin label to the negative value $s=-1/2$.}
\begin{align}
    \Delta_{q}(\mathsf{C}) &= e^{\eta S_1^3} \Bigg( \frac{2 \sinh(\eta S_1^3) \sinh(\eta S_2^3)}{\sinh^2(\eta)} \cosh^2\left( \frac{\eta}{2} \right) \nonumber \\
    &\quad - \frac{2 \cosh(\eta S_1^3) \cosh(\eta S_2^3)}{\sinh^2(\eta)} \sinh^2\left( \frac{\eta}{2} \right) - S_1^+ S_2^- - S_1^- S_2^+ \Bigg) e^{-\eta S_2^3}, \label{eq:q-coproduct-quadratic-casimir}
\end{align}
where the subscript $1$ or $2$ on the generators simply indicates the space of the tensor product on which they are acting.
Clearly, the $U_{q}(\mathfrak{sl}_2)$ generators are connected to those of $\mathfrak{sl}_2$ in the $q \to 1$ limit,
\begin{align}
    S^3 \to J^3, \quad S^{\pm} \to J^{\pm} + \mathcal{O}(q-1)^2. 
\end{align}
Therefore, in this limit the $q$-deformed projectors $\mathsf{P}_{12,j}$ turn into the $\mathfrak{sl}_2$ projectors $P_{12,j}$. Furthermore, the eigenvalues become the harmonic numbers,
\begin{align}
    2 \sinh{\eta} \prod_{k=1}^j  \frac{\cosh{k \eta}}{\sinh{k \eta }} \rightarrow 2 \prod_{k=1}^j \frac{1}{k} = 2 h(j),
\end{align}
thus recovering the $XXX_{-1/2}$ spin-chain~\eqref{eq: xxx-h12}.

The total Hamiltonian $H_{XXZ}$ is not invariant under the global $U_{q}(\mathfrak{sl}_2)$ symmetry. The boundary Hamiltonian density $\mathrm{h}_{L1}$ breaks the symmetry under $\Delta^{(L-1)}(S^\pm)$. However, the $XXZ$ Hamiltonian does commute with $\Delta^{(L-1)}(S^3)$ and therefore possesses a $\mathfrak{u}_1$ symmetry that allows one to construct the eigenstates by fixing the number of excited spins. The eigenstates are essentially given by the same expression~\eqref{eq:undef-solution} as for those of the $XXX$ spin-chain, but with an $S$-matrix and a dispersion relation of the form 
\begin{align}
    S_{xxz}(\lambda_1,\lambda_2) &= \frac{\sinh{(\lambda_1-\lambda_2-\eta)}}{\sinh{(\lambda_1-\lambda_2+\eta)}}, \\
    \epsilon_\eta(\lambda) &= \frac{2(\cosh^2{\eta}-1)}{\cosh{\eta}-\cosh{2\lambda}}, \label{eq:XXZ-energy}
\end{align}
where the pseudorapidity is related to the momenta by the following equation $\eta$-dependent relation
\begin{align}
    p_{\eta}(\lambda) = \frac{1}{i} \log\frac{\sinh{(\lambda+\eta/2)}}{\sinh{(\lambda-\eta/2)}}. \label{eq:xxz-momentum}
\end{align}
We refer to appendix~\ref{ap:dispersion-relation-s-matrix-xxz} for more details.
In order to recover the $S$-matrix~\eqref{eq:undef-solution} and the dispersion relation~\eqref{eq:undef-energy} of the $XXX_{-1/2}$ model, it is necessary to simultaneously rescale the anisotropy parameter and the pseudorapidity as $\eta \to \epsilon \eta$ and $\lambda \to \epsilon \lambda$, then  take the limit $\epsilon \to 0$, and finally set $\eta \to i$.

\subsection{The scaling limit of the $XXZ_{-1/2}$ spin-chain}
\label{subsec:sc-lim-xxz-ham}
In this subsection we show that a Jordanian deformation of the $XXX_{-1/2}$ spin-chain can be obtained from the $XXZ_{-1/2}$ model. In particular, we will prove that the Jordanian twist implementing the deformation is in a $r$-symmetric form, which means that its first order expansion in the parameter deformation is automatically antisymmetric in the two spaces and coincides with the Jordanian solution of the classical Yang-Baxter equation.

To this end, consider the operator $M = e^{\xi J^{-}}$, where $\xi$ is a free parameter that, after applying the scaling limit, will be identified with the Jordanian parameter deformation. Importantly, the operator $M$ is constructed with the $\mathfrak{sl}_2$ generator $J^-$ and \emph{not} with the $U_{q}(\mathfrak{sl}_2)$ generator $S^-$. We perform a similarity transformation on the $XXZ_{-1/2}$ Hamiltonian, $H_{XXZ} \to \Ad_{M(\xi)}^{\otimes L} H_{XXZ}$. The Hamiltonian density is modified as follows,
\begin{align}
    \mathsf{h}_{12} \to  2 \sinh{\eta} \sum_{j=0}^\infty \left(\prod_{k=1}^j  \frac{\cosh{k \eta}}{\sinh{k \eta }}\right) \Ad_{M(\xi)}^{\otimes 2} \mathsf{P}_{12,j},
\end{align}
where the $q$-deformed projectors change accordingly to
\begin{align}
    \Ad_{M(\xi)}^{\otimes 2} \mathsf{P}_{12,j} = \prod_{\substack{r=0 \\r \neq j}}^\infty \frac{\Ad_{M(\xi)}^{\otimes 2} \Delta_{q}(\mathsf{C})-\left[r+1\right]_q\left[r\right]_q}{\left[j+1\right]_q\left[j\right]_q-\left[r+1\right]_q\left[r\right]_q}.
\end{align}
Next, we apply a scaling limit $\epsilon \to 0$ to the anisotropic factor and the pseudorapidity\footnote{The rescaling of the pseudorapidity will be relevant when discussing the limit of the eigenvectors in the following section.} together with a singular scaling of the $\xi$ parameter,
\begin{align}
    \eta \to \epsilon \eta, \quad \lambda \to \epsilon \lambda, \quad \xi \to \frac{\xi}{ \eta \epsilon}. \label{eq:rescale}
\end{align}
In this limit, we obtain the following Hamiltonian density
\begin{align}
    \mathsf{h}_{12} \rightarrow \hat{h}_{12}=\sum_{j=0}^\infty 2 h(j) \hat{P}_{12,j}, \label{eq:scaling-h12}
\end{align}
where
\begin{align}
    \hat{P}_{12,j} = \prod_{\substack{r=0 \\r \neq j}}^\infty \frac{\hat{\Delta}(C)-(r+1)r}{(j+1)j-(r+1)r}, \quad \text{with} \quad 
    \hat{\Delta}(C) = \lim_{\epsilon\to 0} \Ad_{M(\xi/\eta \epsilon)}^{\otimes 2} \Delta_{q}(\mathsf{C}).
\end{align}
The operators $\hat{P}_{12,j}$ are the projectors to the modules $\hat{V}_{j} = \lim_{\epsilon \to 0} M(\xi / \eta \epsilon )^{\otimes 2} D_{j}$. In this limit, the quadratic Casimir becomes diagonal on $\hat{V}_{j}$, with eigenvalues given by $j (j+1)$. We conclude that the Hamiltonian~\eqref{eq:scaling-h12} is a deformation of the $XXX_{-1/2}$ Hamiltonian density that preserves the eigenvalues and the multiplicity of its spectrum. Therefore, both Hamiltonian densities should be related by a similarity transformation
\begin{align}
    \hat{h}_{12} = \hat{F}_{12} h_{12} \hat{F}_{12}^{-1} \ \iff \ \hat{\Delta}(C) = \hat{F}_{12} \Delta(C) \hat{F}_{12}^{-1}.
\end{align}
That is to say, starting from a Hopf algebra given by the $q$-deformed $U_q(\mathfrak{sl}_2)$ algebra we have ended up with a twisted version of the $\mathfrak{sl}_2$ Yangian, $\mathcal{Y}_{\xi}(\mathfrak{sl}_2)$, which is constructed by twisting the coproduct~\eqref{eq:sl2-coproduct} with the adjoint action of the twist operator $\hat{F}_{12}$.

According to a theorem by Drinfeld~\cite{drinfeld1983constant}, every twist is in one to one correspondence with a solution of the classical Yang-Baxter equation. In particular, after antisymmetrising the first non-trivial order that is found from  expanding the twist in the deformation parameter, one obtains a solution of the classical Yang-Baxter equation. Therefore, in order to identify the type of twist of $\hat{F}_{12}$ it is enough to compute the first order of the coproduct of the Casimir $\hat{\Delta}(C)$, which fixes the first order of $\hat{F}_{12}$ by the equation
\begin{align}
    \hat{\Delta}^{(1)}(C) = [\hat{F}_{12}^{(1)},\Delta(C)],
\end{align}
where the superscripts denote the order of the expansion
\begin{align}
    \hat{F}_{12} = \mathbb{I} \otimes \mathbb{I} + \xi \hat{F}^{(1)}_{12} + O(\xi^2) \quad \text{and}\quad \hat{\Delta}(C) = \Delta(C) + \xi \hat{\Delta}^{(1)}(C) + O(\xi^2).
\end{align}
To obtain the order $O(\xi)$ of $\hat{\Delta}(C)$, first notice that the adjoint action by $M$ of the $q$-deformed coproduct of the Casimir takes the form
\begin{align}
    \Ad_{M(\xi)}^{\otimes 2} \Delta_{q}(\mathsf{C}) = \sum_{n,m=0 }^\infty \frac{(-1)^{m}\xi^{n+m}}{m!n!} \left(\Delta(J^-)\right)^n  \Delta_{q}(\mathsf{C})\left(\Delta(J^-)\right)^m, \label{eq:adM-Cq}
\end{align}
If we now introduce the rescaling~\eqref{eq:rescale}, the $q$-deformed Casimir~\eqref{eq:q-coproduct-quadratic-casimir} reads,
\begin{align}
    \Delta_{q}(\mathsf{C}) \rightarrow \Delta(C) + \eta \left(J^3_1 \Delta(C) - \Delta(C)J^3_2\right) \epsilon + O(\epsilon^2).
\end{align}
Therefore, the first order of~\eqref{eq:adM-Cq} rescales as follows,
\begin{align}
    \xi[\Delta(J^-),\Delta_{q}(\mathsf{C})] \rightarrow &\frac{\xi}{\eta \epsilon}[\Delta(J^-),\Delta(C)] + \xi [\Delta(J^-),J^3_1 \Delta(C) - \Delta(C)J^3_2] + O(\epsilon) = \nonumber \\
    &=\xi \left(J^-_1 \Delta(C) - \Delta(C)J^-_2\right) + O(\epsilon).
\end{align}
The $O(\frac{1}{\epsilon})$ divergences cancel thanks to the fact that $C$ is Casimir, and then in the $\epsilon \to 0$ limit we obtain a finite result $\hat{\Delta}^{(1)}(C) = J^-_1 \Delta(C) - \Delta(C)J^-_2$. In order to extract an expression for the first order expansion of the twist, we need to rewrite $\hat{\Delta}^{(1)}(C)$ in the form of a commutator. Expanding $\hat{\Delta}^{(1)}(C)$ in terms of monomials of $a^\dagger$ and $a$ one can verify that
\begin{align}
    \left(J^- \otimes \mathbb{I}\right) \Delta(C) - \Delta(C)\left(\mathbb{I} \otimes J^-\right) = [J^- \wedge J^3, \Delta(C) ]. \label{eq:id-r-symetric-twsit}
\end{align}
We then conclude that $\hat{F}^{(1)}_{12} = J^- \wedge J^3$, which coincides precisely with the classical Jordanian $r$-matrix. Therefore,  the scaling limit of the $XXZ_{-1/2}$ model transformed by the adjoint action by $M$ leads to a Jordanian deformed $XXX_{-1/2}$ spin-chain, with a twist $\hat{F}_{12}$ which is $r$-symmetric. This twist belongs to the same equivalence class as the Jordanian twist of~\eqref{eq:twist}. This implies that there exists an operator $w$ that relates the two twists as~\cite{Giaquinto:1994jx,tolstoy2007twistedquantumdeformationslorentz}
\begin{align}
    \hat{F}_{12} = w^{-1} \otimes w^{-1} F_{12} \Delta(w). \label{eq:for-first-order-symmetric-twist}
\end{align}
Consequently, the two Hamiltonians are related by a similarity transformation given by the $L$-tensor product of the operator $w$,
\begin{align}
    \Tilde{H}_{XXX}= \Ad_w^{\otimes L} \hat{H}_{XXX}.
\end{align}
Knowing the first order of $\hat{F}_{12}$, it is possible to obtain the first order of $w$ by solving the equation
\begin{align}
    \hat{F}^{(1)}_{12}-F^{(1)}_{12} = \Delta(w^{(1)})-\left(w^{(1)} \otimes \mathbb{I}+\mathbb{I}\otimes w^{(1)}\right).
\end{align}
The left-hand side of the previous equation reads as $J^- \otimes J^3 + J^3 \otimes J^-$, therefore a possible solution to the order $O(\xi)$ of $w$ is given by 
\begin{equation}
    w^{(1)} = J^3 J^-. \label{eq:s.t-twisted-ham}
\end{equation}
It is interesting to compare the result obtained in this subsection with that in the spin $1/2$ representation~\cite{Kulish2009}. In the $1/2$ representation, the Hamiltonian obtained in the scaling limit procedure is equal to a Jordanian twisted Hamiltonian with the non-$r$-symmetric twist~\eqref{eq:twist}. This result can be understood as follows. In the $1/2$ representation and in the basis~\eqref{eq:sl2}, the $\mathfrak{sl}_2$ generators can be written in terms of the Pauli matrices
\begin{align}
    J^3 = \frac{1}{2}\sigma^3, \quad J^+ = - \sigma^+, \quad J^- = \sigma^-.
\end{align}
Then, it is straightforward to verify that the right-hand side of~\eqref{eq:for-first-order-symmetric-twist} can be recast as a commutator involving the coefficient at order $O(\xi)$ of the twist~\eqref{eq:twist}
\begin{align}
    [\sigma^-\wedge\frac{1}{2}\sigma^3,\Delta(C)] = [-\sigma^3\otimes\sigma^-,\Delta(C)].
\end{align}
Therefore, in the spin $1/2$ representation the $r$-symmetric twist is physically indistinguishable from the twist~\eqref{eq:twist}, as both give rise to the same Hamiltonian.

\subsection{(Generalised) eigenvectors of the Jordanian  $XXX_{-1/2}$ Hamiltonian from the $XXZ_{-1/2}$ eigenvectors.}
\label{subsec:sc-limit-eig}
In this subsection, we explain how to obtain the (generalised) eigenvectors of the Jordanian twisted Hamiltonian $\hat{H}_{XXX}$ as a scaling limit of the eigenvectors of the $XXZ_{-1/2}$ model. \footnote{See~\cite{NietoGarcia:2021kgh} for a discussion of the construction of the (generalized) eigenvectors of a non-diagonalisable Hamiltonian as a limit of the eigenvectors of a diagonalisable one in a model different from ours.} The structure of the eigenvectors coincides with those of $\tilde{H}_{XXX}$ found in section~\ref{section:CBA}. Indeed, the $J^-$ in the operator $M$ annihilates excitations down to the vacuum; hence, the eigenstates will be labelled by a maximum number of excitations, inherited from the number of excitations of the $XXZ$ eigenvector from which we start.

As we will see, at the level of the eigenstates, in general this limit is singular. However, we find that we can consistently subtract the divergences to obtain  finite  eigenvectors. We will illustrate this procedure by considering states with up to one and two excitations.

\subsubsection*{Example I: at most one excitation}
The Bethe state for a single excitation in the $XXZ_{-1/2}$ model  is given by
\begin{align}
    \ket{\Psi_{xxz}}_{(1)}= \sum_{1 \leq x \leq L} \Psi_{xxz}(x) |x), \label{eq:xxz_state-1ex}
\end{align}
where $\Psi_{xxz}(x)$ is the $XXZ_{-1/2}$ Bethe wavefunction for a one-excitation state
\begin{align}
    \Psi_{xxz}(x)=\left(\frac{\sinh{(\lambda+\eta/2)}}{\sinh{(\lambda-\eta/2)}}\right)^x.
\end{align}
For the moment,  we will consider the pseudorapidity to be a continuum variable not necessarily quantised by the periodic boundary conditions.
Under the similarity transformation with $M=e^{\xi J^-}$ and the rescaling~\eqref{eq:rescale}, the state~\eqref{eq:xxz_state-1ex} transforms as follows
\begin{align}
   \ket{\Psi_{xxz}}_{(1)} &\rightarrow M (\xi/\epsilon \eta)^{\otimes L} \ket{\Psi_{xxz}^\epsilon}_{(1)} = \nonumber \\
   &=\sum_{1 \leq x \leq L} \left(\frac{\sinh{(\epsilon(\lambda+\eta/2))}}{\sinh{(\epsilon(\lambda-\eta/2))}}\right)^x\left(|x) + \left(\frac{\xi}{\epsilon \eta}\right) \ket{0}^{\otimes L}\right), \label{eq:M-trasnform-state}
\end{align}
where $\ket{\Psi_{xxz}^\epsilon}_{(1)}$ denotes the rescaled $XXZ_{-1/2}$ eigenvector. Notice that the state above is singular in the limit $\epsilon \to 0$. To find a regular eigenstate, we write down the expansion in powers of $\epsilon$ of the eigenvalue equation for~\eqref{eq:M-trasnform-state}. First, the $XXZ_{-1/2}$ Bethe wavefunction and the energy~\eqref{eq:XXZ-energy} are even functions in the variable $\epsilon$ with a zero-order coefficient which coincides with their counterparts in the $XXX_{-1/2}$ chain. Thus, we obtain the following equation
\begin{align}
    &\left(\hat{H}_{XXX} + \epsilon H_{XXZ_{\epsilon}}^{(1)} + O(\epsilon^2)\right) \left(\left(\frac{\xi}{\epsilon \eta}\right) \left(\sum_{1 \leq x \leq L}\Psi_{xxx}(x) \right)\ket{0}^{\otimes L} + \ket{\Psi_{xxx}}_{(1)}+O(\epsilon)\right) = \nonumber \\
    &= \left(E_{xxx} + O(\epsilon^2)\right)\left(\left(\frac{\xi}{\epsilon \eta}\right) \left(\sum_{1 \leq x \leq L}\Psi_{xxx}(x) \right)\ket{0}^{\otimes L} + \ket{\Psi_{xxx}}_{(1)}+O(\epsilon)\right), \label{eq:eigenvalue-eq1}
\end{align}
where $H_{XXZ_{\epsilon}}^{(1)}$ denotes the order $O(\epsilon)$ in the expansion of $\Ad_{M\left(\xi/\epsilon \eta \right)}^{\otimes L} H_{XXZ}$, $\ket{\Psi_{xxx}}$ the undeformed wavefunction~\eqref{eq:undef-solution} and $\Psi_{xxx}$ its associated wavefunction. Since $\epsilon$ is a  variable that has been introduced by hand, the equation above must be satisfied at each order in $\epsilon$. In particular, projecting at order zero and $O\left(\frac{1}{\epsilon}\right)$ yields
\begin{align}
    O\left(\frac{1}{\epsilon}\right)&: \hat{H}_{XXX} \left(\sum_{1 \leq x \leq L}\Psi_{xxx}(x) \right)\ket{0}^{\otimes L} = E_{xxx}  \left(\sum_{1 \leq x \leq L}\Psi_{xxx}(x) \right)\ket{0}^{\otimes L},  \label{eq:for-1/epsilon} \\
    O\left(\epsilon^0\right)&: \hat{H}_{XXX} \ket{\Psi_{xxx}}_{(1)} + H_{XXZ_{\epsilon}}^{(1)} \left(\frac{\xi}{\eta}\right) \left(\sum_{1 \leq x \leq L}\Psi_{xxx}(x) \right)\ket{0}^{\otimes L} = \nonumber \\
    &\;= E_{xxx} \ket{\Psi_{xxx}}_{(1)} \label{eq:for-eigenvector1}
\end{align}
Notice that the action of the Hamiltonian  $\Ad_{M\left(\xi/\epsilon \eta \right)}^{\otimes L} H_{XXZ}$ on the vacuum is zero, therefore each order of its expansion in $\epsilon$ will obviously annihilate the vacuum state. With this, equation~\eqref{eq:for-eigenvector1} implies that $\ket{\Psi_{xxx}}_{(1)}$ is an eigenstate with eigenvalue $E_{xxx}$ of the Hamiltonian $\hat{H}_{XXX}$. This state can be obtained by subtracting the $O(\frac{1}{\epsilon})$ singularities from the Bethe state with one excitation of the $XXZ_{-1/2}$ model,
\begin{align}
    \ket{\Psi_{xxx}}_{(1)} = \lim_{\epsilon \to 0} \left(M (\xi/\epsilon \eta)^{\otimes L} \ket{\Psi_{xxz}^\epsilon}_{(1)}-\left(\frac{\xi}{\epsilon \eta}\right) \left(\sum_{1 \leq x \leq L}\Psi_{xxx}(x) \right)\ket{0}^{\otimes L}\right),
\end{align}
Additionally, due to the periodicity of the Hamiltonian $\hat{H}_{XXX}$ , the state $\ket{\Psi_{xxx}}_{(1)}$ must be compatible with the periodic boundary conditions. In this way, we recover the undeformed $XXX_{-1/2}$ Bethe equation in the variable $\lambda$.

On the other hand, the equation~\eqref{eq:for-1/epsilon}  can be recast as
\begin{align}
    0=E_{xxx}  \left(\sum_{1 \leq x \leq L}\Psi_{xxx}(x) \right)\ket{0}^{\otimes L},
\end{align}
If the momentum is zero, the equation above is automatically satisfied since the energy of a magnon with vanishing momentum is zero. Suppose now that the momentum is non-zero. We have seen that physical solutions of~\eqref{eq:for-eigenvector1} require the pseudomomentum to satisfy the undeformed $XXX_{-1/2}$ Bethe equation, namely $e^{i k L} =1$. With this, it is immediate to prove that the coefficient multiplying the vacuum vanishes,
\begin{align}
    \left(\sum_{1 \leq x \leq L} e^{i k x}\right) = \frac{e^{i k} \left(e^{i k L} - 1\right)}{e^{ik}-1} = 0, \quad \text{for} \quad k \neq 0. \label{eq:gemetric-sum-wf1}
\end{align}
As a result, the eigenvalue equation~\eqref{eq:for-1/epsilon} is satisfied for all the allowed values of the  momentum. Moreover, it implies that we can add $\left(\sum_{1 \leq x \leq L}\Psi_{xxx}(x) \right)\ket{0}^{\otimes L}$ to the state $\ket{\Psi_{xxx}}_{(1)}$. Indeed, if $k \neq 0$  the contribution vanishes, while if $k=0$ the contribution is an eigenstate with the same eigenvalue as $\ket{\Psi_{xxx}}_{(1)}$. Therefore, the most general eigenstate of $\hat{H}_{XXX}$ is the linear combination
\begin{align}
    \ket{\Psi_{xxx}}_{(1)} + \alpha \left(\sum_{1 \leq x \leq L}\Psi_{xxx}(x) \right)\ket{0}^{\otimes L}, \quad \text{with} \quad e^{ikL}=1. \label{eq:for-eigenstate1}
\end{align}
with $\alpha$ a free parameter. By applying an appropriate change of basis, it is possible to choose any value for the variable $\alpha$, including zero. This solution  is consistent with the result found in section~\ref{section:CBA}. Specifically, applying the similarity transformation given by $w=\mathbb{I} + \xi J^3J^-$ to the state~\eqref{eq:for-eigenstate1} with $\alpha = -\frac{\xi}{2}$, we recover the eigenstate of the Hamiltonian $\tilde{H}_{XXX}$, i.e. the state  $\ket{\Psi}_{(1)}$ in~\eqref{eq:psi1-def-xxx}. 
    
\subsubsection*{Example II: at most two excitations}
In the $XXZ_{-1/2}$ model, the Bethe state for two excitations is given by
\begin{align}
    \ket{\Psi_{xxz}}_{(2)} = \sum_{1 \leq x_1 \leq x_2 \leq L} \Psi_{xxz}(x_1,x_2) |x_1,x_2) \label{eq:xxz-state-2ex}
\end{align}
where $\Psi_{xxz}(x_1,x_2)$ is the $XXZ_{-1/2}$ wavefunction for  two excitations
\begin{align}
    &\Psi_{xxz}(x_1,x_2) = \left(\frac{\sinh{(\lambda_1+\eta/2)}}{\sinh{(\lambda_1-\eta/2)}}\right)^{x_1} \left(\frac{\sinh{(\lambda_2+\eta/2)}}{\sinh{(\lambda_2-\eta/2)}}\right)^{x_2} + \nonumber \\
    &\quad +\frac{\sinh{(\lambda_2-\lambda_1-\eta)}}{\sinh{(\lambda_2-\lambda_1+\eta)}} \left(\frac{\sinh{(\lambda_2+\eta/2)}}{\sinh{(\lambda_2-\eta/2)}}\right)^{x_1} \left(\frac{\sinh{(\lambda_1+\eta/2)}}{\sinh{(\lambda_1-\eta/2)}}\right)^{x_2}
\end{align}
Under the similarity transformation $M=e^{\xi J^-}$ and the rescaling~\eqref{eq:rescale}, the state~\eqref{eq:xxz-state-2ex} transforms as follows
\begin{align}
    &\ket{\Psi_{xxz}}_{(2)} \rightarrow M (\xi/\epsilon \eta)^{\otimes L} \ket{\Psi_{xxz}^\epsilon}_{(2)} = \nonumber \\
    &\quad =\sum_{1 \leq x_1 \leq x_2 \leq L} \Psi_{xxz}^\epsilon(x_1,x_2) \left( |x_1,x_2) + \left(\frac{\xi}{\epsilon \eta}\right) \left(|x_1) + |x_2)\right) + \left(\frac{\xi}{\epsilon \eta}\right)^2 \ket{0}^{\otimes L}\right),
\end{align}
where $ \Psi_{xxz}^\epsilon(x_1,x_2)$ denotes the $XXZ$ wavefunction transformed by the rescaling~\eqref{eq:rescale}. Again, the state above is singular in the limit $\epsilon \to 0$. To address this, we expand the eigenvalue equation in powers of $\epsilon$. Taking into account that the rescaled $XXZ_{-1/2}$ wavefunction and energy are even polynomials in $\epsilon$, we obtain the following equations for the order $O(1/\epsilon^2)$, $O(1/\epsilon)$ and $O(\epsilon^0)$
\begin{align}
    O(1/\epsilon^2):& \hat{H}_{XXX} \sum_{1 \leq x_1\leq x_2\leq L} \Psi_{xxx}(x_1,x_2) \ket{0}^{\otimes L} =\nonumber \\
    &=E_{xxx} \left(\sum_{1 \leq x_1\leq x_2\leq L} \Psi_{xxx}(x_1,x_2)\right) \ket{0}^{\otimes L}, \label{eq:for1/e^2-2}\\
    O(1/\epsilon):& \hat{H}_{XXX}\sum_{1 \leq x_1\leq x_2\leq L} \Psi_{xxx}(x_1,x_2)  \left(|x_1) + |x_2)\right)= \nonumber \\
   &= E_{xxx} \sum_{1 \leq x_1<x_2\leq L} \Psi_{xxx}(x_1,x_2)  \left(|x_1) + |x_2)\right) \label{eq:for1/e-2}
\end{align}
\begin{align}
    O(\epsilon^0):& \hat{H}_{XXX} \left(\ket{\Psi_{xxx}}_{(2)} +  \left(\frac{\xi}{\eta}\right)^2 \left(\sum_{1 \leq x_1<x_2\leq L} \Psi_{xxz}^{(2)}(x_1,x_2)\right) \ket{0}^{\otimes L}\right)  +\nonumber \\
   &+\left(\frac{\xi}{\eta}\right) H_{XXZ_{\epsilon}}^{(1)}\sum_{1 \leq x_1\leq x_2\leq L} \Psi_{xxx}(x_1,x_2) \left(|x_1) + |x_2)\right)  = \nonumber \\
    &=E_{xxx} \left(\ket{\Psi_{xxx}}_{(2)} + \left(\frac{\xi}{\eta}\right)^2 \left(\sum_{1 \leq x_1<x_2\leq L} \Psi_{xxz}^{(2)}(x_1,x_2)\right) \ket{0}^{\otimes L}\right), \label{eq:for-eigenstate2}
\end{align}
where $\Psi_{xxz}^{(2)}(x_1,x_2)$ is the second-order coefficient in the Taylor expansion of $\Psi_{xxz}^\epsilon(x_1,x_2)$,
\begin{align}
     \Psi_{xxz}^{(2)}(x_1,x_2)= \left(\frac{d^2}{d\epsilon^2} \Psi_{xxz}^{\epsilon}(x_1,x_2)\right)_{\epsilon = 0}.
\end{align}
We need to compute the action of $H_{XXZ_{\epsilon}}^{(1)}$ on a single-excitation state. First, one can check (see the appendix~\ref{ap:action-xxzham_(1,0)-and-(0,1)}) that the action of the Hamiltonian density on all the possible two-site states with one excitation is given by 
\begin{align}
    \mathsf{h}_{12} \ket{10} &= q \ket{10}-\ket{01} \label{eq: action on (1,0)}\\
    \mathsf{h}_{12} \ket{01} &= \frac{1}{q}\ket{01}-\ket{10}. \label{eq:action on (0,1)}
\end{align}
Therefore, the action of the total Hamiltonian on a state with one excitation takes the form
\begin{align}
    H_{XXZ} |x) =  \left(q+\frac{1}{q}\right) |x) - |x-1) - |x+1).
\end{align}
Then, the action of $\Ad_{M\left(\xi \right)}^{\otimes L} H_{XXZ}$ on $|x)$ is given by
\begin{align}
    \Ad_{M\left(\xi \right)}^{\otimes L} H_{XXZ} |x)= H_{XXZ} |x) +   \xi \left(q+\frac{1}{q}-2\right) \ket{0}^{\otimes L}.
\end{align}
If we introduce the rescaling~\eqref{eq:rescale} and expand the expression above in powers of $\epsilon$, we obtain
\begin{align}
    \Ad_{M\left(\xi /\eta \epsilon  \right)}^{\otimes L} H_{XXZ} |x) =\hat{H}_{XXX} |x) + \epsilon \xi \eta \ket{0}^{\otimes L} + O(\epsilon^2), 
\end{align}
where the action of $\hat{H}_{XXZ}$ on $|x)$ coincides with the action of the undeformed Hamiltonian,
\begin{align}
    \hat{H}_{XXX} |x) = 2 |x) - |x-1) - |x+1). \label{eq:action-h-on-x}
\end{align}
From this it follows that
\begin{align}
    H_{XXZ_{\epsilon}}^{(1)} |x) =  \xi \eta \ket{0}^{\otimes L}.
\end{align}
Consequently, the term involving the action of $H_{XXZ_{\epsilon}}^{(1)} $ in~\eqref{eq:for-eigenstate2} can be recast as the following term proportional to the vacuum
\begin{align}
    &H_{XXZ_{\epsilon}}^{(1)}\sum_{1 \leq x_1\leq x_2\leq L} \Psi_{xxx}(x_1,x_2) \left(|x_1) + |x_2)\right) = \nonumber \\
    &=2 \xi \eta \left(\sum_{1 \leq x_1\leq x_2\leq L} \Psi_{xxx}(x_1,x_2)\right) \ket{0}^{\otimes L},
\end{align}
Notice that in order to extract an eigenvector from the equation~\eqref{eq:for-eigenstate2}, we need that  $\sum_{1 \leq x_1\leq x_2\leq L} \Psi_{xxx}(x_1,x_2)$ vanishes. Let us assume for a moment that this is the case. Then, the state
\begin{align}
    \ket{\Psi_{xxx}} +  \left(\frac{\xi}{\eta}\right)^2 \left(\sum_{1 \leq x_1\leq x_2\leq L} \Psi_{xxz}^{(2)}(x_1,x_2)\right) \ket{0}^{\otimes L} \label{eq:xxzt-eigen-2ex}
\end{align}
would be an eigenstate of the Hamiltonian $\hat{H}_{XXX}$ with an eigenvalue given by the undeformed $XXX_{-1/2}$ energy. The state~\eqref{eq:xxzt-eigen-2ex} can be obtained from the $XXZ_{-1/2}$ eigenstate~\eqref{eq:xxz-state-2ex} by subtracting the $O(1/\epsilon^2)$ and $O(1/\epsilon)$ singularities,
\begin{align}
    \lim_{\epsilon \to 0} \left(M (\xi/\epsilon \eta)^{\otimes L} \ket{\Psi_{xxz}^\epsilon}_{(2)} - \left(\frac{\xi}{\epsilon \eta}\right) \sum_{1 \leq x_1 < x_2 \leq L} \Psi_{xxx}(x_1,x_2)   \left(|x_1) + |x_2)\right)\right. \nonumber \\
    \left. - \left(\frac{\xi}{\epsilon \eta}\right)^2 \sum_{1 \leq x_1 \leq x_2 \leq L} \Psi_{xxz}(x_1,x_2)  \ket{0}^{\otimes L}\right).
\end{align}
Moreover, since $\hat{H}_{XXX}$ is periodic, we need to impose the same condition on the state~\eqref{eq:xxzt-eigen-2ex}. The vacuum is periodic by construction, therefore we simply require periodicity on $\ket{\Psi_{xxx}}$, which leads to the undeformed $XXX_{-1/2}$ Bethe equations.

Let us now examine the conditions under which $\sum_{1 \leq x_1\leq x_2\leq L} \Psi_{xxx}(x_1,x_2)$ vanishes. First suppose that the two momenta $k_1$ and $k_2$ are non zero. We have seen that the momenta of the $XXX_{-1/2}$ wavefunction must satisfy the Bethe equations
\begin{align}
    e^{i k_1 L} = S_{xxx}(k_1,k_2), \quad  e^{i k_2 L} = S_{xxx}(k_2,k_1).
\end{align}
Under this conditions one can prove the desired identity (see  appendix~\ref{ap:Null-sum-identities}) 
\begin{align}
    \sum_{1 \leq x_1\leq x_2\leq L} \Psi_{xxx}(x_1,x_2) = 0. \label{eq:null-sum-wf-1}
\end{align}
Suppose now that $k_1=k_2=0$. For this value of the momenta, the undeformed energy $E_{xxx}$ is zero. Moreover, the combination $\sum_{1 \leq x_1\leq x_2\leq L} \Psi_{xxx}(x_1,x_2)$ simplifies to
\begin{align}
    \sum_{1 \leq x_1\leq x_2\leq L} \Psi_{xxx}(x_1,x_2) = L(L+1). 
\end{align}
With this, the equation~\eqref{eq:for-eigenstate2} can be rewritten as follows
\begin{align}
    \hat{H}_{XXX}{\ket{\Psi_{xxx}}_{(2)}} = -2\xi^2 L(L+1) \ket{0}^{\otimes L}.
\end{align}
This is clearly not an eigenvalue equation. Nevertheless, $\hat{H}_{XXX}^2\ket{\Psi_{xxx}}_{(2)}=0$, which implies that the undeformed state $\ket{\Psi_{xxx}}_{(2)}$ with $k_1=k_2=0$, although not an eigenvector, is a generalised eigenvector of the Hamiltonian $\hat{H}_{XXX}$ in the Jordan chain of the vacuum, i.e.~it is associated to the eigenvalue $0$ of the Hamiltonian. This result matches with that found in section~\ref{section:CBA} for the Hamiltonian $\tilde{H}_{XXX}$.

For the sake of completeness, let us verify that the equations~\eqref{eq:for1/e^2-2} and~\eqref{eq:for1/e-2} hold. The equation~\eqref{eq:for1/e^2-2} can be rewritten as
\begin{align}
    0 = E_{xxx} \left(\sum_{1 \leq x_1 \leq x_2 \leq L}\Psi_{xxx}(x_1,x_2)\right).     
\end{align}
If the two momenta are zero, then the energy $E_{xxx}$ is zero and the right hand side of the equation above automatically vanishes. If the momenta are non-zero then, due to~\eqref{eq:null-sum-wf-1}, also the right hand side does vanish.

Consider now the equation~\eqref{eq:for1/e-2}. First, notice that 
\begin{align}
  \sum_{1 \leq x_1\leq x_2\leq L} \Psi_{xxx}(x_1,x_2) \left(|x_1) + |x_2) \right) = \sum_{1 \leq x_1\leq L} \Phi(x_1) |x_1), \label{eq:state-1/e}
\end{align}
with
\begin{align}
    \Phi(x_1) = \sum_{x_2 =x_1}^{L} \Psi_{xxx}(x_1,x_2)+\sum_{x_2=1}^{x_1} \Psi_{xxx}(x_2,x_1)  \label{eq:wf-state-1/e}
\end{align}
Now, using~\eqref{eq:action-h-on-x}, the left-hand side of~\eqref{eq:for1/e-2}~can be recast as follows
\begin{align}
    \sum_{1 \leq x_1 \leq L} \left(2 \Phi(x_1) -\Phi(x_1+1) - \Phi(x_1-1) \right) \ket{x_1}.
\end{align}
Moreover, the shifted function $\Phi(x_1+1)$ can be simplified as
\begin{align}
    \Phi(x_1+1) &= \sum_{x_2 =x_1}^{L-1} \Psi_{xxx}(x_1+1,x_2+1)+\sum_{x_2=0}^{x_1} \Psi_{xxx}(x_2+1,x_1+1) = \nonumber \\
    &= e^{i(k_1+k_2)} \Phi(x_1),
\end{align}
and identically with $\Phi(x_1-1)$,
\begin{align}
     \Phi(x_1+1) &= \sum_{x_2 =x_1}^{L+1} \Psi_{xxx}(x_1-1,x_2-1)+\sum_{x_2=2}^{x_1} \Psi_{xxx}(x_2+1,x_1+1) = \nonumber \\
    &= e^{-i(k_1+k_2)} \Phi(x_1).
\end{align}
With this,
\begin{align}
    \hat{H}_{XXX} \sum_{1 \leq x_1\leq L} \Phi(x_1) |x_1) = \epsilon(k_1+k_2) \sum_{1 \leq x_1\leq L} \Phi(x_1) |x_1),
\end{align}
where $\epsilon(k)$ is the undeformed $XXX_{-1/2}$ dispersion relation~\eqref{eq:undef-energy}. This implies that the state~\eqref{eq:state-1/e} is a one-excitation eigenstate of the Hamiltonian $\hat{H}_{XXX}$ with a momentum equal to the sum of the two momenta inherited from $\ket{\Psi_{xxx}}_{(2)}$. In addition, the equation~\eqref{eq:for1/e-2} can be simplified as
\begin{align}
    \left(\epsilon(k_1+k_2)-\epsilon(k_1)-\epsilon(k_2)\right) \sum_{1 \leq x_1\leq L} \Phi(x_1) |x_1) = 0.
\end{align}
If $\epsilon(k_1+k_2)=\epsilon(k_1)+\epsilon(k_2)$, which corresponds to the values of the momenta $k_1=0$ or $k_2=0$ or $k_1+k_2=\pi$, the equation above is automatically satisfied. Otherwise, the only possibility is that the state~\eqref{eq:state-1/e} vanishes. Indeed, one can verify (see the appendix~\ref{ap:Null-sum-identities} for the demonstration) that whenever any of the two momenta $k_1$ or $k_2$ is non-zero, the undeformed Bethe equation implies
\begin{align}
    \Phi(x_1) = 0\quad \forall x_1 \in [1,L] \label{eq:null-sum-wf-2}.
\end{align}

It is instructive to compare the eigenstate~\eqref{eq:xxzt-eigen-2ex} with the one obtained for the Hamiltonian $\tilde{H}_{XXX}$~\eqref{eq:twsited-eigenstate-2ex}. Since we know the first order correction of the similarity transformation relating $\hat{H}_{XXX}$ to $\tilde{H}_{XXX}$~\eqref{eq:s.t-twisted-ham}, we can  compare the one-excitation correction of the eigenstates. In particular, starting from~\eqref{eq:xxzt-eigen-2ex} we must obtain the one-excitation component of $\ket{\Psi}_{(2)}$ in~\eqref{eq:twsited-eigenstate-2ex}. Acting with~\eqref{eq:s.t-twisted-ham} on $\ket{\Psi_{xxx}}_{(2)}$, we get
\begin{align}
    w ^{\otimes L}\ket{\Psi_{xxx}}_{(2)} &= \ket{\Psi_{xxx}}_{(2)} + 3 \xi \sum_{1 \leq x \leq L} \Psi_{xxx}(x,x) |x) + \nonumber \\
    &\quad +\frac{\xi}{2}\sum_{1 \leq x_1 < x_2 \leq L} \Psi_{xxx}(x,x) \left(|x_1) + |x_2)\right) + O(\xi^2).
\end{align}
If $k_1$ and $k_2$ are non-zero, using~\eqref{eq:null-sum-wf-2}, we can recast the last term in the sum as
\begin{align}
    \frac{\xi}{2}\sum_{1 \leq x_1 > x_2 \leq L} \Psi_{xxx}(x,x) \left(|x_1) + |x_2)\right) = -\xi \sum_{1 \leq x \leq L} \Psi_{xxx}(x,x) |x).
\end{align}
With this, we can write
\begin{align}
    w ^{\otimes L}\ket{\Psi_{xxx}}_{(2)} &= \ket{\Psi_{xxx}}_{(2)} + 2 \xi \sum_{1 \leq x \leq L} \Psi_{xxx}(x,x) |x) + O(\xi^2),
\end{align}
and we recover~\eqref{eq:first-order-twisted-eigenstate-2ex}.

If $k_2=0$ (the same argument applies if $k_1=0$) it is straightforward to verify from~\eqref{eq:wf-state-1/e} that
\begin{align}
    \sum_{1 \leq x_1 < x_2 \leq L} \Psi_{xxx}(x,x) \left(|x_1) + |x_2)\right) = L \sum_{1 \leq x_1 \leq L} e^{i k_1 x_1} |x_1).
\end{align}
Therefore,
\begin{align}
     w ^{\otimes L}\ket{\Psi_{xxx}}_{(2)} &= \ket{\Psi_{xxx}}_{(2)} + \xi \left(\frac{L}{2}+3\right) \sum_{1 \leq x_1 \leq L} e^{i k_1 x_1} |x_1) + O(\xi^2).
\end{align}
As already discussed in section~\ref{section:CBA}, the addition of the state $\sum_{1 \leq x_1 \leq L} e^{i k_1 x_1} |x_1)$ to $\ket{\Psi_{xxx}}_{(2)}$, with $k_2=0$, is still an eigenvector of $\tilde{H}_{XXX}$ since both are eigenstates of $\tilde{H}_{XXX}$ with the same eigenvalue $\epsilon(k_1)$.

\section{Algebraic Bethe Ansatz for the Jordanian $XXX_{-1/2}$ spin-chain.}\label{sec:ABA}
In this section, we discuss the application of the Algebraic Bethe Ansatz (ABA) to the Jordanian deformation of the $XXX_{-1/2}$ model.
Let us first briefly review the standard construction of the Algebraic Bethe Ansatz~\cite{Faddeev:1996iy}. The central object is the monodromy matrix, defined as
\begin{align}
    T_{1/2}(u) = R_{1/2,L}(u) \cdots R_{1/2,1}(u) \label{eq:trasnfer-1/2}
\end{align}
where the index $1/2$ labels the auxiliary space, which is taken to be the fundamental $\mathfrak{sl}_2$ module in the spin $1/2$ representation and that is isomorphic to $\mathbb{C}^2$. As an operator acting on $\mathbb{C}^2$, the monodromy matrix~\eqref{eq:trasnfer-1/2} can be represented as
\begin{align}
    T_{1/2}(u) = \left(
\begin{array}{cc}
 A(u) & B(u) \\
 C(u) & D(u) \\
\end{array}
\right),
\end{align}
where the operators $\{A,B,C,D\}$, which act on the physical Hilbert space of the spin-chain, span the so-called Yang-Baxter algebra of the model.

Note that in the definition of the monodromy matrix~\eqref{eq:xxx-trasnfer-matrix} used to derive the conserved charges, including the $XXX_{-1/2}$ Hamiltonian, the auxiliary space was taken to be the fundamental module of $\mathfrak{sl}_2$ in the spin$=-1/2$ representation. This is due to the fact that, in order to construct the conserved charges, it is necessary for the $R$-matrices  defining the transfer matrix to be regular, i.e. there must exist a value of the spectral parameter $u$ for which the $R$-matrices reduce to the permutation operator. Clearly, this is not possible if the two Hilbert spaces over which the $R$-matrices act are different. 

Conversely, the monodromy matrix~\eqref{eq:trasnfer-1/2} is used to obtain the Bethe equations and eigenstates of the Hamiltonian. In the standard methods of the ABA, one constructs the eigenstates of the trace of the monodromy matrix~\eqref{eq:trasnfer-1/2} by acting with the $C(u)$ operators on some reference state of the model, which for the $XXX_{-1/2}$ model is taken to be the vacuum $\ket{0}^{L}$. Theses states are also eigenstates of the Hamiltonian, since one can verify that $[tr_{1/2}T_{1/2}(u), tr_{a}T_{a}(v)]=0$  for every value of the spectral parameter $u$ and $v$, and with $T_{a}$ the transfer matrix defined over the auxiliary space in the $-1/2$ representation (see equation~\eqref{eq:xxx-trasnfer-matrix}).

Let us consider now the $XXX_{-1/2}$ deformed Hamiltonian~\eqref{eq:deformed-ham}. As already discussed in the previous sections, this Hamiltonian is not diagonalisable. To the best of our knowledge, a satisfactory approach for applying the ABA methods to non-diagonalisable Hamiltonians has not yet been developed. See however the recent interesting works on some models of this type~\cite{Ipsen:2018fmu,Ahn:2020zly,Ahn:2022snr,NietoGarcia:2023jeb}. Furthermore, even if it were possible to obtain the complete set of (generalised) eigenvectors of the trace of the monodromy matrix of the model, in general it is not obvious how to infer from this the set of the (generalised) eigenstates of the Hamiltonian. This is because,  for mutually commuting non-diagonalisable matrices, it is not guaranteed that they share a common basis of (generalised) eigenvectors, or even that they have the same number of Jordan blocks and of the same rank (see e.g.~\cite{NietoGarcia:2023jeb} for a related discussion in a setup similar to ours).

Suppose we insist in constructing the eigenvectors of the trace of the transfer matrix following the standard route of the ABA. To this end,  let us denote by $\tilde{T}_{1/2}(u)$ the monodromy matrix of the deformed $XXX_{-1/2}$ spin-chain with auxiliary space in the $1/2$ representation of the fundamental module of $\mathfrak{sl}_2$,
\begin{align}
    \tilde{T}_{1/2}(u) = \Tilde{R}_{1/2,L}(u) \cdots \Tilde{R}_{1/2,1}(u)= \left(
\begin{array}{cc}
 \tilde{A}(u) & \tilde{B}(u) \\
 \tilde{C}(u) & \tilde{D}(u) \\
\end{array}
\right), \label{eq:monodromy-twisted-1/2}
\end{align}
where $\Tilde{R}_{1/2,n}(u)$ takes the form
\begin{align}
    \Tilde{R}_{1/2,n}(u) = \left(
\begin{array}{cc}
 k^{+}_{n} & \frac{1}{1-u}J^{-}_{n}e^{\frac{-w_{n}}{2}} \\
 -2\xi J^{3}_{n}k^{+}_{n} - \frac{1}{1-u}J^{+}_{n}e^{\frac{w_{n}}{2}} & -2\xi\frac{1}{1-u} J^{3}_{n} J^{-}_{n} e^{\frac{-w_{n}}{2}} + k^{-}_{n} \\
\end{array}
\right), \label{eq:r-matrix-1/2}
\end{align}
with 
\begin{align}
    k^{+}_{n} &= \left(\frac{1-2u}{2(1-u)}+\frac{1}{1-u}J^3_{n}\right)e^{\frac{w_{n}}{2}}, \\
    k^{-}_{n} &= \left(\frac{1-2u}{2(1-u)}-\frac{1}{1-u}J^3_{n}\right)e^{\frac{-w_{n}}{2}}, \\
    w_n &= \log\left(1 + 2\xi J^-_n\right).
\end{align}
The first step is to  compute the commutation relations of the elements $\tilde{A},\tilde{B},\tilde{C},\tilde{D}$ of the Yang-Baxter algebra of the deformed model. This was already done in~\cite{Kulish_1997}. We write down their result following our conventions. We will only need the following commutation relations,
\begin{align}
    \tilde{A}(u) \tilde{C}(v) &= \alpha(u-v) \tilde{C}(v) \tilde{A}(u) + \beta(u-v) \tilde{C}(u) \tilde{A}(v) + \xi \left(\tilde{D}(v) \tilde{A}(u)- \tilde{C}(v) \tilde{B}(u) \right. \nonumber \\
    & \quad \left. - \tilde{A}(u) \tilde{A}(v)\right) + \xi^2 \tilde{D}(v) \tilde{B}(u), \nonumber \\
    \tilde{D}(u) \tilde{C}(v) &= \alpha(v-u) \tilde{C}(v) \tilde{D}(u) + \beta(v-u) \tilde{C}(u) \tilde{D}(v) + \xi \left(\tilde{A}(v) \tilde{D}(u)- \tilde{C}(v) \tilde{B}(u) \right. \nonumber \\
    &\quad \left. - \tilde{D}(u) \tilde{D}(v)\right) + \xi^2 \tilde{A}(v) \tilde{B}(u),\nonumber  \\
    \tilde{B}(u) \tilde{C}(v) &= \tilde{C}(v) \tilde{B}(u) + \beta(u-v) \left(\tilde{D}(u) \tilde{A}(v)-\tilde{D}(v) \tilde{A}(u) \right) + \xi \left(\tilde{D}(v) \tilde{B}(u)  \right.  \\  &\quad \left. + \tilde{B}(u) \tilde{A}(v)\right), \nonumber
\end{align}
\begin{align}
    \tilde{C}(u) \tilde{C}(v) &= \tilde{C}(v) \tilde{C}(u) + \frac{\xi}{\alpha(u-v)} \left(\tilde{D}(v) \tilde{C}(u)-\tilde{C}(v) \tilde{D}(u)+\tilde{C}(u) \tilde{A}(v) \right. \nonumber \\
    &\quad \left.- \tilde{A}(u) \tilde{C}(v)\right) +  \frac{\xi^2}{\alpha(u-v)}\left(\tilde{D}(v) \tilde{D}(u)-\tilde{A}(u) \tilde{A}(v)\right) \nonumber, \label{eq:YBA-commutation}
\end{align}
where we have defined
\begin{align}
    \alpha(u-v) = 1 - \beta(u-v), \quad 
    \beta(u-v) = \frac{1}{u-v}.
\end{align}
As already discussed in section~\ref{section:CBA}, the restriction of the deformed Hamiltonian~\eqref{eq:deformed-ham} to the subspace spanned by states with at most one excitation is diagonalisable and coincides with the $XXX_{-1/2}$ Hamiltonian. Therefore, in this subsector, we must be able to apply the standard ABA. Indeed, from~\eqref{eq:r-matrix-1/2} it is immediate to verify that
\begin{align}
    \tilde{A}(u) \ket{0}^{L} &= \ket{0}^{L}, \\
     \tilde{D}(u)\ket{0}^{L} &= d(u)\ket{0}^{L} \quad \text{with} \quad  d(u) = \left(\frac{u}{u-1}\right)^L,\\
      \tilde{B}(u)\ket{0}^{L} &= 0.
\end{align}
With this and using the commutation relations~\eqref{eq:YBA-commutation}, one can write
\begin{align}
    & \tilde{\tau}(u) \tilde{C}(v) \ket{0}^{L} = \left(\alpha(u-v)+\alpha(v-u) d(u) \right) \tilde{C}(v) \ket{0}^{L} + \nonumber\\
    &\quad +\beta(u-v) \left(1-d(v)\right) \tilde{C}(u) \ket{0}^{L} + \xi \left(1-d(v)\right) \left(d(u) -1\right) \ket{0}^L,
\end{align}
where $\tilde{\tau}(u)$ is the trace of the monodromy matrix~\eqref{eq:monodromy-twisted-1/2},
\begin{align}
    \tilde{\tau}(u) = \tilde{A}(u) + \tilde{D}(u).
\end{align}
Imposing that the state $\tilde{C}(v) \ket{0}^{L}$ is an eigenstate of $\tilde{\tau}(u)$, leads to the undeformed Bethe equations for the one-excitation sector\footnote{This form of the Bethe equations coincides with~\eqref{eq:xxx-Bethe-eqs} if the pseudorapidity is identify with the spectral parameter through the relation $v = i\lambda + 1/2$.}
\begin{align}
    \left(\frac{v}{v-1}\right)^L = 1.
\end{align}
Let us restrict now to the subspace  with at most two excitations. We sketch the form of the standard ABA in this case. Consider the state constructed with two $\tilde{C}$ operators. Acting with the trace of the monodromy matrix~\eqref{eq:monodromy-twisted-1/2} leads schematically to
\begin{align}
    &\tilde{\tau}(u)\tilde{C}(v_1) \tilde{C}(v_2) \ket{0}^{L} = \left[\alpha(u-v_1) \alpha(u-v_2) + \alpha(v_1-u) \alpha(v_2-u) d(u)\right] \tilde{C}(v_1) \tilde{C}(v_2) \ket{0}^{L} \nonumber \\
    &+ \gamma_{1}(v_1,v_2,u) \tilde{C}(u) \tilde{C}(v_1) \ket{0}^{L} +\gamma_{2}(v_1,v_2,u) \tilde{C}(u) \tilde{C}(v_2) \ket{0}^{L} + \xi \gamma_{3}(v_1,v_2,u) \tilde{C}(u)  \ket{0}^{L}+ \nonumber \\
    &+\xi \gamma_{4}(v_1,v_2,u) \tilde{C}(v_1)  \ket{0}^{L}+\xi \gamma_{5}(v_1,v_2,u) \tilde{C}(v_2)  \ket{0}^{L} + \xi^2 \gamma_{6}(v_1,v_2,u)\ket{0}^{L}, \label{eq:ABA-2ex}
\end{align}
where the $\gamma_i's$ are functions obtained after applying the commutation relations~\eqref{eq:YBA-commutation}. The functions $\gamma_1$ and $\gamma_2$ coincide with the coefficients that we would obtain in the ABA for the undeformed and untwisted model. Imposing that they vanish leads to the undeformed Bethe equations,
\begin{align}
    \left(\frac{v_1}{v_1-1}\right)^{L} = \frac{v_1-v_2-1}{v_1-v_2+1}, \quad\quad \left(\frac{v_2}{v_2-1}\right)^{L} = \frac{v_2-v_1-1}{v_2-v_1+1}. \label{eq:xxx-Bethe-eqs-2ex}
\end{align}
Now, however, imposing the undeformed Bethe equations is not sufficient to cancel the remaining $\gamma_{i}'s$. To understand this, note that the relation~\eqref{eq:trasnfer-matrix} allows us to write the operators $\tilde{A},\tilde{B},\tilde{C},\tilde{D}$ in terms of their counterparts in the undeformed and untwisted theory. In fact, it is not difficult to see that~\eqref{eq:trasnfer-matrix} is valid for every representation of the auxiliary space. In particular, taking the spin $1/2$ representation leads to 
\begin{align}
    \tilde{A} &= \Omega A \Omega^{-1} e^{\frac{1}{2} \tilde{\Delta}^{(L-1)}(w)},\\
    \tilde{B} &= \Omega B \Omega^{-1} e^{-\frac{1}{2} \tilde{\Delta}^{(L-1)}(w)}, \\
    \tilde{C} &= \left(-2\xi \Omega \Delta^{(L-1)}(J^3) A \Omega^{-1}+\Omega C \Omega^{-1}\right) e^{\frac{1}{2} \tilde{\Delta}^{(L-1)}(w)}, \\
    \tilde{D} &= \left(-2\xi \Omega \Delta^{(L-1)}(J^3) B \Omega^{-1}+\Omega D \Omega^{-1}\right) e^{-\frac{1}{2} \tilde{\Delta}^{(L-1)}(w)}.
\end{align}
This implies that in the basis of eigenvectors of the undeformed and untwisted model (i.e. in the states constructed by acting with the $C(u)$ operator on $\ket{0}^{L}$), $\tilde{\tau}(u)$ is an upper triangular matrix that in the diagonal has the eigenvalues of the undeformed and untwisted model. Therefore, the eigenvalues of $\tilde{\tau}(u)$ coincide with the eigenvalues of the undeformed and untwisted model. In particular, the eigenvalues of the restriction of $\tilde{\tau}(u)$ to the subspace  with at most two excitations are given by
\begin{align}
    \alpha(u-v_1) \alpha(u-v_2) + \alpha(v_1-u) \alpha(v_2-u) d(u),
\end{align}
with $v_1$ and $v_2$ solutions of the Bethe equations~\eqref{eq:xxx-Bethe-eqs-2ex}. Therefore, the application of the ABA to the deformed $XXX_{-1/2}$ spin-chain must lead to the undeformed Bethe equations. Observe that the coefficient that multiplies $\tilde{C}(v_1) \tilde{C}(v_2) \ket{0}^{L} $ on the right-hand side of~\eqref{eq:ABA-2ex} coincides with the correct eigenvalue provided we impose the undeformed Bethe equations~\eqref{eq:xxx-Bethe-eqs-2ex}. However, as already mentioned, the Bethe equations are not enough to guarantee that $\tilde{C}(v_1) \tilde{C}(v_2) \ket{0}^{L} $ is an eigenstate. To claim that $\tilde{C}(v_1) \tilde{C}(v_2) \ket{0}^{L} $ is an eigenstate of $\tilde{\tau}(u)$ one should impose the vanishing of the remaining extra terms. This, in principle, imposes additional conditions on $v_1,v_2$. When these are met, the vector is an eigenstate, otherwise it is not.

In order to test the above statement, we compute the matrix form of the restriction of $\tilde{\tau}(u)$ to the subspace spanned by states with at most two excitations, and for a spin-chain of length two. Using~\eqref{eq:r-matrix-1/2}, we obtain the following matrix in the basis~\eqref{eq:basis-2ex},
\begin{equation}
    \left(
\begin{array}{cccccc}
 \frac{2 u^2-2 u+1}{(u-1)^2} & 0 & 0 & \frac{\xi ^2 \left(u^2-4 u+2\right)}{2 (u-1)^2} & \frac{\xi ^2 \left(u^2-4 u+2\right)}{2 (u-1)^2} & \frac{2 \xi ^2 u}{u-1} \\
 0 & \frac{2 \left(u^2-u+1\right)}{(u-1)^2} & -\frac{1}{(u-1)^2} & \frac{2 \xi }{(u-1)^2} & \frac{2 \xi }{(u-1)^2} & -\frac{4 \xi }{(u-1)^2} \\
 0 & -\frac{1}{(u-1)^2} & \frac{2 \left(u^2-u+1\right)}{(u-1)^2} & \frac{2 \xi }{(u-1)^2} & \frac{2 \xi }{(u-1)^2} & -\frac{4 \xi }{(u-1)^2} \\
 0 & 0 & 0 & \frac{2 u^2-2 u+3}{(u-1)^2} & 0 & -\frac{2}{(u-1)^2} \\
 0 & 0 & 0 & 0 & \frac{2 u^2-2 u+3}{(u-1)^2} & -\frac{2}{(u-1)^2} \\
 0 & 0 & 0 & -\frac{2}{(u-1)^2} & -\frac{2}{(u-1)^2} & \frac{2 u^2-2 u+5}{(u-1)^2} \\
\end{array}
\right) .\label{eq:transfer-matrix-2ex}
\end{equation}
Its Jordan normal form is given by
\begin{align}
    \left(
\begin{array}{cccccc}
\frac{2 u^2-2 u+1}{(u-1)^2} & 1 & 0 & 0 & 0 &0\\
0 & \frac{2 u^2-2 u+1}{(u-1)^2} & 0 & 0 & 0 &0\\ 
0&0&\frac{2 u^2-2 u+1}{(u-1)^2} & 0 & 0 & 0  \\
 0 & 0 & 0 & \frac{2 u^2-2 u+3}{(u-1)^2} & 0 & 0 \\
 0 & 0 & 0 & 0 & \frac{2 u^2-2 u+3}{(u-1)^2} & 0 \\
 0 & 0 & 0 & 0 & 0 & \frac{2 u^2-2 u+7}{(u-1)^2} \\
\end{array}
\right). \label{eq:jordan-monodromy}
\end{align}
We see that the eigenvalues have no $\xi$-dependence and coincide with those of the undeformed and untwisted model. Interestingly,~\eqref{eq:jordan-monodromy} possesses the same number of Jordan blocks and with the same rank as the deformed Hamiltonian~\eqref{eq:Jordan-form-ham-length2-2ex}. In Table~\ref{tab:gen-eigen-monodromy-length2-2ex} we list the set of (generalised) eigenvectors of~\eqref{eq:transfer-matrix-2ex}.
\begin{table}[h!]
    \centering
    \begin{tabular}{c|c}
 Eigenvalue & (Generalised) eigenvector \\
 \hline
 $\frac{2 u^2-2 u+1}{(u-1)^2}$ & $\ket{00}$ \\
 \hline
 $\frac{2 u^2-2 u+1}{(u-1)^2}$& $\frac{(u-1)^2}{\xi ^2 \left(3 u^2-6 u+2\right)}\left(\ket{20} + \ket{02}+ \ket{11}\right)$ \\
 \hline
 $\frac{2 u^2-2 u+1}{(u-1)^2}$&$\ket{10} + \ket{01}$ \\
 \hline
 $\frac{2 u^2-2 u+3}{(u-1)^2}$ &  $\ket{10}-\ket{01}$ \\
 \hline
 $\frac{2 u^2-2 u+3}{(u-1)^2}$ &  $\ket{20}-\ket{02}$ \\
 \hline
 $\frac{2 u^2-2 u+7}{(u-1)^2}$ &  $3 \left(\ket{20} + \ket{02}\right) - 6 \ket{11}+6 \xi \left(\ket{10}+\ket{01}\right) + \xi ^2 \left(1-\frac{3}{2} u^2\right) \ket{00}$ \\  
    \end{tabular}
\caption{Eigenvalues and (generalised) eigenvectors of the matrix~\eqref{eq:transfer-matrix-2ex} for non zero $\xi$. The generalised eigenvector is the state $\frac{(u-1)^2}{\xi ^2 \left(3 u^2-6 u+2\right)}\left(\ket{20} + \ket{02}+ \ket{11}\right)$.}
\label{tab:gen-eigen-monodromy-length2-2ex}
\end{table}

For the (generalised) eigenvectors of~\eqref{eq:transfer-matrix-2ex}, observe that their components with the maximum number of excitations coincide with the eigenstates of the undeformed and untwisted model listed in Table~\ref{tab:eigen-ham-length2-2ex} (up to some normalisation factor that is singular in the $\xi \to 0$ limit for the generalised eigenstate). Moreover, we see that some of the (generalised) eigenstates contain factors of $u$. Because of this explicit $u$-dependence, it is clear that the family of states $\tilde{C}(v_1) \tilde{C}(v_2) \ket{0}^{L} $ cannot generate the full set of (generalised) eigenstates of $\tilde{\tau}(u)$, unlike what usually happens  in the ABA. Finally,  setting $u = 0$ we recover the eigenvectors of the deformed Hamiltonian (see Table~\ref{tab:gen-eigen-ham-length2-2ex}). We remark that this is a non-trivial result, since in general, when considering two  non-diagonalisable matrices that commute with each other, not only they may not have a common basis of (generalised) eigenvectors, they may  not even have the same number of Jordan blocks and of the same rank.

\section{Generalization to arbitrary non-compact spin values.}
\label{sec:gen-arbitrary-spin}
In this section, we argue that all the properties of the Jordanian deformed $XXX_{-1/2}$ model remain valid for  arbitrary values of the non-compact spin.
Let us consider the $XXX_{j}$ spin-chain for general negative values of the spin $j$. In this representation the $\mathfrak{sl}_2$ generators are given by
\begin{align}
    J^+=a^\dagger(-2j+a^\dagger a), \quad J^-=a, \quad J^3 = -j + a^\dagger a, \label{eq:arb-rep}
\end{align}
where $[a,a^\dagger]=1$ and $j$ is a negative number. In addition, the coproduct of the quadratic Casimir can be represented as follows
\begin{align}
    \Delta(C) = -(a_1^\dagger-a_2^\dagger)^2a_1 a_2 - 2j (a_1^\dagger-a_2^\dagger)(a_1-a_2) + 2j(2j+1).\label{eq:arb-casi}
\end{align}
The spectrum of the $XXX_{j}$ chain is given by~\cite{Hao:2019cfu}
\begin{align}
    E= \sum_{p=1}^N \frac{2 |j|}{j^2+\lambda_p^2},
\end{align}
where the set of Bethe roots $\{\lambda_j\}$ satisfy the following Bethe equations
\begin{align}
    \left(\frac{\lambda_p+|j|i}{\lambda_p-|j| i}\right)^L=\prod_{\substack{r=0 \\r \neq p}}^N S_{xxx}(\lambda_p,\lambda_r),
\end{align}
with an $S_{xxx}$-matrix that coincides with that of the $j=-1/2$ representation~\eqref{eq:undef-solution}.

Now, we deform the $XXX_j$ model by the Jordanian twist~\eqref{eq:twist}. First, notice that the derivation in appendix~\ref{ap:boundary-twist-ham} of the equivalence between the Jordanian deformed model and the undeformed model with twisted  boundary conditions is independent of the $\mathfrak{sl}_2$ representation. Therefore it applies to the spin-chain for arbitrary values of the non-compact spin. 
Moreover, in the representation~\eqref{eq:arb-rep}, the action of the Jordanian twist on a two-site state is given by
\begin{align}
    F_{12} \ket{n_1,n_2} = \sum_{k=0}^{n_2}\left(-2\xi \right)^k\binom{n_2}{k} \left(|j| +n_1\right)^{\braket{k}} \ket{n_1,n_2-k}.
\end{align}
Therefore, once again, the twist does not preserve the number of excitations. In fact, the action of the deformed Hamiltonian on a state with $n$ excitations leads to a superposition of states with a decreasing number of excitations down to the vacuum, with  coefficients that depend on the spin-value $j$ and with a zeroth-order term in $\xi$ that coincides with the action of the undeformed Hamiltonian. 

Using the same argument as explained for the $j=-1/2$ representation, then, we conclude that the Jordanian-deformed $XXX_{j}$ chain for negative values of $j$ is not diagonalizable, with a spectrum that coincides with that of the undeformed model. 
In other words, in the basis of eigenstates of the undeformed model, the deformed Hamiltonian is an upper-triangular operator, with diagonal elements equal to the eigenvalues of the undeformed Hamiltonian. Furthermore, since the operator~\eqref{eq:twist-boundary} that implements the twisted boundary conditions in the undeformed yet twisted image is non-trivial, the deformed Hamiltonian must be non-diagonalizable, otherwise it would be similar to the undeformed $XXX_j$.

Every (generalized) eigenvector of the deformed model is in one-to-one correspondence with an eigenvector of the undeformed spin-chain. More precisely, the (generalized) eigenvectors of the deformed model are given by the eigenvectors of the undeformed spin-chain plus a superposition of states with fewer excitations down to the vacuum, with coefficients that depend on the deformation parameter $\xi$ and the spin value $j$. 
The (generalized) eigenvectors can be obtained by applying the same method described in section~\ref{section:CBA}. In fact, starting from an eigenvector of the undeformed theory, we may solve the deformed eigenvalue equation by recursively projecting onto states with a fixed number of excitations, until we are left with an equation  for the coefficient that multiplies the vacuum state.

Finally, let us discuss about the possibility of obtaining the Jordanian-deformed $XXX_{j}$ model by applying a singular scaling limit to the $XXZ_j$ model. 
The $XXZ_j$ Hamiltonian for arbitrary negative values of the spin was constructed in~\cite{Frassek:2019isa,Bytsko:2001uh}. If we apply the similarity transformation given by the operator
\begin{align}
    M = e^{\xi J^-},
\end{align}
together with the scaling limit~\eqref{eq:rescale}, it is straightforward to verify that the first-order term in $\xi$ of the resulting Hamiltonian coincides with the one obtained in the subsection~\ref{subsec:sc-lim-xxz-ham}
\begin{align}
     \left(J^- \otimes \mathbb{I}\right) \Delta(C) - \Delta(C)\left(\mathbb{I} \otimes J^-\right).
\end{align}
This can  be shown to satisfy the identity~\eqref{eq:id-r-symetric-twsit} also for the representation~\eqref{eq:arb-rep} and for the coproduct of the Casimir~\eqref{eq:arb-casi}. Therefore, it is possible to obtain an $r-$symmetric Jordanian deformation of the $XXX_j$ Hamiltonian from its $q$-deformation, along with its corresponding (generalized) eigenvectors, by following the same method explained in subsection~\ref{subsec:sc-limit-eig} for the $j=-1/2$ representation.

\section{Conclusions}\label{sec:conclusions}
In this paper we have constructed the Jordanian deformation/twist of the non-compact $\mathfrak{sl}_2$ spin-chain. This is an integrable model with a continuos parameter $\xi$, and it reduces to the usual $XXX$ model when $\xi\to 0$. We discussed the equivalence of two possible interpretations, namely the deformed model with periodic boundary conditions and the undeformed model with twisted boundary conditions. We also explained how to obtain the Jordanian deformation from a scaling limit of the $q$-deformed model (i.e.~the $U_q(\mathfrak{sl}_2)$-invariant $XXZ$ model). We found that the Hamiltonian of the Jordanian deformed/twisted model is non-diagonalisable, meaning that it admits a decomposition with non-trivial Jordan blocks. We discussed how to obtain the (generalised) eigenvectors from the coordinate and algebraic Bethe ansatz, as well as from the scaling limit of $XXZ$. Importantly, the eigenvalues of the Jordanian deformed/twisted Hamiltonian do not depend on the deformation parameter $\xi$, in other words they coincide with the eigenvalues of the original undeformed and periodic Hamiltonian. The dependence on $\xi$ only enters in the expression for the (generalised) eigenvectors.

The non-diagonalisability in the Jordanian case is related to the fact that the Jordanian twist introduces new terms that cause the Hamiltonian to be non-hermitian. Importantly, for our motivations, the non-hermiticity of the spin-chain Hamiltonian does not necessarily signal a sickness of the string sigma model or of the gauge theory that would emerge from the Jordanian deformation/twist of AdS/CFT. The Jordanian-deformed  string sigma-model  certainly has a real action, and something similar may be expected for the dual gauge theory. In fact, dealing with non-hermitian Hamiltonians is not necessarily a drawback as long as their eigenvalues are real, as it is the case for the spin-chain model that we studied. Interestingly, other examples of non-hermitian integrable spin-chains have been studied in the literature. In particular, ``eclectic spin-chains'' were studied with the motivation of accessing a strongly-twisted limit of $\mathcal{N}=4$ super Yang-Mills~\cite{Ipsen:2018fmu,Ahn:2020zly,Ahn:2022snr}. Integrable spin-chains with sites having a two-dimensional Hilbert space were classified in~\cite{DeLeeuw:2019gxe}, and it was found that some of them are non-diagonalisable. We refer to~\cite{NietoGarcia:2023jeb} for a detailed discussion on one of them. In general, the integrability methods are not yet mature in the non-diagonalisable setup, in the sense that it is often challenging to use integrability-based techniques to write down the complete set of generalised eigenvectors. We observed similar issues in our model when applying the coordinate and algebraic Bethe ansatz methods. It would be interesting to see if it is possible to identify a universal strategy for non-diagonalisable integrable spin-chains.

We carried out this work as a first step in the understanding of how to incorporate the Jordanian deformation/twist on gauge theories appearing in the context of the AdS/CFT correspondence, in particular on $\mathcal{N}=4$ super Yang-Mills.
The next step that we will hopefully take in the near future will be to apply the Jordanian deformation/twist to the full spin-chain that is invariant under $\mathfrak{psu}(2,2|4)$, and that describes the one-loop planar spectrum of $\mathcal{N}=4$ super Yang-Mills. Importantly, the Hamiltonian of that spin-chain encodes the anomalous dimensions of local single-trace operators of the gauge theory, and for this reason it is invariant under the full $\mathfrak{psu}(2,2|4)$ symmetry. Therefore, it will be possible to construct all the spin-chain realisations of the Jordanian twists classified in~\cite{Borsato:2022ubq}. This should be contrasted with the S-matrix of the undeformed model, that is instead invariant under a smaller $\mathfrak{su}(2|2)^2$ symmetry.

Considering the full $\mathfrak{psu}(2,2|4)$-invariant spin-chain, and therefore going beyond the sector considered here, is particularly important to have a meaningful embedding in string theory. It is known, in fact, that in certain cases Jordanian deformations of $AdS_5\times S^5$ can give rise to backgrounds that do not satisfy the supergravity equations~\cite{Hoare:2016hwh}.
In general, to give rise to a solution of supergravity, the deformation must satisfy a unimodular condition~\cite{Borsato:2016ose}. For Jordanian deformations this is possible only if fermionic generators participate in the deformation, to complement the role played by the bosonic generators $h=J^3, e=-J^-$~\cite{vanTongeren:2019dlq}. Then, one can reproduce for example the supergravity  solution of~\cite{Kawaguchi:2014fca}. The full list of unimodular Jordanian deformations is given in~\cite{Borsato:2022ubq}.
 
Recently, Jordanian deformations of string sigma-models were shown to be in tension with the usual lore of integrability, according to which the number of particles should be conserved in a scattering process of an integrable model. In particular, after fixing light-cone gauge on the worldsheet of a Jordanian deformation of $AdS_5\times S^5$, it was found that the worldsheet scattering presents cubic couplings leading to particle production~\cite{Borsato:2024sru}. That paper is actually not the first instance of particle production in the context of integrable models, see for example~\cite{Kozlowski:2016too} where the conservation of the number of particles is replaced by the conservation of a different quantum number. Nevertheless, it is fair to say that the particle production in the Jordanian-deformed string sigma model certainly requires more understanding because it clashes with the usual intuition on integrability. Actually, the construction of the (generalised) eigenvectors performed in the present paper can perhaps give us some clues on this problem. In fact, the (generalised) eigenvectors of the Jordanian spin-chain Hamiltonian are, in general, a superposition of vectors containing \emph{different} numbers of excitations. This is a consequence of the fact that the twist is constructed with the lowering operator $J^-$, and the raising operator $J^+$ does not appear in the twist. That means that multiparticle states with a fixed number of excitations do \emph{not} span closed subsectors of the scattering problem. It is clear, then, that even in the Jordanian spin-chain there is a non-vanishing transition amplitude between states with a different number of excitations. Despite this, it does not seem to imply that integrability is lost. Perhaps, also in this case, the correct interpretation may simply be that the number of asymptotic excitations is not a good quantum number to identify sectors of the scattering problem, and it may be possible that a different quantum number will play that role.

\section*{Acknowledgements}
We thank Tim Meier, Juan Miguel Nieto Garc\'ia, Alessandro Torrielli and especially Stijn van Tongeren for useful discussions.
The work of RB was supported by  RYC2021-032371-I, funded by MCIN/AEI/10.13039/501100011033 and by the European Union ``NextGenerationEU''/PRTR).
The work of MGF was funded by Xunta de Galicia through the ``Programa de
axudas á etapa predoutoral da Xunta de Galicia'' (Consellería de Cultura, Educación
e Universidade) with reference code ED481A-2024-096. 
We also acknowledge the grants 2023-PG083 (with reference code ED431F 2023/19 funded by Xunta de Galicia),  PID2023-152148NB-I00 (funded by AEI-Spain), the Mar\'ia de Maeztu grant CEX2023-001318-M (funded by MICIU/AEI /10.13039/501100011033), the CIGUS Network of Research Centres, and the European Union.

\appendix

\section{Proof of the equivalence between the  deformed Hamiltonian and the undeformed one with twisted boundary conditions}
\label{ap:boundary-twist-ham}
In this appendix, we demonstrate the equivalence between the deformed Hamiltonian $\tilde{H}_{XXX}$~\eqref{eq:deformed-ham} and the undeformed one with twisted boundary conditions $\mathbb{H}_{XXX}$~\eqref{eq:ham-non-periodic}. In particular, we prove the following equation,
\begin{align}
    \tilde{H}_{XXX} = \Omega \mathbb{H}_{XXX} \Omega^{-1},\label{eq:equivalence-models}
\end{align}
with $\Omega$ defined in~\eqref{eq:global-intertwiner}. Let us first consider the identity~\eqref{eq:equivalence-models} involving the bulk terms, i.e the terms of the total Hamiltonian without the boundary hamiltonian density. That is to say, we want to show that
\begin{equation}
    \Tilde{h}_{m,m+1} = \Omega h_{m,m+1} \Omega^{-1}, \quad m = 1,\ldots,L-1. \label{eq:equivalence-bulk-app}
\end{equation}
In order to prove the last equation, we need to use the following factorisation properties of the Jordanian twist
\begin{align}
    \left(\Delta \otimes id\right)\left(F\right) &= F_{13} F_{23}, \\
    \left(id \otimes \Tilde{\Delta} \right)\left(F\right) &= F_{12} F_{13},
\end{align}
 that, together with the quasi-cocommutativity property of the $R$-matrix, allow one to write
\begin{align}
    R_{12} F_{13} F_{23} &= F_{23} F_{13} R_{12}, \label{eq:rff1} \\
    \Tilde{R}_{23} F_{12} F_{13} & = F_{13} F_{12} \Tilde{R}_{23}. \label{eq:rff2}
\end{align}

Now, notice that the terms $\left(F_{m+2,m+3}\cdots F_{m+2,L}\right)\cdots F_{L-1,L}$ in $\Omega$ commute with $h_{m,m+1}$ and thus can be cancelled with the corresponding inverse terms in $\Omega^{-1}$, therefore the right-hand side of~\eqref{eq:equivalence-bulk-app} can be recast as follows
\begin{align}
    \Omega h_{m,m+1} \Omega^{-1} &= \left(F_{12}\cdots F_{1L}\right)\cdots \left(F_{m+1,m+2}\cdots F_{m+1,L}\right) h_{m,m+1} \nonumber \\
    &\times \left(F_{m+1,L}^{-1} \cdots F_{m+1,m+2}^{-1}\right)\cdots\left(F_{1L}^{-1}\cdots F_{12}^{-1}\right).
\end{align}
Subsequently, we write explicitly the Hamiltonian density as the derivative of the $R$-matrix,  $h_{m,m+1} = P_{m,m+1}R'_{m,m+1}(0)$, where we use the notation $R'_{m,m+1}(0) =\left( \frac{\mathrm{d}}{\mathrm{d}u} R_{m,m+1}(u)\right)_{u=0}$. Then, using the identity $F_{ab}P_{bc} = P_{bc} F_{ac}$, we can move the permutation operator next to the $F_{m,m+1}$ term
\begin{align}
    \Omega h_{m,m+1} \Omega^{-1} = \cdots \left(F_{m-1,m}\cdots F_{m-1,L}\right) \Lambda \left(F_{m-1,L}^{-1} \cdots F_{m-1,m}^{-1} \right) \cdots,
\end{align}
with $\Lambda$ defined as follows
\begin{align}
    \Lambda=&F_{m,m+1} P_{m,m+1}\left(F_{m+1,m+2}\dots F_{m+1,L}\right)\left(F_{m,m+2}\cdots F_{mL}\right) \times \nonumber \\
    &\times R'_{m,m+1}(0) \left(F_{m+1,L}^{-1} \cdots F_{m+1,m+2}^{-1}\right) 
   \times \left(F_{mL}^{-1} \cdots F_{m,m+1}^{-1}\right).
\end{align}
Let us focus on the operator $\Lambda$. Notice that we can commute the twist $F_{m+1,L}$ with the $F$'s sitting to its right until we move it next to $F_{mL}$
\begin{align}
   \Lambda=&F_{m,m+1} P_{m,m+1}\left(F_{m+1,m+2}\dots F_{m+1,L-1}\right)\left(F_{m,m+2}\cdots F_{m,L-1}\right)F_{m+1,L} F_{mL}  \times \nonumber \\ 
   & \times R'_{m,m+1}(0) \left(F_{m+1,L}^{-1} \cdots F_{m+1,m+2}^{-1}\right) \left(F_{mL}^{-1} \cdots F_{m,m+1}^{-1}\right). 
\end{align}
Now, we use the relation~\eqref{eq:rff1}, so that
\begin{equation}
      F_{m+1,L} F_{mL} R'_{m,m+1}(0) = R'_{m,m+1}(0) F_{mL} F_{m+1,L},
\end{equation}
Then, we cancel the twist $F_{m+1,L}$ with its inverse, while $F_{mL}$ can be commuted until it cancels with $F_{mL}^{-1}$,
\begin{align}
   \Lambda=&F_{m,m+1} P_{m,m+1}\left(F_{m+1,m+2}\dots F_{m+1,L-1}\right)\left(F_{m,m+2}\cdots F_{m,L-1}\right) R'_{m,m+1}(0)  \times \nonumber \\
   & \times \left(F_{m+1,L-1}^{-1} \cdots F_{m+1,m+2}^{-1}\right) \left(F_{m,L-1}^{-1} \cdots F_{m,m+1}^{-1}\right). 
\end{align}
We can iterate this process until we are left with
\begin{align}
    \Lambda=F_{m,m+1} P_{m,m+1} R'_{m,m+1}(0) F_{m,m+1}^{-1} (0) =\Tilde{R}'_{m,m+1}(0) .
\end{align}
Therefore,
\begin{align}
    \Omega h_{m,m+1} \Omega^{-1} = \cdots\left(F_{m-1,m}\cdots F_{m-1,L}\right)  P_{m,m+1} \Tilde{R}'_{m,m+1}(0)\left(F_{m-1,L}^{-1}\cdots F_{m-1,m}^{-1}\right)\cdots.
\end{align}
Subsequently, we move $P_{m,m+1}$ to the far left, which yields
\begin{align}
    \Omega h_{m,m+1} \Omega^{-1} = P_{m,m+1} \Gamma,
\end{align}
with $\Gamma$ defined as
\begin{align}
    \Gamma =& \left(F_{12}\cdots F_{1,m+1}F_{1m}\cdots F_{1L}\right)\cdots \left(F_{m-1,m+1}F_{m-1,m}\cdots F_{m-1,L}\right) \times \nonumber \\
    &\times  \Tilde{R}'_{m,m+1}(0) \left(F_{m-1,L}^{-1}\cdots F_{m-1,m}^{-1}\right)\cdots\left(F_{1L}^{-1}\cdots F_{12}^{-1}\right).
\end{align}
We now simplify the expression for the operator $\Gamma$. Note that the terms
$F_{m-1,m+2}\cdots F_{m-1,L}$ commute with $\Tilde{R}_{m,m+1}'(0)$ and they can be cancelled with the corresponding inverse terms in $F_{m-1,L}^{-1}\cdots F_{m-1,m}^{-1}$. Therefore $\Gamma$ can be rewritten as 
\begin{align}
    \Gamma=&\left(F_{12}\cdots F_{1,m+1}F_{1m}\cdots F_{1L}\right)\cdots F_{m-1,m+1}F_{m-1,m}\Tilde{R}'_{m,m+1}(0)\times \nonumber\\
   &\times F_{m-1,m+1}^{-1}F_{m-1,m}^{-1} \left(F_{m-2,L}^{-1}\cdots F_{m-2,m-1}^{-1}\right) \cdots \left(F_{1L}^{-1}\cdots F_{12}^{-1}\right).
\end{align}
Then, using the factorisation property~\eqref{eq:rff2},
\begin{align}
    F_{m-1,m+1}F_{m-1,m}\Tilde{R}'_{m,m+1}(0) = \Tilde{R}'_{m,m+1}(0)F_{m-1,m}F_{m-1,m+1},
\end{align}
gives
\begin{align}
    \Gamma &= \left(F_{12}\cdots F_{1,m+1}F_{1,m}\cdots F_{1L}\right)\cdots \left(F_{m-2,m-1}F_{m-2,m+1}F_{m-2,m}\cdots \right. \nonumber \\
    &\left. \cdots F_{m-2,L}\right)  \Tilde{R}'_{m,m+1}(0) \left(F_{m-2,L}^{-1}\cdots F_{m-2,m-1}^{-1}\right) \cdots \left(F_{1L}^{-1}\cdots F_{12}^{-1}\right).
\end{align}
This process can be iteratively applied until we are left with
\begin{align}
    \Gamma = \Tilde{R}'_{m,m+1}(0).
\end{align}
Therefore,
\begin{align}
    \Omega h_{m,m+1} \Omega^{-1} = P_{m,m+1} \Tilde{R}'_{m,m+1}(0) = \tilde{h}_{m,m+1}
\end{align}
which proves the equation~\eqref{eq:equivalence-bulk-app}.

Once we prove the relation~\eqref{eq:equivalence-bulk-app} involving the bulk terms of the Hamiltonian, the boundary operator~\eqref{eq:twist-boundary} is automatically fixed, which leads to the identity~\eqref{eq:equivalence-models}. In fact,
\begin{align}
    \tilde{H}_{xxx} &= \Omega \sum_{m=1}^{L-1} h_{m,m+1}\Omega^{-1} + \tilde{h}_{L1} =\Omega\left(\sum_{m=1}^{L-1} h_{m,m+1}+\Omega^{-1} F_{L1} h_{L1} F_{L1}^{-1}\Omega \right)\Omega^{-1} =\nonumber \\
    &=\Omega \mathbb{H}_{XXX} \Omega^{-1}
\end{align}
which proves the equivalence between the Jordanian deformed $XXX_{-1/2}$ Hamiltonian and the undeformed one with twisted boundary conditions.
\section{Details on the deformed transfer matrix}
\label{ap:twisted-transfer-matrix}
In this appendix, we derived the identity~\eqref{eq:trasnfer-matrix}.
To do so, notice that the relation~\eqref{eq:rff1} implies the following expression for the inverse of the twist
\begin{equation}
    F_{13}^{-1} F_{23}^{-1} R_{12} = R_{12} F_{23}^{-1} F_{13}^{-1}. \label{eq:ffr}
\end{equation}
Moreover, the equation ~\eqref{eq:rff2} can be rewritten as follows
\begin{equation}
    \Tilde{R}_{23} F_{12} = F_{13} F_{12} \Tilde{R}_{23} F_{13}^{-1}. \label{eq:rf}
\end{equation}
With this, let us consider first the product of two twisted $R$-matrices,
\begin{align}
    \Tilde{R}_{a2} \Tilde{R}_{a1} = \Tilde{R}_{a2} F_{1a} R_{a1} F_{a1}^{-1}. 
\end{align}
The identity~\eqref{eq:rf} allows one to write
\begin{align}
    \Tilde{R}_{a2} \Tilde{R}_{a1} = F_{12} F_{1a} F_{2a} R_{a2} F_{a2}^{-1} F_{12}^{-1} R_{a1} F_{a1}^{-1},
\end{align}
which together with the relation~\eqref{eq:ffr} implies
\begin{align}
    \Tilde{R}_{a2} \Tilde{R}_{a1} = F_{12}  F_{1a} F_{2a} R_{a2} R_{a1} F_{12}^{-1} F_{a2}^{-1} F_{a1}^{-1}.
\end{align}
Suppose we add a third $R$-matrix in the multiplication. Then, by applying the identity~\eqref{eq:rf} two consecutive times, it follows
\begin{align}
    \Tilde{R}_{a3} F_{1a} F_{2a} R_{a2} R_{a1} = F_{13} F_{23} F_{1a} F_{2a} F_{3a} R_{a3} F_{a3}^{-1} F_{23}^{-1} F_{13}^{-1}  R_{a2} R_{a1}.
\end{align}
If we now apply~\eqref{eq:ffr} twice, we obtain
\begin{align}
    \Tilde{R}_{a3} F_{1a} F_{2a} R_{a2} R_{a1} = F_{13} F_{23} F_{1a} F_{2a} F_{3a} R_{a3} R_{a2} R_{a1} F_{13}^{-1} F_{23}^{-1} F_{a3}^{-1},
\end{align}
which implies,
\begin{align}
    \Tilde{R}_{a3} \Tilde{R}_{a2} \Tilde{R}_{a1} = F_{12} F_{13} F_{23} F_{1a} F_{2a} F_{3a} R_{a3} R_{a2} R_{a1} F_{23}^{-1} F_{13}^{-1}F_{12}^{-1}F_{a3}^{-1}F_{a2}^{-1} F_{a1}^{-1}.
\end{align}
We can iterate this procedure. Every time we add an operator $R_{aj}$ and move all the $R$-matrices together, we obtain a factor $\left(F_{1j}F_{2j}\cdots F_{j-1,j}\right)F_{ja}$ on the left and a factor $\left(F_{j-1,j}^{-1}\cdots F_{2j}^{-1}F_{1j}^{-1}\right)F_{aj}^{-1}$
on the right. Therefore,
\begin{align}
    \Tilde{T}_a =\Omega F_{a} T_{a} \Omega^{-1}\left(F_{a}\right)_{op}^{-1}, 
\end{align}
where
\begin{align}
    \Omega &= \left(F_{12}\cdots F_{1L}\right) \left(F_{23}\cdots F_{2L}\right)\cdots F_{L-1,L} = \nonumber \\
    &= \left(F_{12}\right) \left(F_{13}F_{23}\right)\cdots \left(F_{1L}\cdots F_{L-1,L}\right), \\
    F_a &=F_{1a}\cdots F_{La}, \\   \left(F_{a}\right)_{op}^{-1} &=F_{aL}^{-1}\cdots F_{a1}^{-1}. 
\end{align}

\section{Derivation of the dispersion relation and $S$-matrix for the $XXZ_{-1/2}$ spin-chain}
\label{ap:dispersion-relation-s-matrix-xxz}
In this appendix, we derive the expression for the dispersion relation and the S-matrix for the $XXZ_{-1/2}$ model in our conventions. To this end, we will need the action of~\eqref{eq:xxz-ham-density} on all two-site states of one and two excitations. We take the result from the generic action of all two-site states found in~\cite{Frassek:2019isa} with a normalisation factor of $q(1-q^2)$ (see the appendix~\ref{ap:action-xxzham_(1,0)-and-(0,1)}),
\begin{align}
    \mathsf{h}_{12} \ket{10} &= q \ket{10}-\ket{01}, \quad \mathsf{h}_{12} \ket{01} = \frac{1}{q}\ket{01}-\ket{10}, \nonumber\\
    \mathsf{h}_{12} \ket{11} &= \frac{1+q^2}{q} \ket{11}-\ket{20} - \ket{02}, \nonumber\\
    \mathsf{h}_{12} \ket{20} &= \frac{q+2q^3}{1+q^2} \ket{20}-\ket{11}-\frac{q}{1+q^2}\ket{02}, \nonumber\\
    \mathsf{h}_{12} \ket{02} &= \frac{2+q^2}{q+q^3} \ket{02}-\ket{11} -\frac{q}{1+q^2} \ket{20} .\label{eq:xxz-action-up-2}
\end{align}
To derive the form of the dispersion relation, let us consider first a state with one excitation of the form,
\begin{align}
    \ket{\Psi_{xxz}}_{(1)} = \sum_{1 \leq x \leq L} e^{i k x}|x) \quad \text{with} \quad e^{i k L} = 1.
\end{align}
Using the action~\eqref{eq:xxz-action-up-2} on the states $\ket{10}$ and $\ket{01}$, together with the periodicity condition $e^{i k L} = 1$, it follows
\begin{align}
    H_{XXZ} \ket{\Psi_{xxz}}_{(1)} &= \sum_{1 \leq x \leq L} e^{ikx} \left(\left(q+\frac{1}{q}\right) |x) - |x+1)-|x-1)\right) = \nonumber \\
    &\quad = 2\left(\cosh{\eta}-\cos{k}\right)\sum_{1 \leq x \leq L} e^{ikx}  |x)
\end{align}
where in the last step, we have used the parametrisation of $q$ in terms of the anisotropy parameter $q=e^{\eta}$. With this, we have that the dispersion relation $\epsilon_\eta(k)$ is given by 
\begin{align}
    \epsilon_\eta(k) = 2\left(\cosh{\eta}-\cos{k}\right). \label{eq:xxz-dispersion-relation}
\end{align}
If we introduce the pseudorapidity~\eqref{eq:xxz-momentum}, the energy~\eqref{eq:xxz-dispersion-relation} can be rewritten as follows 
\begin{align}
    \epsilon_\eta(\lambda) = 2\left(\frac{\cosh^2{\eta}-1}{\cosh{\eta}-\cosh{2\lambda}}\right).
\end{align}
In order to obtain the form of the $S$-matrix we consider the following state with two excitations
\begin{align}
    \ket{\Psi_{xxz}}_{(2)} = \sum_{1 \leq x_1 \leq x_2 \leq L} \Psi_{xxz}(x_1,x_2) |x_1,x_2), \label{eq:xxz-state2ex}
\end{align}
with
\begin{align}
    \Psi_{xxz}(x_1,x_2)=e^{i(k_1 x_1+k_2 x_2)} + S_{xxz}(k_2,k_1) e^{i(k_2 x_1 + k_1 x_2)}. \label{eq:xxz-wf-2ex}
\end{align}
Now we impose~\eqref{eq:xxz-state2ex} to be an eigenstate of $H_{XXZ}$. Using~\eqref{eq:xxz-action-up-2}, together with the periodicity condition $\Psi_{xxz}(x_1,x_2+L) = \Psi_{xxz}(x_2+L,x_1)$, gives the following equations
\begin{align}
    &\text{if} \quad x_1 = x_2 =x: \nonumber \\
    &\quad E_{xxz} \Psi_{xxz}(x,x) = \frac{2\left(1+q^2+q^4\right)}{q+q^3} \Psi_{xxz}(x,x)- \frac{q}{1+q^2}\Psi_{xxz}(x+1,x+1) \nonumber \\
    & \quad \quad - \frac{q}{1+q^2}\Psi_{xxz}(x-1,x-1)-\Psi_{xxz}(x-1,x) - \Psi_{xxz}(x,x+1). \label{eq:xxz-x1=x2}\\
    &\text{if} \quad x_1 < x_2: \nonumber \\
    &\quad E_{xxz} \Psi_{xxz}(x_1,x_2) = 2 \frac{1+q^2}{q} \Psi_{xxz}(x_1,x_2) - \Psi_{xxz}(x_1-1,x_2) \nonumber \\
    &\quad \quad - \Psi_{xxz}(x_1+1,x_2) - \Psi_{xxz}(x_1,x_2-1) - \Psi_{xxz}(x_1,x_2+1). \label{eq:xxz-x1<x2}
\end{align}
Substituting the wavefunction~\eqref{eq:xxz-wf-2ex} in the equation~\eqref{eq:xxz-x1<x2}, it is immediate to verify that the energy of~\eqref{eq:xxz-state2ex} is equal to the sum of the energy of two single excitations with momenta $k_1$ and $k_2$, 
\begin{align}
    E_{xxz} = \epsilon_\eta(k_1) + \epsilon_\eta(k_2). \label{eq:xxz-energy-2ex}
\end{align}
On the other hand, from the equation~\eqref{eq:xxz-x1=x2} we can obtain the $S$-matrix. Indeed, substituting back the wavefunction~\eqref{eq:xxz-wf-2ex} and the energy~\eqref{eq:xxz-energy-2ex}  gives
\begin{align}
    C(k_1,k_2) + C(k_2,k_1) S_{xxz}(k_2,k_1) = 0,
\end{align}
where we have defined
\begin{align}
    C(k_1,k_2) &= \frac{q \left(1+e^{-i(k_1+k_2)}\right) }{1+q^2} \left(1+ e^{i (k_1 + k_2)} - 2\cosh{\eta} e^{i k_1}\right),\label{eq:def-C}
\end{align}
With this, the $S$-matrix takes the form
\begin{align}
    S_{xxz}(k_1,k_2) = - \frac{1+ e^{i (k_1 + k_2)} - 2\cosh{\eta}\, e^{i k_2}}{1+ e^{i (k_1 + k_2)} - 2\cosh{\eta}\, e^{i k_1}}. \label{eq:xxz-S-matrix}
\end{align}
If we introduce the pseudorapidity~\eqref{eq:xxz-momentum}, the S-matrix~\eqref{eq:xxz-S-matrix} can we rewritten as follows
\begin{align}
    S_{xxz}(\lambda_1,\lambda_2) &= \frac{\sinh{(\lambda_1-\lambda_2-\eta)}}{\sinh{(\lambda_1-\lambda_2+\eta)}}.\label{eq:xxz-S-matrix-rap}
\end{align}
If we compare the dispersion relation~\eqref{eq:xxz-dispersion-relation} and the $S$-matrix~\eqref{eq:xxz-S-matrix} with that of the spin-chain $XXZ_{1/2}$ in the $1/2$ representation (see e.g. \cite{Lamers:2015dfa,Faddeev:1996iy}), we see that the energy for the two representations coincides, while the $S$-matrices are related by the interchange of the momenta $k_1$, $k_2$ in~\eqref{eq:xxz-S-matrix} or by sending $\eta\to-\eta$ in~\eqref{eq:xxz-S-matrix-rap} while keeping the rapidities fixed. In other words, the S-matrix in the $-1/2$ representation is the inverse of the $S$-matrix in the $1/2$ representation.

\section{Some details on the Hamiltonian density of $XXZ_{-1/2}$}
\label{ap:action-xxzham_(1,0)-and-(0,1)}
In this appendix, we compute the action of the $XXZ_{-1/2}$ Hamiltonian density~\eqref{eq:xxz-ham-density} on the two-site states $\ket{10}$ and $\ket{01}$, to prove equations~\eqref{eq: action on (1,0)} and~\eqref{eq:action on (0,1)}.  
To this end, we need to write $\ket{10}$ and $\ket{01}$ as a linear combination of states belonging to the modules $D_{j+1}$ in~\eqref{eq:irreducible-decomposition-xxz}. The form of their lowest-weight states $\ket{\chi_{j}}$ was proven in~\cite{Frassek:2019isa}. In particular, up to some normalisation constant, the first two lowest-weight states takes the form
\begin{align}
    \ket{\chi_0} &= \ket{00}, \\
    \ket{\chi_1} &= q \ket{10}-\ket{01}.
\end{align}
Moreover, from equation~\eqref{eq:xxz_cop_s+} it follows that the action of $\Delta_q(S^+)$ on $\ket{\chi_0}$ is given by
\begin{align}
    \Delta_q(S^+)\ket{\chi_0} = q^{-1/2} \ket{10} + q^{1/2} \ket{01}.
\end{align}
With this, we can recast $\ket{10}$ and $\ket{01}$ as the following linear combinations,
\begin{align}
    \ket{10}&=\frac{1}{q+q^{-1}} \left(q^{-1/2}\Delta_q(S^+)\ket{\chi_0} + \ket{\chi_1}\right),\\
    \ket{01}&=\frac{1}{q^2+1} \left(q^{3/2}\Delta_q(S^+)\ket{\chi_0}-\ket{\chi_1}\right).
\end{align}
Since the eigenvalue of the Hamiltonian on the vacuum is zero, it is clear that the Hamiltonian density~\eqref{eq:xxz-ham-density} acting on $\ket{10}$ and $\ket{01}$ will project onto the state $\ket{\chi_1}$,
\begin{align}
    \mathsf{h}_{12} \ket{10} &= \ket{\chi_1} = q \ket{10}-\ket{01}, \\
     \mathsf{h}_{12} \ket{01} &= -\frac{1}{q}\ket{\chi_1} = \frac{1}{q}\ket{01}-\ket{10}.
\end{align}
The result derived in this appendix coincides with the action on a generic two-site state given in~\cite{Frassek:2019isa}, provided we multiply their result by the normalisation factor $q(1-q^2)$.

\section{Proof of the equations~\eqref{eq:null-sum-wf-1} and~\eqref{eq:null-sum-wf-2}}
\label{ap:Null-sum-identities}
In this appendix, we prove the identities~\eqref{eq:null-sum-wf-1} and~\eqref{eq:null-sum-wf-2}. Let us consider first the equation~\eqref{eq:null-sum-wf-1}. If $k_2=0$ (the same argument works for $k_1=0$), we have 
\begin{align}
    \sum_{1 \leq x_1 \leq x_2 \leq L} \Psi_{xxx}(x_1,x_2) &=  \sum_{1 \leq x_1 \leq x_2 \leq L} \left(e^{i k_1 x_1} + e^{i k_1 x_2}\right) = \nonumber \\
    &= (L-1)\sum_{1 \leq x_1 \leq L}e^{i k_1 x_1} = 0  \quad \text{with} \quad e^{ik_1 L} =1,
\end{align}
where in the last equality we have performed the geometric sum in $x_1$ as in~\eqref{eq:gemetric-sum-wf1}. 

Suppose now that neither of the two momenta, $k_1$ nor $k_2$ is zero. Then, performing the geometric sum in $x_1$ and $x_2$ yields
\begin{align}
     \sum_{1 \leq x_1 \leq x_2 \leq L} \Psi_{xxx}(x_1,x_2) &=\frac{G(k_1,k_2) + S_{xxx}(k_2,k_1) G(k_2,k_1)}{\left(e^{i k_1}-1\right)\left(e^{i k_2}-1\right)\left(e^{i (k_1+k_2)}-1\right)} , \label{eq:expansion-null-sum-wf1}
\end{align}
where we have defined
\begin{align}
    G(k_1,k_2) &= e^{i(2k_1 + k_2)}\left(1-e^{i(k_1+k_2)L}\right) + e^{i (k_1 + k_2 + k_2 L)} - e^{i (2 k_1 + k_2 (2 + L))} \nonumber \\
    &\quad - e^{i (k_1 + k_2)}+e^{i (k_1 + k_2) (2 + L)}. \label{eq:definition-of-G}
\end{align}
For $k_1$ and $k_2$ different from zero the denominator of~\eqref{eq:expansion-null-sum-wf1} never vanishes \footnote{Remember that in our convention the momenta take values in the range $[0,2\pi)$.}. Therefore, in order to prove~\eqref{eq:null-sum-wf-1} we must verify that
\begin{align}
    G(k_1,k_2) + S_{xxx}(k_2,k_1) G(k_2,k_1) = 0
\end{align}
From the undeformed $XXX_{-1/2}$ Bethe equations
\begin{align}
    e^{ik_1 L} = S_{xxx}(k_1,k_2) \quad \text{and} \quad e^{ik_2L} = S_{xxx}(k_2,k_1) \label{eq:Bethe-eqs-2ex}
\end{align}
we can simplify~\eqref{eq:definition-of-G} as follows
\begin{align}
    G(k_1,k_2) = e^{2i(k_1+k_2)}-e^{i(k_1+k_2)} + S_{xxx}(k_2,k_1) \left(e^{i(k_1+k_2)} - e^{2i(k_1+k_2)}\right).
\end{align}
With this, using the identity $S_{xxx}(k_1,k_2) S_{xxx}(k_2,k_1) = 1$, it follows that
\begin{align}
   S_{xxx}(k_2,k_1) G(k_2,k_1) &=  e^{i(k_1+k_2)} - e^{2i(k_1+k_2)}+ S_{xxx}(k_2,k_1)\left(e^{2i(k_1+k_2)}-e^{i(k_1+k_2)}\right) = \nonumber \\
   &=-G(k_1,k_2),
\end{align}
which proves~\eqref{eq:null-sum-wf-1}.

Let us now prove identity~\eqref{eq:null-sum-wf-2}. Performing the geometric sum of of $x_2$ in the definition of $\Phi(x_1)$~\eqref{eq:wf-state-1/e}, we obtain
\begin{align}
     \Phi(x_1)&=\frac{D(k_1,k_2) + S_{xxx}(k_2,k_1) D(k_2,k_1)}{\left(e^{i k_1}-1\right)\left(e^{i k_2}-1\right)}, \label{eq:expansion-null-sum-wf2}
\end{align}
where we have defined
\begin{align}
    D(k_1,k_2) &= e^{i (k_1 + k_2) x_1} + e^{i (k_1 + k_2) (x_1+1)} + e^{i (k_1 (1+x_1)+ k_2 (1+L)) } + e^{i (k_1 + k_2 x_1)} \nonumber \\
    & \quad -e^{i (k_1 x_1 + k_2 (1+L))}-e^{i (k_1 + k_2(1 + x_1))} - 2  e^{i (k_1(1 + x_1) + k_2 x_1)} .\label{eq:definition-of-D}
\end{align}
For $k_1$ and $k_2$ non-zero, the denominator of~\eqref{eq:expansion-null-sum-wf2} does not vanish. Therefore, it is enough to prove that the numerator of~\eqref{eq:expansion-null-sum-wf2} is zero. Using~\eqref{eq:Bethe-eqs-2ex}, one can rewrite~\eqref{eq:definition-of-D} as follows
\begin{align}
    D(k_1,k_2) = e^{i (k_1 + k_2) x_1} + e^{i (k_1 + k_2) (x_1+1)} + e^{i (k_1 + k_2 x_1)}-e^{i (k_1 + k_2(1 + x_1))} \nonumber \\
    - 2  e^{i (k_1(1 + x_1) + k_2 x_1)} + S_{xxx}(k_2,k_1) \left(e^{i (k_1 (1+x_1)+ k_2 )}-e^{i (k_1 x_1 + k_2 )}\right).
\end{align}
Moreover, using $S_{xxx}(k_1,k_2) S_{xxx}(k_2,k_1) = 1$, we have
\begin{align}
     D(k_1,k_2) &+ S_{xxx}(k_2,k_1) D(k_2,k_1) = e^{i (k_1 + k_2) x_1}\left(1+e^{i(k_1+k_2)}-2e^{ik_1}\right) + \nonumber \\
    &+ e^{i (k_1 + k_2) x_1}S_{xxx}(k_2,k_1) \left(1+e^{i(k_1+k_2)}-2e^{ik_2}\right) = 0. 
\end{align}
where in the last step we have used the definition of the $XXX_{-1/2}$ S-matrix~\eqref{eq:undef-solution}. This proves the equation~\eqref{eq:null-sum-wf-2}.

\bibliographystyle{nb}
\bibliography{biblio}{}

\end{document}